\documentclass[aps,prx,twocolumn,notitlepage]{revtex4-2} 

\usepackage[T1]{fontenc}
\usepackage{amsmath,amsfonts,amssymb,amsthm,bbm,bm, braket}
\usepackage{wasysym}
\usepackage{comment}
\usepackage{scalerel}
\usepackage{enumerate}
\usepackage{graphicx}
\usepackage{mathtools}
\usepackage{ifthen}
\usepackage{tensor}
\usepackage{tikz}
\usepackage{tikz-network}
\usetikzlibrary{patterns,decorations.pathreplacing}

\theoremstyle{break}

\usepackage{color}
\definecolor{myred}{RGB}{232,102,102}
\definecolor{myblue}{RGB}{187,187,255}
\definecolor{myorange}{RGB}{255,165,0}
\definecolor{mygrey}{RGB}{105,105,105}
\definecolor{OliveGreen}{RGB}{85,107,47}
\definecolor{NavyBlue}{RGB}{0,0,128}
\definecolor{mygreen}{RGB}{34,139,34}
\definecolor{myY}{RGB}{220,255,203}
\definecolor{myYO}{RGB}{255, 220, 151}
\usepackage{tensor}
\newcommand{\be}{\begin{equation}}
\newcommand{\ee}{\end{equation}}
\newcommand{\ba}{\begin{aligned}}
\newcommand{\ea}{\end{aligned}}
\newcommand{\bw}{\begin{widetext}}
\newcommand{\ew}{\end{widetext}}
\newcommand{\1}{\mathbbm{1}}

\theoremstyle{plain}
\newtheorem{property}{Property}
\theoremstyle{plain}

\usepackage[colorlinks,bookmarks=false,citecolor=NavyBlue,linkcolor=OliveGreen,urlcolor=blue]{hyperref}

\hypersetup{
pdftitle={Correlations in Perturbed Dual-Unitary Circuits: Efficient Path-Integral Formula},
pdfsubject={Statistical mechanics},
pdfauthor={Pavel Kos, Bruno Bertini, and Tomaz Prosen},
pdfkeywords={correlations, dual-unitary circuits, quantum many-body systems}
}

\newcommand{\mcirc}{\mathbin{\scalerel*{\fullmoon}{G}}}

\newcommand{\mcircf}{\mathbin{\scalerel*{\newmoon}{G}}}

\newcommand{\du}{{\rm du}}

\newcommand{\Wgateorange}[2]{
\draw[very thick] (#1-0.5, #2 +0.5) -- (#1+0.5,#2-0.5);
\draw[very thick] (#1-0.5,#2-0.5) -- (#1+0.5,#2+0.5);
\draw[thick, fill=myorange, rounded corners=2pt] (#1-0.25,#2+0.25) rectangle (#1+0.25,#2-0.25);
\draw[thick] (#1,#2+0.15) -- (#1+0.15,#2+0.15) -- (#1+0.15,#2);
}
\newcommand{\Wgategreen}[2]{
\draw[very thick] (#1-0.5, #2 +0.5) -- (#1+0.5,#2-0.5);
\draw[very thick] (#1-0.5,#2-0.5) -- (#1+0.5,#2+0.5);
\draw[ thick, fill=mygreen, rounded corners=2pt] (#1-0.25,#2+0.25) rectangle (#1+0.25,#2-0.25);
\draw[thick] (#1,#2+0.15) -- (#1+0.15,#2+0.15) -- (#1+0.15,#2);
}

\newcommand{\lineW}{.8mm}

\newcommand{\oGate}[2]{
\draw[thin, fill=myY, rounded corners=0pt] (#1-0.5,#2+0.5) rectangle (#1+0.5,#2-0.5);}

\newcommand{\aGate}[2]{
\draw[thin, fill=myY, rounded corners=0pt] (#1-0.5,#2+0.5) rectangle (#1+0.5,#2-0.5);
\draw[line width=\lineW] (#1-0.5,#2) -- (#1+0.5,#2);}

\newcommand{\cGate}[2]{
\draw[thin, fill=myY, rounded corners=0pt] (#1-0.5,#2+0.5) rectangle (#1+0.5,#2-0.5);
\draw[line width=\lineW] (#1,#2-0.5) -- (#1,#2+0.5);}

\newcommand{\gGate}[2]{
\draw[thin, fill=myY, rounded corners=0pt] (#1-0.5,#2+0.5) rectangle (#1+0.5,#2-0.5);
\draw[line width=\lineW] (#1,#2-0.5) -- (#1,#2+0.5);
\draw[line width=\lineW] (#1-0.5,#2) -- (#1+0.5,#2);}

\newcommand{\bGate}[2]{
\draw[thin, fill=myY, rounded corners=0pt] (#1-0.5,#2+0.5) rectangle (#1+0.5,#2-0.5);
\draw[line width=\lineW] (#1,#2-0.5) -- (#1,#2);
\draw[line width=\lineW] (#1-0.5,#2) -- (#1+0.5,#2);}

\newcommand{\dGate}[2]{
\draw[thin, fill=myY, rounded corners=0pt] (#1-0.5,#2+0.5) rectangle (#1+0.5,#2-0.5);
\draw[line width=\lineW] (#1,#2-0.5) -- (#1,#2+0.5);
\draw[line width=\lineW] (#1-0.5,#2) -- (#1,#2);}

\newcommand{\fGate}[2]{
\draw[thin, fill=myY, rounded corners=0pt] (#1-0.5,#2+0.5) rectangle (#1+0.5,#2-0.5);
\draw[line width=\lineW] (#1,#2) -- (#1,#2+0.5);
\draw[line width=\lineW] (#1-0.5,#2) -- (#1+0.5,#2);}

\newcommand{\eGate}[2]{
\draw[thin, fill=myY, rounded corners=0pt] (#1-0.5,#2+0.5) rectangle (#1+0.5,#2-0.5);
\draw[line width=\lineW] (#1,#2-0.5) -- (#1,#2+0.5);
\draw[line width=\lineW] (#1,#2) -- (#1+0.5,#2);}

\newcommand{\doGate}[2]{
\draw[thin, fill=myY, rounded corners=0pt] (#1-0.5,#2+0.5) rectangle (#1+0.5,#2-0.5);
\draw[line width=\lineW] (#1,#2-0.5) -- (#1,#2) -- (#1 +0.5,#2);}

\newcommand{\dtGate}[2]{
\draw[thin, fill=myY, rounded corners=0pt] (#1-0.5,#2+0.5) rectangle (#1+0.5,#2-0.5);
\draw[line width=\lineW] (#1-0.5,#2) -- (#1,#2) -- (#1 ,#2+0.5);}

\newcommand{\sGate}[4]{
\draw[thin, fill=myY, rounded corners=0pt] (#1-0.5,#2+0.5) rectangle (#1+0.5,#2-0.5);
\Text[x=#1, y=#2, , anchor = center] {$  s_{#4,#3}$};
}

\newcommand{\oGateO}[2]{
\draw[thin, fill=myYO, rounded corners=0pt] (#1-0.5,#2+0.5) rectangle (#1+0.5,#2-0.5);}

\newcommand{\aGateO}[2]{
\draw[thin, fill=myYO, rounded corners=0pt] (#1-0.5,#2+0.5) rectangle (#1+0.5,#2-0.5);
\draw[line width=\lineW] (#1-0.5,#2) -- (#1+0.5,#2);}

\newcommand{\cGateO}[2]{
\draw[thin, fill=myYO, rounded corners=0pt] (#1-0.5,#2+0.5) rectangle (#1+0.5,#2-0.5);
\draw[line width=\lineW] (#1,#2-0.5) -- (#1,#2+0.5);}

\newcommand{\bGateO}[2]{
\draw[thin, fill=myYO, rounded corners=0pt] (#1-0.5,#2+0.5) rectangle (#1+0.5,#2-0.5);
\draw[line width=\lineW] (#1,#2-0.5) -- (#1,#2);
\draw[line width=\lineW] (#1-0.5,#2) -- (#1+0.5,#2);}

\newcommand{\dGateO}[2]{
\draw[thin, fill=myYO, rounded corners=0pt] (#1-0.5,#2+0.5) rectangle (#1+0.5,#2-0.5);
\draw[line width=\lineW] (#1,#2-0.5) -- (#1,#2+0.5);
\draw[line width=\lineW] (#1-0.5,#2) -- (#1,#2);}

\newcommand{\fGateO}[2]{
\draw[thin, fill=myYO, rounded corners=0pt] (#1-0.5,#2+0.5) rectangle (#1+0.5,#2-0.5);
\draw[line width=\lineW] (#1,#2) -- (#1,#2+0.5);
\draw[line width=\lineW] (#1-0.5,#2) -- (#1+0.5,#2);}

\newcommand{\eGateO}[2]{
\draw[thin, fill=myYO, rounded corners=0pt] (#1-0.5,#2+0.5) rectangle (#1+0.5,#2-0.5);
\draw[line width=\lineW] (#1,#2-0.5) -- (#1,#2+0.5);
\draw[line width=\lineW] (#1,#2) -- (#1+0.5,#2);}

\begin{document}

\title{Correlations in Perturbed Dual-Unitary Circuits: Efficient Path-Integral Formula}
\author{Pavel Kos, Bruno Bertini, and Toma\v z Prosen}

\affiliation{Department of Physics, Faculty of Mathematics and Physics, University of Ljubljana, Jadranska 19, SI-1000 Ljubljana, Slovenia}

\begin{abstract}
Interacting many-body systems with explicitly accessible spatio-temporal correlation functions are extremely rare, especially in the absence of Bethe-ansatz or Yang-Baxter integrability. Recently, we identified a remarkable class of such systems and termed them dual-unitary quantum circuits. These are brickwork type local quantum circuits whose dynamics are unitary in both time and space. The spatio-temporal correlation functions of these systems turn out to be non-trivial only at the edge of the causal light cone and can be computed in terms of one-dimensional transfer matrices. Dual-unitarity, however, requires fine-tuning and the degree of generality of the observed dynamical features remained unclear. Here we address this question by studying perturbed dual-unitary quantum circuits. 
Considering arbitrary perturbations of the local gates, we prove that for a particular class of unperturbed elementary dual-unitary gates the correlation functions are still expressed in terms of one-dimensional transfer matrices. These matrices, however, are now contracted over generic paths connecting the origin to a fixed end point inside the causal light cone. The correlation function is given as a sum over all such paths. Our statement is rigorous in the ``dilute limit'', where only a small fraction of the gates is perturbed, and in the presence of random longitudinal fields, but we provide theoretical arguments and stringent numerical checks supporting its validity even in the clean case (no randomness) and when all gates are perturbed. As a byproduct of our analysis, in the case of random longitudinal fields --- which turns out to be equivalent to certain classical Markov chains --- we find four types of non-dual-unitary (and non-integrable) interacting many-body systems where the correlation functions are exactly solvable and given --- without approximations --- by the path-sum formula. 
\end{abstract}

\date{\today}

\maketitle

\tableofcontents

\section{Introduction}
Understanding the dynamics of extended quantum many-body systems with local interactions is the core problem of nonequilibrium statistical mechanics, with a wide range of applications
ranging from condensed matter physics to high energy theory and quantum gravity. In particular, the set of two-point spatio-temporal correlation functions of local observables can be considered as the prime quantifier of the dynamics. For example, they can be used in the framework of linear response theory~\cite{altland,mahanbook} to express coefficients, such as conductivities and kinematic viscosities, that describe macroscopic transport properties. 

A major obstacle is that computing dynamical correlation functions in interacting systems is notoriously hard.
This is true for numerical simulations of correlations in real time, which are typically exponentially hard (in physical simulation time) \cite{Daley,Karrasch} (see also \cite{NumericalOperatorEvolution}), and even more so for analytical computations.
 Even in Bethe-ansatz integrable systems the task turns out to be daunting: while recent breakthroughs allowed for the calculation of dynamical correlations in the late-time, hydrodynamic, regime~\cite{JSTAT16, DoyonLectures, TransportReview}, the computation of two-point correlations at arbitrary intermediate times remained out of reach. This leaves us with non-interacting, quasi-free, or Gaussian theories, as the only general class of systems where dynamical correlations can be analytically computed in general.
Dynamical correlations in coupled (interacting) theories are then usually formulated in terms of (Keldysh) diagrammatic many-body perturbation theory \cite{altland}. In these approaches correlations are written as power series in the coupling constant around the underlaying free theory. Such perturbative series, however, generically have a vanishing radius of convergence --- especially in the thermodynamic limit --- and their relevance for determining the actual physical behaviour of correlation functions in extended systems is questionable. One typically finds an even qualitative change in the behaviour of correlations and transport coefficients at finite temperatures when going from free to strongly coupled theories. For example, while free theories usually behave as ballistic conductors, strongly coupled ones are expected to display diffusive transport, which is thus a non-perturbative effect. In fact, establishing the microscopic origin of diffusion remains one of the main challenges of statistical physics \cite{Lebowitz,Dhar}.

Very recently, we proposed a new class of locally interacting many-body systems, which allows for an exact computation of spatio-temporal correlation functions of local observables \cite{BKP:dualunitary} and of some other indicators of quantum chaos and scrambling of quantum information~\cite{BKP:kickedIsing, PBCP20,BKP:OEergodicandmixing,BKP:OEsolitons,BP:tripartite,GoLa19, ClLa20, BKP:entropy, Aravinda, G:DUcorrelations}. These systems are special brickwork type quantum circuits (see, e.g., Ref.~\cite{Keyserlingk, Nahum:operatorspreadingRU, RandomCircuitsEnt}) called ``dual-unitary'' quantum circuits and are characterised by unitary evolution, not only in the direction of time propagation, but also in the orthogonal direction of space propagation. Owing to this property, dual-unitary systems allow for a conceptual exchange of space and time axes and, in a vague sense, are discrete-space-time analogues of conformal field theories. In particular, the correlations in these systems can propagate only at the maximal speed and along one-dimensional straight paths in the space-time. 
One may think of these models as being `statistically exactly solvable', and yet, their dynamics is generically ergodic \cite{BKP:dualunitary} and quantum chaotic, both from the point of view of spectral statistics \cite{BKP:kickedIsing} and dynamical complexity~\cite{BKP:OEergodicandmixing, BP:tripartite}. This is similar to the situation in classical dynamical systems~\cite{cornfeld}, where correlation functions of certain strongly chaotic single-particle models, such as e.g. Arnold cat maps or Baker maps, are exactly computable \cite{arnold,ott} despite their deterministic trajectories being unpredictable and ``incomputable'' in the algorithmic complexity sense.
Similarly, for generic quantum chaotic dual-unitary dynamics, the Heisenberg evolution of local operators is exponentially hard to simulate (classically) \cite{BKP:OEergodicandmixing,BP:tripartite} even though the two-point functions at infinite temperature are computable exactly (i.e., with vanishing algorithmic complexity).

Dual unitarity, however, requires fine tuning of coupling parameters in the elementary two-body gates defining the systems and it is thus clearly not stable under generic perturbations. Nonetheless, one might wonder whether the properties of ergodicity and quantum chaos of dual-unitary circuits are {\em structurally stable} in a similar way as in classical ergodic theory. There, one can prove that linear automorphisms on the tori (Arnold cat maps) are topologically (structurally) stable under perturbations \cite{arnold2}. A quantitative measure of such stability may be a uniform continuity of two-point time correlation functions with respect to the perturbation parameter.

While one may imagine integrable and free extended quantum many-body systems as isolated points or low dimensional sub-manifolds (clearly structurally unstable in the above sense) in some high dimensional manifold of all appropriate (say, translationally invariant, locally interacting, etc.) systems, the chaotic and ergodic models may represent a finite domain with positive measure. As a matter of fact, the prevailing opinion of the scientific community is that almost all models are ergodic, depending on the precise definition of the ergodic class. 
 
In this paper, we make the first step in addressing the question of structural stability by studying correlation functions in perturbed dual-unitary circuits. We consider generic perturbations on two different types of dual-unitary circuits: (i) noisy circuits on qubits: there are random local longitudinal fields at each space-time point over which we average; (ii) clean circuits on qudits.
In the case (i), we rigorously show that there exists a class of generic (not fine-tuned) dual-unitary circuits that is ``stable'' under perturbations. More precisely, when these systems are perturbed, correlations continue to propagate along one-dimensional paths in the space-time,  though, these paths do not need to be straight anymore. As a result, all points of the causal light cone acquire non trivial correlations. 
To find the correlations between any two points one has to sum the contribution of all ``skeleton diagrams'', i.e., all allowed space-time paths connecting them. Although our rigorous result holds only in the limit of vanishing density of perturbed gates (or, at fixed, but arbitrary large order in perturbation theory), 
we provide further theoretical arguments and numerical evidence for its validity even when \emph{all} gates are perturbed by a small enough perturbation. Moreover, in this setting we discover four new types of circuits where the two-point correlation functions are exactly (without approximations) given by the sum of skeleton diagrams. Finally, we formulate two simple analytical conditions for identifying the stable subclass of dual-unitary circuits also in the clean case (ii). Considering qubit gates we show numerically that  the stable subclass exists and explicitly demonstrate that, for sufficiently small perturbation strength, the sum of skeleton diagrams approximates exact two-point correlation functions with arbitrary precision. Finally, we contrast these findings with the case of \emph{random perturbations}, where the light cone structure of dual-unitary dynamical correlations \cite{BKP:dualunitary} is preserved (stable) and the only effect of the perturbations is an additional exponential damping along light rays. 

Alongside the dynamics of unitary quantum circuits we also discuss dynamics of a set of classical bistochastic many-body Markov chains, which can be written as local ``Markov circuits''. Indeed --- as we show in the paper --- the formal treatment of these two classes of systems is completely analogous. In particular, in Markov circuits the property of ``dual-unitarity'' is replaced by that of ``dual-bistochasticity'', meaning that the circuit is bistochastic when propagating in both space and time directions. Remarkably, the average over random longitudinal fields discussed at point (i) above maps quantum unitary circuits into Markov circuits. This means that our aforementioned rigorous results find a direct application also in the case of interacting classical stochastic dynamics.

The rest of the paper is structured as follows. In Sec.~\ref{sec:techintro} we describe the technical setting of the problem and introduce the necessary diagrammatic notation, while in Sec.~\ref{sec:perturbativeexpansion} we present the basic strategy and the main results of the paper. In 
Sec.~\ref{sec:U(1)avg} we consider fixed perturbations in noisy or classical Markov systems. The validity of our conclusions for clean (non-random) perturbed dual-unitary circuits is demonstrated Sec.~\ref{sec:genericnumerics} by combining analytical arguments with numerical results. Finally, Sec.~\ref{sec:conclusions} contains our conclusions. Some further discussions, technical details, and proofs are relegated to the appendices.

\section{Setting of the Problem}
\label{sec:techintro}
We consider one-dimensional many-body systems composed of $2L$ qudits, each of them with $d$ internal states. In other words, we examine a chain where at each site there is a quantum system with Hilbert space $\mathcal H \simeq \mathbb C^d$ and denote an orthonormal basis of $\mathcal H$ by  
\be
\label{eq:localbasis}
\mathcal B = \{\ket{j},\quad j=0,\ldots,d-1\}\, .
\ee
The time evolution in this system is generated by discrete applications of ``Floquet operators'' $\mathbb U$ of the form    
\begin{align}
&\mathbb U=\bigotimes_{x\in \mathbb Z_L} U_{x,x+1/2} \bigotimes_{x\in \mathbb Z_L+1/2} U_{x,x+1/2}\notag\\
& =
\begin{tikzpicture}[baseline=(current  bounding  box.center), scale=0.55]
\foreach \i in{1.5}
{
\draw[thick] (0.5,2*\i-0.5-3.5) arc (45:-90:0.15);
\draw[thick] (-10+0.5+0,2*\i-0.5-3.5) arc (-135:-270:0.15);
}
\foreach \i in {0,1}
{
\Text[x=1.25,y=-2+2*\i]{\scriptsize$\i$}
}
\foreach \i in {1}
{
\Text[x=1.25,y=-2+\i]{\small$\frac{\i}{2}$}
}
\foreach \i in {1,3,5}
{
\Text[x=-7.5+\i+1,y=-2.6]{\small$\frac{\i}{2}$}
}
\foreach \i in {1}
{
\Text[x=-7.8-2*\i+2,y=-2.6]{\small$\,-\frac{\i}{2}$}
}
\foreach \i in {0,1,2}
{
\Text[x=-7.5+2*\i+1,y=-2.6]{\scriptsize${\i}$}
}
\foreach \i in {3}
{
\Text[x=-7.5+2*\i+1,y=-2.6]{\scriptsize${\i}$}
}
\Text[x=-8.8,y=-2.6]{$\cdots$}
\Text[x=.5,y=-2.6]{$\cdots$}
\foreach \i in {3,...,13}
{
\draw[thick, gray, dashed] (-12.5+\i,-2.1) -- (-12.5+\i,0.3);
}
\foreach \i in {3,...,5}
{
\draw[thick, gray, dashed] (-9.75,3-\i) -- (.75,3-\i);
}
\foreach \i in {0,...,4}
{
\draw[thick] (-.5-2*\i,-1) -- (0.525-2*\i,0.025);
\draw[thick] (-0.525-2*\i,0.025) -- (0.5-2*\i,-1);
\draw[thick, fill=myred, rounded corners=2pt] (-0.25-2*\i,-0.25) rectangle (.25-2*\i,-0.75);
\draw[thick] (-2*\i,-0.35) -- (-2*\i+0.15,-0.35) -- (-2*\i+0.15,-0.5);%
}
\foreach \jj[evaluate=\jj as \j using -2*(ceil(\jj/2)-\jj/2)] in {0}
\foreach \i in {1,...,5}
{
\draw[thick] (.5-2*\i-1*\j,-2-1*\jj) -- (1-2*\i-1*\j,-1.5-\jj);
\draw[thick] (1-2*\i-1*\j,-1.5-1*\jj) -- (1.5-2*\i-1*\j,-2-\jj);
\draw[thick] (.5-2*\i-1*\j,-1-1*\jj) -- (1-2*\i-1*\j,-1.5-\jj);
\draw[thick] (1-2*\i-1*\j,-1.5-1*\jj) -- (1.5-2*\i-1*\j,-1-\jj);
\draw[thick, fill=myred, rounded corners=2pt] (0.75-2*\i-1*\j,-1.75-\jj) rectangle (1.25-2*\i-1*\j,-1.25-\jj);
\draw[thick] (-2*\i+1,-1.35-\jj) -- (-2*\i+1.15,-1.35-\jj) -- (-2*\i+1.15,-1.5-\jj);%
}
\Text[x=1.8,y=-1]{.}
\Text[x=-8.8,y=-2.6]{$\cdots$}
\Text[x=.5,y=-2.6]{$\cdots$}
\end{tikzpicture}
\end{align}
Here we considered periodic boundary conditions, we assumed translational invariance, and we represented diagrammatically the unitary operator $U_{x,x+1/2}=U$ (``local gate''), $x\in\{-{L}/{2},\dots,-{1}/{2},0,{1}/{2},1,\ldots, {(L-1)}/{2}\}$, as
\be
\label{eq:Ugate}
U=\begin{tikzpicture}[baseline=(current  bounding  box.center), scale=1]
\draw[ thick] (-4.25,0.5) -- (-3.25,-0.5);
\draw[ thick] (-4.25,-0.5) -- (-3.25,0.5);
\draw[ thick, fill=myred, rounded corners=2pt] (-4,0.25) rectangle (-3.5,-0.25);
\draw[thick] (-3.75,0.15) -- (-3.75+0.15,0.15) -- (-3.75+0.15,0);
\Text[x=-4.25,y=-0.75]{}
\end{tikzpicture}.
\ee
Finally, we adopted the convention of time running upwards, namely in a product $A B$ the symbol for the operator $B$ is depicted \emph{below} $A$. Introducing the diagrammatic representation
\be
U^\dag=\begin{tikzpicture}[baseline=(current  bounding  box.center), scale=1]
\draw[ thick] (-4.25,0.5) -- (-3.25,-0.5);
\draw[ thick] (-4.25,-0.5) -- (-3.25,0.5);
\draw[ thick, fill=myblue, rounded corners=2pt] (-4,0.25) rectangle (-3.5,-0.25);
\draw[thick] (-3.75,0.15) -- (-3.75+0.15,0.15) -- (-3.75+0.15,0);
\Text[x=-4.25,y=-0.75]{}
\end{tikzpicture},
\ee
 we can express the unitarity of the gates in terms of the following diagrammatic rules
\be
\label{eq:unitarity}
\begin{split}
UU^\dag &=\begin{tikzpicture}[baseline=(current  bounding  box.center), scale=1]
\def\ep{1.3};
\def\eps{0.8};
\draw[thick] (-2.25,1) -- (-1.75,0.5);
\draw[thick] (-1.75,0.5) -- (-1.25,1);
\draw[thick] (-2.25,-1) -- (-1.75,-0.5);
\draw[thick] (-1.75,-0.5) -- (-1.25,-1);
\draw[thick] (-1.9,0.35) to[out=170, in=-170] (-1.9,-0.35);
\draw[thick] (-1.6,0.35) to[out=10, in=-10] (-1.6,-0.35);
\draw[thick, fill=myred, rounded corners=2pt] (-2,0.25) rectangle (-1.5,0.75);
\draw[thick, fill=myblue, rounded corners=2pt] (-2,-0.25) rectangle (-1.5,-0.75);
\draw[thick] (-1.75,0.65) -- (-1.6,0.65) -- (-1.6,0.5);
\draw[thick] (-1.75,-0.35) -- (-1.6,-0.35) -- (-1.6,-0.5);
\Text[x=-2.25,y=-0.5]{}
\end{tikzpicture}
=
\begin{tikzpicture}[baseline=(current  bounding  box.center), scale=1]
\draw[ thick] (.7,-.75) -- (.7,.75) (1.3,-0.75) -- (1.3,0.75) (.7,-.75) -- (.45,.-1) (1.3,-.75) -- (1.55,.-1) (.7,.75) -- (.45,1) (1.3,.75) -- (1.55,1);
\Text[x=0.7,y=-0.5]{}
\end{tikzpicture}=\1,\\
&\\
U^\dag U &= 
\begin{tikzpicture}[baseline=(current  bounding  box.center), scale=1]
\def\ep{1.3};
\def\eps{0.8};
\draw[ thick] (-2.25,1) -- (-1.75,0.5);
\draw[ thick] (-1.75,0.5) -- (-1.25,1);
\draw[ thick] (-2.25,-1) -- (-1.75,-0.5);
\draw[ thick] (-1.75,-0.5) -- (-1.25,-1);
\draw[ thick] (-1.9,0.35) to[out=170, in=-170] (-1.9,-0.35);
\draw[ thick] (-1.6,0.35) to[out=10, in=-10] (-1.6,-0.35);
\draw[thick, fill=myblue, rounded corners=2pt] (-2,0.25) rectangle (-1.5,0.75);
\draw[thick, fill=myred, rounded corners=2pt] (-2,-0.25) rectangle (-1.5,-0.75);
\draw[thick] (-1.75,0.65) -- (-1.6,0.65) -- (-1.6,0.5);
\draw[thick] (-1.75,-0.35) -- (-1.6,-0.35) -- (-1.6,-0.5);
\Text[x=-2.25,y=-0.5]{}
\end{tikzpicture}
=
\begin{tikzpicture}[baseline=(current  bounding  box.center), scale=1]
\draw[ thick] (.7,-.75) -- (.7,.75) (1.3,-0.75) -- (1.3,0.75) (.7,-.75) -- (.45,.-1) (1.3,-.75) -- (1.55,.-1) (.7,.75) -- (.45,1) (1.3,.75) -- (1.55,1);
\Text[x=0.7,y=-0.5]{}
\end{tikzpicture}
=\1\,.
\end{split}
\ee

\subsection{Dynamical Correlations in the Folded Picture}
\label{sec:foldedpicture}

In this work we will adopt the so called ``folded'' representation of the circuit, a standard trick in  tensor-network theory~\cite{BHVC09,MuCB12} that has recently found many applications in the studies of local quantum circuits~\cite{BKP:OEsolitons, BKP:OEergodicandmixing, ZN:nonrandommembrane, BP:tripartite, Nahum:operatorspreadingRU, JHN:OperatorEntanglement, Keyserlingk, ZN:statmech, NumericalOperatorEvolution}. Here we consider the simplest possible folding mapping that consists of turning operators into states of a quantum circuit with larger local Hilbert space by ``bending'' their lower legs on top of the upper ones. More specifically, the folding mapping turns operators in ${\rm End}((\mathbb C^{d})^{\otimes 2 L})$ into states in $(\mathbb C^{d^2})^{\otimes 2 L}$. For example, considering local operators and  following exactly the notation of \cite{BKP:OEergodicandmixing} we have
\bw
\be
\frac{1}{d^{L-1/2}} a_x=
\frac{1}{d^{L-1/2}} 
\begin{tikzpicture}[baseline=(current  bounding  box.center), scale=0.8]
\def\eps{0.5};
\draw[thick] (-3.5,0.5) -- (-3.5,-0.5);
\draw[thick, fill=black] (-3.5,0) circle (0.1cm); 
\Text[x=-3.5,y=-0.65]{}
\Text[x=-3.2,y=0]{$a$}
\foreach \i in {-2,...,2}
\draw[thick] (-3.5+0.5*\i,0.5) -- (-3.5+0.5*\i,-0.5);
\Text[x=-5,y=0]{$\dots$}
\Text[x=-2,y=0]{$\dots$}
\end{tikzpicture}
\longmapsto
\,\,
\frac{1}{d^{L-1/2}} 
\begin{tikzpicture}[baseline=(current  bounding  box.center), scale=0.8]
\def\eps{0.5};
\Text[x=-1.75,y=-0.65]{$a$}
\draw[thick, fill=black] (-1.75,-0.3) circle (0.1cm); 
\foreach \i in {-2,...,2}{
\draw[thick] (-2+0.75*\i,0.25) -- (-2+0.75*\i,-0.251);
\draw[thick] (-1.5+0.75*\i,0.4) -- (-1.5+0.75*\i,-0.101);
\draw[thick] (-2+0.75*\i,-0.25) to[out=-85,in=-80] (-1.505+0.75*\i,-0.1);
}
\Text[x=-4,y=0]{$\dots$}
\Text[x=.5,y=0]{$\dots$}
\Text[x=-3.5,y=0.5]{}
\end{tikzpicture}
= \ket{a_x}\,.
\label{eq:opfolding}
\ee
Note that in this paper we will always consider operators that are Hilbert-Schmidt normalised, which lead to normalised states. Moreover, introducing 
\begin{equation}
\ket{\mcirc}:=\frac{1}{\sqrt{d}}\,\,
\begin{tikzpicture}[baseline=(current  bounding  box.center), scale=0.8]
\draw[thick] (-2,0.25) -- (-2,-0.251);
\draw[thick] (-1.5,0.4) -- (-1.5,-0.101);
\draw[thick] (-2,-0.25) to[out=-85,in=-80] ( (-1.505,-0.1);
\Text[x=-1.75,y=-0.65]{}
\end{tikzpicture}
\,\equiv\,\,
\begin{tikzpicture}[baseline=(current  bounding  box.center), scale=0.8]
\draw[very thick] (-0.15,0.25) -- (-0.15,-0.251);
\draw[thick, fill=white] (-.15,-0.25) circle (0.1cm); 
\Text[x=-0.15,y=-0.65]{}
\end{tikzpicture}\,,
\qquad\qquad 
\ket{a}:=
\begin{tikzpicture}[baseline=(current  bounding  box.center), scale=0.8]
\draw[thick] (-2,0.25) -- (-2,-0.251);
\draw[thick] (-1.5,0.4) -- (-1.5,-0.101);
\draw[thick] (-2,-0.25) to[out=-85,in=-80] ( (-1.505,-0.1);
\draw[thick, fill=black] (-1.75,-0.3) circle (0.1cm); 
\Text[x=-1.75,y=-0.65]{$a$}
\end{tikzpicture}
\equiv
\begin{tikzpicture}[baseline=(current  bounding  box.center), scale=0.8]
\draw[very thick] (-0.15,0.25) -- (-0.15,-0.251);
\draw[thick, fill=black] (-.15,-0.25) circle (0.1cm); 
\Text[x=-0.15,y=-0.65]{$a$}
\end{tikzpicture}\,,
\label{eq:operatorstate}
\end{equation}
we can rewrite the r.h.s. of \eqref{eq:opfolding} as 
\be
\ket{a_x} = \ket{\mcirc}^{\otimes (L+2x)}\otimes\ket{a}\otimes\ket{\mcirc}^{\otimes (L-1-2x)}\,.
\ee
The folding mapping can also be applied to time-evolved operators in Heisenberg picture. Considering again the example of local operators we have 
\be
\label{eq:opevolution}
\frac{1}{d^{L-1/2}} a_x(t)=
\begin{tikzpicture}[baseline=(current  bounding  box.center), scale=0.55]
\def\eps{0};
\def\shift{11}
\def\shifty{-2.5}
\draw [thick, decorate, decoration={brace,amplitude=4pt,mirror,raise=4pt},yshift=0pt]
(-1.5,0) -- (-1.5,6) node [black,midway,xshift=.5cm] {$t$};
\draw [thick, decorate, decoration={brace,amplitude=4pt,raise=4pt},yshift=0pt]
(-6.5,5.8) -- (-3.5,5.8) node [black,midway,yshift=.5cm] {$x$};
\foreach \i in{2,...,4}
{
\draw[thick] (-1.5,2*\i-0.5-8.5) arc (45:-90:0.15);
\draw[thick] (-10+0.5+0,2*\i-0.5-8.5) arc (-135:-270:0.15);}
\foreach \i in{2,...,3}
{
\draw[thick] (-1.5,1+2*\i-0.5-8.5) arc (-45:90:0.15);
\draw[thick] (-10+0.5,1+2*\i-0.5-8.5) arc (135:270:0.15);
}
\foreach \i in{3,...,4}
{
\draw[thick] (-1.5,2*\i-0.5-3.5) arc (45:-90:0.15);
\draw[thick] (-10+0.5+0,2*\i-0.5-3.5) arc (-135:-270:0.15);}
\foreach \i in{2,...,4}
{
\draw[thick] (-1.5,1+2*\i-0.5-3.5) arc (-45:90:0.15);
\draw[thick] (-10+0.5,1+2*\i-0.5-3.5) arc (135:270:0.15);
}
\foreach \i in {1,...,4}
{
\draw[thick] (-.5-2*\i,1) -- (0.5-2*\i,0);
\draw[thick] (-0.5-2*\i,0) -- (0.5-2*\i,1);
\draw[thick] (-.5-2*\i,-1+\eps) -- (0.5-2*\i,\eps);
\draw[thick] (-0.5-2*\i,0) -- (-0.5-2*\i,\eps);
\draw[thick] (-0.5-2*\i,\eps) -- (0.5-2*\i,-1+\eps);
\draw[thick] (-0.5-2*\i+1,0) -- (-0.5-2*\i+1,\eps);
\draw[thick, fill=myblue, rounded corners=1pt] (-0.25-2*\i,0.25) rectangle (.25-2*\i,0.75);
\draw[thick] (-2*\i,0.65) -- (.15-2*\i,.65) -- (.15-2*\i,0.5);
\draw[thick, fill=myred, rounded corners=1pt] (-0.25-2*\i,-0.25+\eps) rectangle (.25-2*\i,-0.75+\eps);
\draw[thick] (-2*\i,-0.35+\eps) -- (.15-2*\i,-.35+\eps) -- (.15-2*\i,-0.5+\eps);
}
\foreach \i in {2,...,5}
{
\draw[thick] (.5-2*\i,6) -- (1-2*\i,5.5);
\draw[thick] (1.5-2*\i,6) -- (1-2*\i,5.5);
\draw[thick] (.5-2*\i,-6+\eps) -- (1-2*\i,-5.5+\eps);
\draw[thick] (1.5-2*\i,-6+\eps) -- (1-2*\i,-5.5+\eps);
}
\foreach \jj[evaluate=\jj as \j using -2*(ceil(\jj/2)-\jj/2)] in {0,...,3}
\foreach \i in {2,...,5}
{
\draw[thick] (.5-2*\i-1*\j,2+1*\jj) -- (1-2*\i-1*\j,1.5+\jj);
\draw[thick] (1-2*\i-1*\j,1.5+1*\jj) -- (1.5-2*\i-1*\j,2+\jj);
}
\foreach \jj[evaluate=\jj as \j using -2*(ceil(\jj/2)-\jj/2)] in {0,...,4}
\foreach \i in {2,...,5}
{
\draw[thick] (.5-2*\i-1*\j,1+1*\jj) -- (1-2*\i-1*\j,1.5+\jj);
\draw[thick] (1-2*\i-1*\j,1.5+1*\jj) -- (1.5-2*\i-1*\j,1+\jj);
\draw[thick, fill=myblue, rounded corners=1pt] (0.75-2*\i-1*\j,1.75+\jj) rectangle (1.25-2*\i-1*\j,1.25+\jj);
\draw[thick] (1-2*\i-1*\j,1.65+1*\jj) -- (1.15-2*\i-1*\j,1.65+1*\jj) -- (1.15-2*\i-1*\j,1.5+1*\jj);
}
\foreach \jj[evaluate=\jj as \j using -2*(ceil(\jj/2)-\jj/2)] in {0,...,3}
\foreach \i in {2,...,5}
{
\draw[thick] (.5-2*\i-1*\j,-2-1*\jj+\eps) -- (1-2*\i-1*\j,-1.5-\jj+\eps);
\draw[thick] (1-2*\i-1*\j,-1.5-1*\jj+\eps) -- (1.5-2*\i-1*\j,-2-\jj+\eps);
}
\foreach \jj[evaluate=\jj as \j using -2*(ceil(\jj/2)-\jj/2)] in {0,...,4}
\foreach \i in {2,...,5}
{
\draw[thick] (.5-2*\i-1*\j,-1-1*\jj+\eps) -- (1-2*\i-1*\j,-1.5-\jj+\eps);
\draw[thick] (1-2*\i-1*\j,-1.5-1*\jj+\eps) -- (1.5-2*\i-1*\j,-1-\jj+\eps);
\draw[thick, fill=myred, rounded corners=1pt] (0.75-2*\i-1*\j,-1.75-\jj+\eps) rectangle (1.25-2*\i-1*\j,-1.25-\jj+\eps);
\draw[thick] (1-2*\i-1*\j,-1.35-1*\jj+\eps) -- (1.15-2*\i-1*\j,-1.35-1*\jj+\eps) -- (1.15-2*\i-1*\j,-1.5-1*\jj+\eps);
}
\draw[thick, fill=black] (-3.5,0) circle (0.1cm); 
\Text[x=-3.3,y=-0.3]{$a$}
\Text[x=1.2, y=-6.8]{}
\end{tikzpicture}\!\!\!\!\!\!\!\!\!\!
\!\!\!\!\!\!\!\!\longmapsto\,\,
\begin{tikzpicture}[baseline=(current  bounding  box.center), scale=0.55]
\def\eps{0};
\def\shift{11}
\def\shifty{-2.5}
\foreach \i in{3,...,4}
{
\draw[very thick] (\shift-1.5,2*\i-0.5-3.5+\shifty) arc (45:-90:0.15);
\draw[very thick] (\shift-10+0.5+0,2*\i-0.5-3.5+\shifty) arc (-135:-270:0.15);}
\foreach \i in{2,...,4}
{
\draw[very thick] (\shift-1.5,1+2*\i-0.5-3.5+\shifty) arc (-45:90:0.15);
\draw[very thick] (\shift-10+0.5,1+2*\i-0.5-3.5+\shifty) arc (135:270:0.15);
}
\foreach \i in {1,...,4}
{
\draw[very thick] (\shift-.5-2*\i,1+\shifty) -- (\shift+0.5-2*\i,0+\shifty);
\draw[very thick] (\shift-0.5-2*\i,0+\shifty) -- (\shift+0.5-2*\i,1+\shifty);
\draw[ thick, fill=mygreen, rounded corners=1pt] (\shift-0.25-2*\i,0.25+\shifty) rectangle (\shift+.25-2*\i,0.75+\shifty);
\draw[thick] (\shift-2*\i,0.65+\shifty) -- (\shift+.15-2*\i,.65+\shifty) -- (\shift+.15-2*\i,0.5+\shifty);
}
\foreach \i in {2,...,5}
{
\draw[very thick] (\shift+.5-2*\i,6+\shifty) -- (\shift+1-2*\i,5.5+\shifty);
\draw[very thick] (\shift+1.5-2*\i,6+\shifty) -- (\shift+1-2*\i,5.5+\shifty);
}
\foreach \jj[evaluate=\jj as \j using -2*(ceil(\jj/2)-\jj/2)] in {0,...,3}
\foreach \i in {2,...,5}
{
\draw[very thick] (\shift+.5-2*\i-1*\j,2+1*\jj+\shifty) -- (\shift+1-2*\i-1*\j,1.5+\jj+\shifty);
\draw[very thick] (\shift+1-2*\i-1*\j,1.5+1*\jj+\shifty) -- (\shift+1.5-2*\i-1*\j,2+\jj+\shifty);
}
\foreach \i in {1,...,4}
{
\draw[very thick] (\shift-.5-2*\i,1+\shifty) -- (\shift+0.5-2*\i,0+\shifty);
\draw[very thick] (\shift-0.5-2*\i,0+\shifty) -- (\shift+0.5-2*\i,1+\shifty);
\draw[ thick, fill=mygreen, rounded corners=1pt] (\shift-0.25-2*\i,0.25+\shifty) rectangle (\shift+.25-2*\i,0.75+\shifty);
\draw[thick] (\shift-2*\i,0.65+\shifty) -- (\shift+.15-2*\i,.65+\shifty) -- (\shift+.15-2*\i,0.5+\shifty);
}
\foreach \jj[evaluate=\jj as \j using -2*(ceil(\jj/2)-\jj/2)] in {0,...,4}
\foreach \i in {2,...,5}
{
\draw[very thick] (\shift+.5-2*\i-1*\j,1+1*\jj+\shifty) -- (\shift+1-2*\i-1*\j,1.5+\jj+\shifty);
\draw[very thick] (\shift+1-2*\i-1*\j,1.5+1*\jj+\shifty) -- (\shift+1.5-2*\i-1*\j,1+\jj+\shifty);
\draw[ thick, fill=mygreen, rounded corners=1pt] (\shift+0.75-2*\i-1*\j,1.75+\jj+\shifty) rectangle (\shift+1.25-2*\i-1*\j,1.25+\jj+\shifty);
\draw[thick] (\shift+1-2*\i-1*\j,1.65+1*\jj+\shifty) -- (\shift+1.15-2*\i-1*\j,1.65+1*\jj+\shifty) -- (\shift+1.15-2*\i-1*\j,1.5+1*\jj+\shifty);
}
\draw[ thick, fill=white] (\shift-8.5,\shifty) circle (0.1cm); 
\draw[ thick, fill=white] (\shift-7.5,\shifty) circle (0.1cm);
\draw[ thick, fill=white] (\shift-6.5,\shifty) circle (0.1cm); 
\draw[ thick, fill=white] (\shift-5.5,\shifty) circle (0.1cm); 
\draw[ thick, fill=white] (\shift-4.5,\shifty) circle (0.1cm); 
\draw[ thick, fill=white] (\shift-3.5,\shifty) circle (0.1cm); 
\draw[ thick, fill=white] (\shift-2.5,\shifty) circle (0.1cm); 
\draw[ thick, fill=white] (\shift-1.5,\shifty) circle (0.1cm);
\draw[thick, fill=black] (\shift-3.5,\shifty) circle (0.1cm); 
\Text[x=-3.3+\shift,y=-0.4+\shifty]{$a$}
\end{tikzpicture}
\equiv
\ket{a_x(t)}\,, 
\ee
\ew
where we introduced the ``doubled gate'' (or the Heisenberg operator gate)
\begin{equation}
\label{eq:doublegate}
W=
\begin{tikzpicture}[baseline=(current  bounding  box.center), scale=1]
\def\eps{0.5};
\Wgategreen{-3.75}{0};
\Text[x=-3.75,y=-0.6]{}
\end{tikzpicture}
=
\begin{tikzpicture}[baseline=(current  bounding  box.center), scale=1]
\draw[thick] (-1.65,0.65) -- (-0.65,-0.35);
\draw[thick] (-1.65,-0.35) -- (-0.65,0.65);
\draw[ thick, fill=myred, rounded corners=2pt] (-1.4,0.4) rectangle (-.9,-0.1);
\draw[thick] (-1.15,0) -- (-1,0) -- (-1,0.15);
\draw[thick] (-2.25,0.5) -- (-1.25,-0.5);
\draw[thick] (-2.25,-0.5) -- (-1.25,0.5);
\draw[ thick, fill=myblue, rounded corners=2pt] (-2,0.25) rectangle (-1.5,-0.25);
\draw[thick] (-1.75,0.15) -- (-1.6,0.15) -- (-1.6,0);
\Text[x=-2.25,y=-0.6]{}
\end{tikzpicture}
= U^\dag\otimes U^{T}\,,
\end{equation}
and denoted by $(\cdot)^T$ the matrix transposition. Note that, on the level of the doubled gate, the relations \eqref{eq:unitarity} read as 
\be
\begin{tikzpicture}[baseline=(current  bounding  box.center), scale=1]
\def\eps{0.5};
\draw[very thick] (-4.25,0.5) -- (-3.25,-0.5);
\draw[very thick] (-4.25,-0.5) -- (-3.25,0.5);
\draw[ thick, fill=mygreen, rounded corners=2pt] (-4,0.25) rectangle (-3.5,-0.25);
\draw[thick] (-3.75,0.15) -- (-3.6,0.15) -- (-3.6,0);
\Text[x=-2.75,y=0.0, anchor = center]{$=$}
\draw[thick, fill=white] (-4.25,-0.5) circle (0.1cm); 
\draw[thick, fill=white] (-3.25,-0.5) circle (0.1cm); 
\draw[very thick] (-2.25,0.5) -- (-2.25,-0.5);
\draw[very thick] (-1.25,-0.5) -- (-1.25,0.5);
\draw[thick, fill=white] (-2.25,-0.5) circle (0.1cm); 
\draw[thick, fill=white] (-1.25,-0.5) circle (0.1cm); 
\end{tikzpicture},
\qquad 
\begin{tikzpicture}[baseline=(current  bounding  box.center), scale=1]
\def\eps{0.5};
\draw[very thick] (-4.25,0.5) -- (-3.25,-0.5);
\draw[very thick] (-4.25,-0.5) -- (-3.25,0.5);
\draw[ thick, fill=mygreen, rounded corners=2pt] (-4,0.25) rectangle (-3.5,-0.25);
\draw[thick] (-3.75,0.15) -- (-3.6,0.15) -- (-3.6,0);
\Text[x=-2.75,y=0.0, anchor = center]{$=$}
\draw[thick, fill=white] (-4.25,0.5) circle (0.1cm); 
\draw[thick, fill=white] (-3.25,0.5) circle (0.1cm); 
\draw[very thick] (-2.25,0.5) -- (-2.25,-0.5);
\draw[very thick] (-1.25,-0.5) -- (-1.25,0.5);
\draw[thick, fill=white] (-2.25,0.5) circle (0.1cm); 
\draw[thick, fill=white] (-1.25,0.5) circle (0.1cm); 
\end{tikzpicture}\,,
\label{eq:unitality}
\ee
namely, they impose a {\em unitality} condition on $W$ ensuring that it maps the ``identity state'' $\ket{\mcirc\mcirc}$ into itself. 

In this setting we consider dynamical correlation functions in the infinite-temperature state, which in the folded picture are represented as 
\be
\braket{b_{x+y} |a_y(t)} =
\begin{tikzpicture}[baseline=(current  bounding  box.center), scale=0.55]
\def\eps{0};
\def\shift{11}
\def\shifty{-2.5}
\Text[x=\shift-3, y=-4]{}
\foreach \i in{3,...,4}
{
\draw[very thick] (\shift-1.5,2*\i-0.5-3.5+\shifty) arc (45:-90:0.15);
\draw[very thick] (\shift-10+0.5+0,2*\i-0.5-3.5+\shifty) arc (-135:-270:0.15);}
\foreach \i in{2,...,4}
{
\draw[very thick] (\shift-1.5,1+2*\i-0.5-3.5+\shifty) arc (-45:90:0.15);
\draw[very thick] (\shift-10+0.5,1+2*\i-0.5-3.5+\shifty) arc (135:270:0.15);
}
\foreach \i in {1,...,4}
{
\draw[very thick] (\shift-.5-2*\i,1+\shifty) -- (\shift+0.5-2*\i,0+\shifty);
\draw[very thick] (\shift-0.5-2*\i,0+\shifty) -- (\shift+0.5-2*\i,1+\shifty);
\draw[ thick, fill=mygreen, rounded corners=1pt] (\shift-0.25-2*\i,0.25+\shifty) rectangle (\shift+.25-2*\i,0.75+\shifty);
\draw[thick] (\shift-2*\i,0.65+\shifty) -- (\shift+.15-2*\i,.65+\shifty) -- (\shift+.15-2*\i,0.5+\shifty);
}
\foreach \i in {2,...,5}
{
\draw[very thick] (\shift+.5-2*\i,6+\shifty) -- (\shift+1-2*\i,5.5+\shifty);
\draw[very thick] (\shift+1.5-2*\i,6+\shifty) -- (\shift+1-2*\i,5.5+\shifty);
}
\foreach \jj[evaluate=\jj as \j using -2*(ceil(\jj/2)-\jj/2)] in {0,...,3}
\foreach \i in {2,...,5}
{
\draw[very thick] (\shift+.5-2*\i-1*\j,2+1*\jj+\shifty) -- (\shift+1-2*\i-1*\j,1.5+\jj+\shifty);
\draw[very thick] (\shift+1-2*\i-1*\j,1.5+1*\jj+\shifty) -- (\shift+1.5-2*\i-1*\j,2+\jj+\shifty);
}
\foreach \i in {1,...,4}
{
\draw[very thick] (\shift-.5-2*\i,1+\shifty) -- (\shift+0.5-2*\i,0+\shifty);
\draw[very thick] (\shift-0.5-2*\i,0+\shifty) -- (\shift+0.5-2*\i,1+\shifty);
\draw[ thick, fill=mygreen, rounded corners=1pt] (\shift-0.25-2*\i,0.25+\shifty) rectangle (\shift+.25-2*\i,0.75+\shifty);
\draw[thick] (\shift-2*\i,0.65+\shifty) -- (\shift+.15-2*\i,.65+\shifty) -- (\shift+.15-2*\i,0.5+\shifty);
}
\foreach \jj[evaluate=\jj as \j using -2*(ceil(\jj/2)-\jj/2)] in {0,...,4}
\foreach \i in {2,...,5}
{
\draw[very thick] (\shift+.5-2*\i-1*\j,1+1*\jj+\shifty) -- (\shift+1-2*\i-1*\j,1.5+\jj+\shifty);
\draw[very thick] (\shift+1-2*\i-1*\j,1.5+1*\jj+\shifty) -- (\shift+1.5-2*\i-1*\j,1+\jj+\shifty);
\draw[ thick, fill=mygreen, rounded corners=1pt] (\shift+0.75-2*\i-1*\j,1.75+\jj+\shifty) rectangle (\shift+1.25-2*\i-1*\j,1.25+\jj+\shifty);
\draw[thick] (\shift+1-2*\i-1*\j,1.65+1*\jj+\shifty) -- (\shift+1.15-2*\i-1*\j,1.65+1*\jj+\shifty) -- (\shift+1.15-2*\i-1*\j,1.5+1*\jj+\shifty);
}
\draw[ thick, fill=white] (\shift-8.5,\shifty) circle (0.1cm); 
\draw[ thick, fill=white] (\shift-7.5,\shifty) circle (0.1cm);
\draw[ thick, fill=white] (\shift-6.5,\shifty) circle (0.1cm); 
\draw[ thick, fill=white] (\shift-5.5,\shifty) circle (0.1cm); 
\draw[ thick, fill=white] (\shift-4.5,\shifty) circle (0.1cm); 
\draw[ thick, fill=white] (\shift-3.5,\shifty) circle (0.1cm); 
\draw[ thick, fill=white] (\shift-2.5,\shifty) circle (0.1cm); 
\draw[ thick, fill=white] (\shift-1.5,\shifty) circle (0.1cm);
\draw[ thick, fill=white] (\shift-8.5,\shifty+6) circle (0.1cm); 
\draw[ thick, fill=white] (\shift-7.5,\shifty+6) circle (0.1cm);
\draw[ thick, fill=white] (\shift-6.5,\shifty+6) circle (0.1cm); 
\draw[ thick, fill=white] (\shift-5.5,\shifty+6) circle (0.1cm); 
\draw[ thick, fill=white] (\shift-4.5,\shifty+6) circle (0.1cm); 
\draw[ thick, fill=white] (\shift-3.5,\shifty+6) circle (0.1cm); 
\draw[ thick, fill=white] (\shift-2.5,\shifty+6) circle (0.1cm); 
\draw[ thick, fill=white] (\shift-9.5,\shifty+6) circle (0.1cm); 
\draw[thick, fill=black] (\shift-6.5,\shifty) circle (0.1cm); 
\draw[thick, fill=black] (\shift-3.5,6+\shifty) circle (0.1cm); 
\Text[x=-6.8+\shift,y=-0.4+\shifty]{$a$}
\Text[x=-3.9+\shift,y=6.5+\shifty]{$b$}
\end{tikzpicture}.
\label{eq:mapcorrelations}
\ee
Whenever $t<2L-|x+2y|$, namely when we are effectively in the thermodynamic limit, non-trivial correlations are contained in a causal light cone. This is readily seen by using the rules \eqref{eq:unitality}, which allow us to rewrite \eqref{eq:mapcorrelations} as follows  
\bw
\be
\braket{b_{x+y}|a_y(t)} =
\begin{tikzpicture}[baseline=(current  bounding  box.center), scale=0.7]
\def\eps{-0.5};
\Wgategreen{-6}{.5};
\foreach \jj[evaluate=\jj as \j using 2*(ceil(\jj/2)-\jj/2), evaluate=\jj as \aa using int(3-\jj/2), evaluate=\jj as \bb using int(4+\jj/2)] in {0,...,1}{
\foreach \i in {\aa,...,\bb}
{
\Wgategreen{1-2*\i-1*\j}{1.5+\jj};
}
{
\draw[thick, fill=white] (1-2*\aa+0.5-\j,1+\jj) circle (0.1cm);
\draw[thick, fill=white] (1-2*\bb-0.5-\j,1+\jj) circle (0.1cm);
}
}
\foreach \jj[evaluate=\jj as \j using 2*(ceil(\jj/2)-\jj/2), evaluate=\jj as \aa using int(3-(1-\jj)/2), evaluate=\jj as \bb using int(4+(1-\jj)/2)] in {0,...,2}{
\foreach \i in {\aa,...,\bb}
{
\Wgategreen{1-2*\i-1*\j+2*\j}{1.5+\jj+2};
}
{
\draw[thick, fill=white] (1-2*\aa+.5+\j,4+\jj) circle (0.1cm);
\draw[thick, fill=white] (1-2*\bb-1.5+\j,3+\jj) circle (0.1cm);
}
}
\draw[thick, fill=black] (-4.5,6) circle (0.1cm); 
\draw[thick, fill=white] (-5.5,6) circle (0.1cm); 
\draw[thick, fill=black] (-6.5,0) circle (0.1cm); 
\draw[thick, fill=white] (-5.5,0) circle (0.1cm); 
\draw[thick, fill=white] (-2.5,3) circle (0.1cm); 
\Text[x=-4.25,y=6.25]{$b$}
\Text[x=-6.5,y=-0.4]{$a$}
\end{tikzpicture}
=
\begin{tikzpicture}[baseline=(current  bounding  box.center), scale=0.7]
\def\eps{-0.5};
\foreach \i in {0,...,2}
\foreach \j in {1,...,-2}{
\draw[very thick] (-4.5+1.5*\j,1.5*\i) -- (-3+1.5*\j,1.5*\i);
\draw[very thick] (-3.75+1.5*\j,1.5*\i-.75) -- (-3.75+1.5*\j,1.5*\i+.75);
\draw[ thick, fill=mygreen, rounded corners=2pt, rotate around ={45: (-3.75+1.5*\j,1.5*\i)}] (-4+1.5*\j,0.25+1.5*\i) rectangle (-3.5+1.5*\j,-0.25+1.5*\i);
\draw[thick, rotate around ={-45: (-3.75+1.5*\j,1.5*\i)}] (-3.75+1.5*\j,.15+1.5*\i) -- (-3.6+1.5*\j,.15+1.5*\i) -- (-3.6+1.5*\j,1.5*\i);
\draw[thick, fill=white] (-3+1.5,1.5*\i) circle (0.1cm); 
\draw[thick, fill=white] (-3-3*1.5,1.5*\i) circle (0.1cm); 
\draw[thick, fill=white] (-3.75+1.5*\j,-0.75) circle (0.1cm); 
\draw[thick, fill=white] (-3.75+1.5*\j,3.75) circle (0.1cm);
}
\draw[thick, fill=black] (-3+1.5,3) circle (0.1cm); 
\draw[thick, fill=black] (-3-3*1.5,0) circle (0.1cm); 
\Text[x=-3+2,y=3]{$b$}
\Text[x=-3.5-3*1.5,y=0]{$a$}
\Text[x=-8.25,y=3]{}
\draw[<-, thick] (-.5,3.35) arc (0:80:1);
\end{tikzpicture}
\label{eq:pictorialcorrelation}\,.
\ee
\ew
Here we conveniently rotated the picture by $45^\circ$ in the clockwise direction and, for concreteness, we depicted correlation functions for integer coordinates $y\in\mathbb Z$ and $x\in\mathbb Z$. 
The cases with half-integer coordinates can be treated in an analogous fashion: if $y\in\mathbb Z+{1}/{2}$ the states at the bottom ($\mcircf$ and $\mcirc$) are exchanged, while when $x+y\in\mathbb Z+{1}/{2}$ those at the top ($\mcircf$ and $\mcirc$) are exchanged.

We see that in this representation correlation functions correspond to partition functions of a statistical mechanical model (with complex weights determined by the tensor $W$) defined on a rectangular lattice of dimensions 
\begin{align}
x_+&= t+ \lceil x \rceil  & &x_-= t+1-\lceil x \rceil, & &y\in\mathbb Z,\label{eq:xpxmyi}\\
x_+&= t+1+ \lceil x \rceil  & &x_-= t-\lceil x \rceil, & &y\in\mathbb Z+\frac{1}{2},
\label{eq:xpxmyho}
\end{align}
where we introduced the \emph{ceiling function} $\lceil \cdot \rceil$, such that $\lceil x \rceil\in \mathbb Z$ and $x\leq \lceil x \rceil<x+1$ for any $x\in \mathbb R$ (this definition applies also for negative $x$). Moreover, we note that for values of $x$ such that $x_\pm\leq0$ the correlations vanish identically. For definiteness, from now on we will always consider $y=0$ unless otherwise stated.

In the representation~\eqref{eq:pictorialcorrelation}, it is natural to think of the correlations in terms of horizontal and vertical transfer matrices 
\begin{equation}
    A_x^{a b} = \begin{tikzpicture}[baseline={([yshift=-0.5ex] current  bounding  box.center)}, scale=0.75]
\foreach \j in {1,2}{
\draw[very thick, rotate around ={45: (-6+1.5*\j,0.5)}] (-.5-6+1.5*\j,0) -- (0.5-6+1.5*\j,1);
\draw[very thick, rotate around ={45: (-6+1.5*\j,0.5)}] (-5.4+1.5*\j,-0.1) -- (-6.6+1.5*\j,1.1);
\draw[thick, fill=mygreen, rounded corners=2pt, rotate around ={45: (-6+1.5*\j,0.5)}] (-0.25-6+1.5*\j,0.25) rectangle (.25-6+1.5*\j,0.75);
\draw[thick, rotate around ={-135: (-6+1.5*\j,0.5)}] (-6+1.5*\j,0.65) -- (-5.85+1.5*\j,0.65) -- (-5.85+1.5*\j,0.5);
}
\Text[x=-1.55,y=0.5]{$\cdots$}
\Text[x=-1.55,y=2.2]{}
\Text[x=2.3,y=1]{$b$}
\Text[x=-5.3,y=1]{$a$}
\foreach \j in {4,5}{
\draw[very thick, rotate around ={45: (-6+1.5*\j,0.5)}] (-.5-6+1.5*\j,0) -- (0.5-6+1.5*\j,1);
\draw[very thick, rotate around ={45: (-6+1.5*\j,0.5)}] (-5.4+1.5*\j,-0.1) -- (-6.6+1.5*\j,1.1);
\draw[thick, fill=mygreen, rounded corners=2pt, rotate around ={45: (-6+1.5*\j,0.5)}] (-0.25-6+1.5*\j,0.25) rectangle (.25-6+1.5*\j,0.75);
\draw[thick, rotate around ={-135: (-6+1.5*\j,0.5)}] (-6+1.5*\j,0.65) -- (-5.85+1.5*\j,0.65) -- (-5.85+1.5*\j,0.5);
}
\draw[thick, fill=black] (-5.25,0.5) circle (0.1cm);
\draw[thick, fill=black] (2.25,0.5) circle (0.1cm);
\draw [thick, decorate, decoration={brace,amplitude=8pt, raise=4pt, mirror},yshift=0pt]
(-4.75,-0.1) -- (1.75,-0.1) node [black,midway,yshift=-0.95cm, anchor = south] {$x$};
\end{tikzpicture},
\label{eq:transferMat}
\end{equation}
\begin{equation}
    C_x^{a b} = \begin{tikzpicture}[baseline={([yshift=-0.5ex] current  bounding  box.center)}, scale=0.75]
\foreach \j in {1,2}{
\draw[very thick, rotate around ={45: (-6+1.5*\j,0.5)}] (-.5-6+1.5*\j,0) -- (0.5-6+1.5*\j,1);
\draw[very thick, rotate around ={45: (-6+1.5*\j,0.5)}] (-5.4+1.5*\j,-0.1) -- (-6.6+1.5*\j,1.1);
\draw[thick, fill=mygreen, rounded corners=2pt, rotate around ={45: (-6+1.5*\j,0.5)}] (-0.25-6+1.5*\j,0.25) rectangle (.25-6+1.5*\j,0.75);
\draw[thick, rotate around ={-45: (-6+1.5*\j,0.5)}] (-6+1.5*\j,0.65) -- (-5.85+1.5*\j,0.65) -- (-5.85+1.5*\j,0.5);
}
\Text[x=-1.55,y=0.5]{$\cdots$}
\Text[x=-1.55,y=2.2]{}
\Text[x=2.3,y=1]{$b$}
\Text[x=-5.3,y=1]{$a$}
\foreach \j in {4,5}{
\draw[very thick, rotate around ={45: (-6+1.5*\j,0.5)}] (-.5-6+1.5*\j,0) -- (0.5-6+1.5*\j,1);
\draw[very thick, rotate around ={45: (-6+1.5*\j,0.5)}] (-5.4+1.5*\j,-0.1) -- (-6.6+1.5*\j,1.1);
\draw[thick, fill=mygreen, rounded corners=2pt, rotate around ={45: (-6+1.5*\j,0.5)}] (-0.25-6+1.5*\j,0.25) rectangle (.25-6+1.5*\j,0.75);
\draw[thick, rotate around ={-45: (-6+1.5*\j,0.5)}] (-6+1.5*\j,0.65) -- (-5.85+1.5*\j,0.65) -- (-5.85+1.5*\j,0.5);
}
\draw[thick, fill=black] (-5.25,0.5) circle (0.1cm);
\draw[thick, fill=black] (2.25,0.5) circle (0.1cm);
\draw [thick, decorate, decoration={brace,amplitude=8pt, raise=4pt, mirror},yshift=0pt]
(-4.75,-0.1) -- (1.75,-0.1) node [black,midway,yshift=-0.95cm, anchor = south] {$x$};
\end{tikzpicture},
\label{eq:transferMat2}
\end{equation}
as follows
\bw
\begin{align}
\label{eq:TMcorrelations}
\braket{b_{x}|a_0(t)}= 
\begin{cases}
\displaystyle \braket{a \mcirc \ldots\mcirc |  (A_{x_-}^{\mcirc\mcirc})^{x_+ -1} A_{x_-}^{\mcirc b}|\mcirc\ldots\mcirc} = \braket{\mcirc\ldots\mcirc b | (C_{x_+}^{\mcirc \mcirc})^{x_- -1} C_{x_+}^{a \mcirc} | \mcirc \ldots\mcirc}\!, & x\in\mathbb Z+\frac{1}{2},\\
 \\
 \displaystyle \braket{a \mcirc \ldots\mcirc |  (A_{x_-}^{\mcirc\mcirc})^{x_+} |\mcirc\ldots\mcirc b} = \braket{\mcirc \ldots\mcirc |  C_{x_+}^{ \mcirc b} (C_{x_+}^{\mcirc \mcirc})^{x_- -2} C_{x_+}^{a \mcirc }|\mcirc\ldots\mcirc}\!, & x\in\mathbb Z 
\end{cases}\,.
\end{align}
\ew
Note that the above transfer matrices fulfil the following two properties\\

\noindent i) $A_x^{a b}$ and $C_x^{a b}$ are \emph{contracting}, i.e.\ their eigenvalues lie on the closed unit disk around $0$ in the complex plane, for all $a,b$. This is a consequence of the unitarity of $W$ and can be established following the derivation in Appendix A of Ref.~\cite{BKP:OEergodicandmixing}.\\
    
\noindent ii) The state $\ket{\mcirc}^{\otimes x}$ is an eigenvector of $A_x^{\mcirc \mcirc}$ and $C_x^{\mcirc\mcirc}$ with eigenvalue one. This is a direct consequence of the unitality relations~\eqref{eq:unitality}. \\

The folding mapping described in this subsection turns the evolution of operators in the quantum circuit defined by the elementary gate $U$ into that of states in a larger (super) quantum circuit defined by the elementary gate $W$. In this language the correlation functions are nothing but matrix elements of powers of the evolution operator 
\be
\mathbb W= \bigotimes_{x\in\mathbb Z_L + 1/2} W_{x,x+1/2} \bigotimes_{x\in\mathbb Z_L} W_{x,x+1/2},
\label{eq:Wevolution}
\ee
between two specific states. In particular those of local operators are matrix elements of $\mathbb W^t$ between two ``one particle states'' composed by the tensor product of $2L-1$ copies of $\ket{\mcirc}$  and one state orthogonal to it ($\ket{a}$ and $\ket{b}$ in \eqref{eq:mapcorrelations}). Matrix elements of this kind can be brought to the form \eqref{eq:pictorialcorrelation}. Even though the gate $W$ is unitary by construction (cf.~\eqref{eq:doublegate}), the unitarity of $W$ is not needed for the simplification: one just needs the unitality property \eqref{eq:unitality}. Therefore, this setting can be used to study more general problems than that of computing correlations in unitary quantum circuits. An example of it is given in Appendix~\ref{sec:markovchains} where we use it to study correlations in ``Markov circuits'', i.e. classical Markov chains where at each half time step the time evolution couples only nearest neighbours (in a brickwork fashion) with bistochastic matrices (see the definition in \eqref{eq:Wbistochastic}).

\subsection{Dual-unitary Gates}
\label{sec:dualunitary}

As observed in Ref.~\cite{BKP:dualunitary}, the correlations drastically simplify whenever the gate, together with~\eqref{eq:unitality}, also fulfils 
\be
\label{eq:dualunitality}
\begin{tikzpicture}[baseline=(current  bounding  box.center), scale=1]
\def\eps{0.5};
\draw[very thick] (-4.25,0.5) -- (-3.25,-0.5);
\draw[very thick] (-4.25,-0.5) -- (-3.25,0.5);
\draw[ thick, fill=myorange, rounded corners=2pt] (-4,0.25) rectangle (-3.5,-0.25);
\draw[thick] (-3.75,0.15) -- (-3.6,0.15) -- (-3.6,0);
\Text[x=-2.75,y=0.0, anchor = center]{$=$}
\draw[thick, fill=white] (-3.25,-0.5) circle (0.1cm); 
\draw[thick, fill=white] (-3.25, 0.5) circle (0.1cm); 
\draw[very thick] (-1.25,0.5) -- (-2.25, 0.5);
\draw[very thick] (-1.25,-0.5) -- (-2.25,-0.5);
\draw[thick, fill=white] (-1.25, 0.5) circle (0.1cm); 
\draw[thick, fill=white] (-1.25,-0.5) circle (0.1cm); 
\Text[x=-1,y=0.0, anchor = center]{,}
\end{tikzpicture}
\qquad\,\, 
\begin{tikzpicture}[baseline=(current  bounding  box.center), scale=1]
\def\eps{0.5};
\draw[very thick] (-4.25,0.5) -- (-3.25,-0.5);
\draw[very thick] (-4.25,-0.5) -- (-3.25,0.5);
\draw[ thick, fill=myorange, rounded corners=2pt] (-4,0.25) rectangle (-3.5,-0.25);
\draw[thick] (-3.75,0.15) -- (-3.6,0.15) -- (-3.6,0);
\Text[x=-2.75,y=0.0, anchor = center]{$=$}
\draw[thick, fill=white] (-4.25,0.5) circle (0.1cm); 
\draw[thick, fill=white] (-4.25,-0.5) circle (0.1cm); 
\draw[very thick] (-2.25,0.5) -- (-1.25,0.5);
\draw[very thick] (-2.25,-0.5) -- (-1.25,-0.5);
\draw[thick, fill=white] (-2.25,-0.5) circle (0.1cm); 
\draw[thick, fill=white] (-2.25,0.5) circle (0.1cm); 
\end{tikzpicture}\,.
\ee
We use a different (orange) color to denote doubled gates fulfilling these two additional conditions. The conditions \eqref{eq:dualunitality} originate from the following requirements on the single (unfolded) gates
\be
\begin{tikzpicture}[baseline=(current  bounding  box.center), scale=0.8]
\def\eps{0.2};
\def\ep{1.3}
\Text[x=-0.125,y=-\ep]{}
\draw[ thick] (-1.75+2,0.5+\eps) -- (-1.25+2,1+\eps);
\draw[ thick] (-1.25+2,0+\eps) -- (-1.75+2,0.5+\eps);
\draw[ thick] (-1.25+2,0-\eps) -- (-1.75+2,-0.5-\eps);
\draw[ thick] (-1.75+2,-0.5-\eps) -- (-1.25+2,-1-\eps);
\draw[ thick] (-1.9+2,0.7+\eps) to[out=170, in=-170] (-1.9+2,-0.7-\eps);
\draw[ thick] (-1.9+2,0.3+\eps) to[out=170, in=-170] (-1.9+2,-0.3-\eps);
\draw[thick, fill=myblue, rounded corners=2pt] (-2+2,0.25+\eps) rectangle (-1.5+2,0.75+\eps);
\draw[thick, fill=myred, rounded corners=2pt] (-2+2,-0.25-\eps) rectangle (-1.5+2,-0.75-\eps);
\draw[thick] (.25,0.65+\eps) -- (.4,0.65+\eps) -- (.4,0.5+\eps);
\draw[thick] (.25,-0.35-\eps) -- (.4,-0.35-\eps) -- (.4,-0.5-\eps);
\draw[ thick] (-1.45+4,0.7+\eps) to[out=170, in=-170] (-1.45+4,-0.7-\eps);
\draw[ thick] (-1.45+4,0.3+\eps) to[out=170, in=-170] (-1.45+4,-0.3-\eps);
\draw[ thick] (-1.1+4,0.7+\eps) -- (-1.45+4,0.7+\eps)(-1.1+4,0.7+\eps) -- (-.75+4,1+\eps);
\draw[ thick] (-1.1+4,0.3+\eps) -- (-1.45+4,0.3+\eps)(-1.1+4,0.3+\eps) -- (-.75+4,+\eps);
\draw[ thick] (-1.1+4,-0.3-\eps) -- (-1.45+4,-0.3-\eps)(-1.1+4,-0.3-\eps) -- (-.75+4,-\eps);
\draw[ thick] (-1.1+4,-0.7-\eps) -- (-1.45+4,-0.7-\eps)(-1.1+4,-0.7-\eps) -- (-.75+4,-1-\eps);
\Text[x=1.45,y=0]{$=$}
\Text[x=3.45,y=0]{,}
\end{tikzpicture}
\qquad 
\begin{tikzpicture}[baseline=(current  bounding  box.center), scale=0.8]
\def\eps{0.2};
\def\ep{1.3}
\Text[x=-0.125,y=-\ep]{}
\draw[ thick] (-2.25+2,1+\eps) -- (-1.75+2,0.5+\eps);
\draw[ thick] (-1.75+2,0.5+\eps) -- (-2.25+2,0+\eps);
\draw[ thick] (-1.75+2,-0.5-\eps) -- (-2.25+2,0-\eps);
\draw[ thick] (-2.25+2,-1-\eps) -- (-1.75+2,-0.5-\eps);
\draw[ thick] (-1.6+2,0.7+\eps) to[out=10, in=-10] (-1.6+2,-0.7-\eps);
\draw[ thick] (-1.6+2,0.3+\eps) to[out=10, in=-10] (-1.6+2,-0.3-\eps);
\draw[thick, fill=myblue, rounded corners=2pt] (-2+2,0.25+\eps) rectangle (-1.5+2,0.75+\eps);
\draw[thick, fill=myred, rounded corners=2pt] (-2+2,-0.25-\eps) rectangle (-1.5+2,-0.75-\eps);
\draw[thick] (.25,0.65+\eps) -- (.4,0.65+\eps) -- (.4,0.5+\eps);
\draw[thick] (.25,-0.35-\eps) -- (.4,-0.35-\eps) -- (.4,-0.5-\eps);
\draw[ thick] (-.7+3.2,0.7+\eps) to[out=10, in=-10] (-.7+3.2,-0.7-\eps);
\draw[ thick] (-.7+3.2,0.3+\eps) to[out=10, in=-10] (-.7+3.2,-0.3-\eps);
\draw[ thick] (-.7+3.2,0.7+\eps) -- (-1.05+3.2,0.7+\eps)(-1.35+3.2,1+\eps) -- (-1.05+3.2,0.7+\eps);
\draw[ thick] (-.7+3.2,0.3+\eps) -- (-1.05+3.2,0.3+\eps)(-1.35+3.2,+\eps) -- (-1.05+3.2,0.3+\eps);
\draw[ thick] (-.7+3.2,-0.3-\eps) -- (-1.05+3.2,-0.3-\eps)(-1.35+3.2,-\eps) -- (-1.05+3.2,-0.3-\eps);
\draw[ thick] (-.7+3.2,-0.7-\eps) -- (-1.05+3.2,-0.7-\eps)(-1.35+3.2,-1-\eps) -- (-1.05+3.2,-0.7-\eps);
\Text[x=1.5,y=0]{$=$}
\end{tikzpicture},
\label{eq:dualunitaritysinglegate}
\ee
and essentially imply that the evolution of the system remains unitary when one exchanges the roles of space and time. Gates with these properties, called ``dual-unitary'', have recently been used to obtain exact results in a number of different problems concerning non-equilibrium dynamics of quantum many-body systems and quantum many-body chaos~\cite{BKP:dualunitary, BKP:entropy, BKP:OEsolitons, PBCP20, BKP:kickedIsing, BKP:OEergodicandmixing, ClLa20, BP:tripartite, GoLa19}. 

The simplification of dynamical correlations can be seen, for instance, by applying the first of \eqref{eq:dualunitality} to the rightmost corner of \eqref{eq:pictorialcorrelation}
\begin{equation}
\braket{b_{x}| a_0(t)}_{\du}\!\! =\!\!\!\!
\begin{tikzpicture}[baseline=(current  bounding  box.center), scale=0.7]
\def\eps{-0.5};
\foreach \i in {0,...,2}
\foreach \j in {0,...,-2}{
\draw[very thick] (-4.5+1.5*\j,1.5*\i) -- (-3+1.5*\j,1.5*\i);
\draw[very thick] (-3.75+1.5*\j,1.5*\i-.75) -- (-3.75+1.5*\j,1.5*\i+.75);
\draw[ thick, fill=myorange, rounded corners=2pt, rotate around ={45: (-3.75+1.5*\j,1.5*\i)}] (-4+1.5*\j,0.25+1.5*\i) rectangle (-3.5+1.5*\j,-0.25+1.5*\i);
\draw[thick, rotate around ={-45: (-3.75+1.5*\j,1.5*\i)}] (-3.75+1.5*\j,.15+1.5*\i) -- (-3.6+1.5*\j,.15+1.5*\i) -- (-3.6+1.5*\j,1.5*\i);
\draw[thick, fill=white] (-3.75+1.5*\j,-0.75) circle (0.1cm); 
\draw[thick, fill=white] (-3.75+1.5*\j,3.75) circle (0.1cm);
}
\foreach \i in {1,...,2}
\foreach \j in {1}{
\draw[very thick] (-4.5+1.5*\j,1.5*\i) -- (-3+1.5*\j,1.5*\i);
\draw[very thick] (-3.75+1.5*\j,1.5*\i-.75) -- (-3.75+1.5*\j,1.5*\i+.75);
\draw[ thick, fill=myorange, rounded corners=2pt, rotate around ={45: (-3.75+1.5*\j,1.5*\i)}] (-4+1.5*\j,0.25+1.5*\i) rectangle (-3.5+1.5*\j,-0.25+1.5*\i);
\draw[thick, rotate around ={-45: (-3.75+1.5*\j,1.5*\i)}] (-3.75+1.5*\j,.15+1.5*\i) -- (-3.6+1.5*\j,.15+1.5*\i) -- (-3.6+1.5*\j,1.5*\i);
\draw[thick, fill=white] (-3+1.5,1.5*\i) circle (0.1cm); 
\draw[thick, fill=white] (-3-3*1.5,1.5*\i) circle (0.1cm); 
\draw[thick, fill=white] (-3.75+1.5*\j,3.75) circle (0.1cm);
}
\draw[thick, fill=black] (-3+1.5,3) circle (0.1cm); 
\draw[thick, fill=black] (-3-3*1.5,0) circle (0.1cm); 
\draw[thick, fill=white] (-3+.75,0.75) circle (0.1cm); 
\draw[thick, fill=white] (-3,0) circle (0.1cm); 
\Text[x=-3+2,y=3]{$b$}
\Text[x=-3.5-3*1.5,y=0]{$a$}
\Text[x=-8.25,y=3]{}
\end{tikzpicture}\,.
\ee
We see that the relation can be applied further and ultimately allows us to bring the diagram in the following form 
\begin{equation}
\braket{b_{x}|a_0(t)}_{\du} \!\! =\!\!\!\!\!\! \begin{tikzpicture}[baseline=(current  bounding  box.center), scale=0.7]
\def\eps{-0.5};
\foreach \i in {2}
\foreach \j in {1,...,-2}{
\draw[very thick] (-4.5+1.5*\j,1.5*\i) -- (-3+1.5*\j,1.5*\i);
\draw[very thick] (-3.75+1.5*\j,1.5*\i-.75) -- (-3.75+1.5*\j,1.5*\i+.75);
\draw[ thick, fill=myorange, rounded corners=2pt, rotate around ={45: (-3.75+1.5*\j,1.5*\i)}] (-4+1.5*\j,0.25+1.5*\i) rectangle (-3.5+1.5*\j,-0.25+1.5*\i);
\draw[thick, rotate around ={-45: (-3.75+1.5*\j,1.5*\i)}] (-3.75+1.5*\j,.15+1.5*\i) -- (-3.6+1.5*\j,.15+1.5*\i) -- (-3.6+1.5*\j,1.5*\i);
\draw[thick, fill=white] (-3.75+1.5*\j,3.75) circle (0.1cm); 
}
\foreach \i in {0,...,1}
\foreach \j in {-2}{
\draw[very thick] (-4.5+1.5*\j,1.5*\i) -- (-3+1.5*\j,1.5*\i);
\draw[very thick] (-3.75+1.5*\j,1.5*\i-.75) -- (-3.75+1.5*\j,1.5*\i+.75);
\draw[ thick, fill=myorange, rounded corners=2pt, rotate around ={45: (-3.75+1.5*\j,1.5*\i)}] (-4+1.5*\j,0.25+1.5*\i) rectangle (-3.5+1.5*\j,-0.25+1.5*\i);
\draw[thick, rotate around ={-45: (-3.75+1.5*\j,1.5*\i)}] (-3.75+1.5*\j,.15+1.5*\i) -- (-3.6+1.5*\j,.15+1.5*\i) -- (-3.6+1.5*\j,1.5*\i);
\draw[thick, fill=white] (-7.5,1.5*\i+1.5) circle (0.1cm); 
\draw[thick, fill=white] (-6,1.5*\i) circle (0.1cm); 
}
\draw[thick, fill=black] (-3+1.5,3) circle (0.1cm); 
\draw[thick, fill=black] (-3-3*1.5,0) circle (0.1cm); 
\draw[thick, fill=white] (-3-3*1.5,1.5) circle (0.1cm);
\draw[thick, fill=white] (-3-3*1.5+.75,-0.75) circle (0.1cm);
\draw[thick, fill=white] (-2.25,2.25) circle (0.1cm);
\draw[thick, fill=white] (-2.25-1.5,2.25) circle (0.1cm);
\draw[thick, fill=white] (-2.25-3,2.25) circle (0.1cm);
\Text[x=-3+2,y=3]{$b$}
\Text[x=-3.5-3*1.5,y=0]{$a$}
\Text[x=-8.25,y=3]{}
\end{tikzpicture}\,.
\ee
By applying the second of \eqref{eq:dualunitality} we see that the correlation factorises, and reduces to 
\begin{equation}
\braket{b_{x}|a_0(t)}_{\du} =
\begin{tikzpicture}[baseline=(current  bounding  box.center), scale=0.7]
\def\eps{-0.5};
\Wgateorange{-6}{.5};
\Wgateorange{-4}{.5};
\draw[thick, fill=black] (-6.5,0) circle (0.1cm); 
\draw[thick, fill=white] (-5.5,0) circle (0.1cm); 
\draw[thick, fill=white] (-4.5,0) circle (0.1cm); 
\draw[thick, fill=white] (-3.5,0) circle (0.1cm); 
\draw[thick, fill=white] (-6.5,1) circle (0.1cm); 
\draw[thick, fill=white] (-5.5,1) circle (0.1cm); 
\draw[thick, fill=white] (-4.5,1) circle (0.1cm); 
\draw[thick, fill=black] (-3.5,1) circle (0.1cm); 
\Text[x=-3.25,y=1.3]{$b$}
\Text[x=-6.5,y=-0.4]{$a$}
\end{tikzpicture}=\begin{tikzpicture}[baseline=(current  bounding  box.center), scale=0.7]
\def\eps{-0.5};
\draw[very thick] (-5.5,0) -- (-5.5,1);
\draw[very thick] (-6.5,0) -- (-6.5,1);
\draw[thick, fill=black] (-6.5,0) circle (0.1cm); 
\draw[thick, fill=white] (-5.5,0) circle (0.1cm); 
\draw[thick, fill=white] (-6.5,1) circle (0.1cm); 
\draw[thick, fill=black] (-5.5,1) circle (0.1cm); 
\Text[x=-5.75,y=1.4]{$b$}
\Text[x=-6.5,y=-0.4]{$a$}
\end{tikzpicture}\,.
\ee
These two terms are zero for $a$ and $b$ orthogonal to the identity operator, i.e.\ if they are traceless. Following this derivation it is easy to see that the only non trivial correlations are obtained in diagrams with \emph{no corners with two neighbouring $\mcirc$}. Namely, when the rectangle in \eqref{eq:pictorialcorrelation} reduces to a line with two bullets $\mcircf$ at the ends. For integer starting points, $y\in\mathbb Z$, this happens when   
\be
x=t\quad \Rightarrow \quad x_+ = 2t\,,\quad x_-=1\,. 
\label{eq:DUyint}
\ee
In this case the diagram reads as 
\begin{align}
\braket{b_{x}|a_0(t)}_{\du} & \!\! =\!\!
\begin{tikzpicture}[baseline=(current  bounding  box.center), scale=0.7]
\def\eps{-0.5};
\foreach \i in {2}
\foreach \j in {1,...,-2}{
\draw[very thick] (-4.5+1.5*\j,1.5*\i) -- (-3+1.5*\j,1.5*\i);
\draw[very thick] (-3.75+1.5*\j,1.5*\i-.75) -- (-3.75+1.5*\j,1.5*\i+.75);
\draw[ thick, fill=myorange, rounded corners=2pt, rotate around ={45: (-3.75+1.5*\j,1.5*\i)}] (-4+1.5*\j,0.25+1.5*\i) rectangle (-3.5+1.5*\j,-0.25+1.5*\i);
\draw[thick, rotate around ={-45: (-3.75+1.5*\j,1.5*\i)}] (-3.75+1.5*\j,.15+1.5*\i) -- (-3.6+1.5*\j,.15+1.5*\i) -- (-3.6+1.5*\j,1.5*\i);
\draw[thick, fill=white] (-3.75+1.5*\j,3.75) circle (0.1cm); 
\draw[thick, fill=white] (-3.75+1.5*\j,2.25) circle (0.1cm); 
}
\draw[thick, fill=black] (-3+1.5,3) circle (0.1cm); 
\draw[thick, fill=black] (-1.5-4*1.5,3) circle (0.1cm); 
\Text[x=-3+1.8,y=3.05]{$b$}
\Text[x=0.1+-2-4*1.5,y=3]{$a$}
\Text[x=-8.25,y=2]{}
\end{tikzpicture}\notag\\
&= \braket{a |  (A_{\du,1}^{\mcirc\mcirc})^{2t} |b}
\label{eq:DUdiagram}
\,,
\end{align}
where in the last step we used the transfer matrix \eqref{eq:transferMat} for a simple one-dimensional system ($x=1$). Moreover, we added the subscript ``$\du$'' to stress that $A_{\du,x}^{\mcirc\mcirc}$ is made from dual-unitary gates.

Analogously, when the starting point is a half odd-integer, say $y=1/2$, the correlation is non-zero only for  
\be
x=-t \qquad \Rightarrow \quad x_+ = 1\,,\quad x_-=2t\,, 
\label{eq:DUysemint}
\ee 
and reads as 
\begin{align}
\braket{b_{x+1/2}| a_{1/2}(t)}_{\du} & \!\! =\!\!\!\!
\begin{tikzpicture}[baseline=(current  bounding  box.center), scale=0.7]
\def\eps{-0.5};
\foreach \i in {2}
\foreach \j in {1,...,-2}{
\draw[very thick] (-4.5+1.5*\j,1.5*\i) -- (-3+1.5*\j,1.5*\i);
\draw[very thick] (-3.75+1.5*\j,1.5*\i-.75) -- (-3.75+1.5*\j,1.5*\i+.75);
\draw[ thick, fill=myorange, rounded corners=2pt, rotate around ={45: (-3.75+1.5*\j,1.5*\i)}] (-4+1.5*\j,0.25+1.5*\i) rectangle (-3.5+1.5*\j,-0.25+1.5*\i);
\draw[thick, rotate around ={45: (-3.75+1.5*\j,1.5*\i)}] (-3.75+1.5*\j,.15+1.5*\i) -- (-3.6+1.5*\j,.15+1.5*\i) -- (-3.6+1.5*\j,1.5*\i);
\draw[thick, fill=white] (-3.75+1.5*\j,3.75) circle (0.1cm); 
\draw[thick, fill=white] (-3.75+1.5*\j,2.25) circle (0.1cm); 
}
\draw[thick, fill=black] (-3+1.5,3) circle (0.1cm); 
\draw[thick, fill=black] (-1.5-4*1.5,3) circle (0.1cm); 
\Text[x=-3+2-0.2,y=3]{$a$}
\Text[x=-2-4*1.5+0.1,y=3]{$b$}
\Text[x=-8.25,y=2]{}
\label{eq:DUdiagram2}
\end{tikzpicture}\notag\\
&=\braket{b | (C_{\du,1}^{\mcirc \mcirc})^{2t}  | a}\,.
\end{align}
In summary, in dual-unitary circuits the correlations are entirely determined by 1d transfer matrices (or equivalently 1-qudit maps) and take the following simple form  
\begin{align}
\braket{b_{x+y}|a_y(t)}_{\du} &=  {\rm mod}(2y+1,2)  \delta_{x-t} \braket{a |  (A_{\du,1}^{\mcirc\mcirc})^{2t} |b} \notag\\ 
&+ {\rm mod}(2y,2) \delta_{x+t}\braket{b | (C_{\du,1}^{\mcirc \mcirc})^{2t}  | a}\,,
\label{eq:DUcorrelations}
\end{align}
where $\delta_x$ denotes the Kronecker delta function and ${\rm mod}(m,n)\equiv m\, {\rm mod}\, n$ is the mod function. As discussed in Ref.~\cite{BKP:dualunitary}, depending on the spectrum of the transfer matrices $A_{\du,1}^{\mcirc \mcirc}$ and $C_{\du,1}^{\mcirc \mcirc}$, these correlations can show four increasing degrees of ergodicity ranging from non-interacting behaviour --- where correlations are all constant --- to the ergodic and mixing one --- where all correlations decay exponentially. In particular, by providing a complete parametrisation of dual-unitary gates for $d=2$, Ref.~\cite{BKP:dualunitary} showed that the ergodic and mixing case is typical (i.e.\ it has measure one in the parametrisation space).

Finally, we stress that if the double gate is defined as in \eqref{eq:doublegate} and $U$ fulfils \eqref{eq:dualunitaritysinglegate}, then also $W$ and $W^\dag$ fulfil an analogous diagrammatic relation. This is however not needed to obtain the results of this subsection. We only need the conditions \eqref{eq:dualunitality}, which we dub ``dual-unitality''. Even though  the two conditions are equivalent when the gate $W$ comes from a folded quantum circuit, Eq.~\eqref{eq:dualunitality} is less restrictive and can hold in a more general setting. For example in Appendix~\ref{sec:markovchains} we show that this property arises in Markov circuits where the local evolution matrix is bistochastic in both space and time directions.

\section{Strategy and Results}
\label{sec:perturbativeexpansion}

The goal of this paper is to develop a perturbative expansion of correlation functions around the dual-unitary point. The idea is to consider circuits with a number of non-dual-unitary gates $U_\eta$ composed of a dual-unitary term $U_\du$ and a non-dual-unitary correction. To have gates that are manifestly unitary we consider perturbations of the multiplicative form
\begin{equation}
    U_\eta = U_{\du} e^{i \eta P}\,,
\label{eq:defect}
\end{equation}
where $P$ is a generic hermitian 2-qudit operator and nonnegative real parameter $\eta$ sets the \emph{strength} of the perturbation. 
The folded gate $W_\eta$ can then be written as 
\be
\label{eq:W}
    W_\eta =   \left(e^{-i \eta P}\otimes e^{i \eta P^T}\right) W_\du.
\ee
Note that any quantum gate can be written in the form \eqref{eq:defect}. Therefore, the expression \eqref{eq:defect} is the most general perturbation preserving the circuit-structure of the time evolution.

When representing correlation functions as partition functions in a lattice of doubled gates (cf.~\eqref{eq:pictorialcorrelation}) one can think of the gates $W_\eta$, for $\eta>0$, as defects. Even though the methods that we develop are applicable to arbitrary distributions of defects, for the sake of clarity in the main text we only consider the case where all defects are the same, namely there is a non-random systematic breaking of dual-unitarity. 
For comparison, we consider in Appendix~\ref{sec:averagegate} the opposite extreme case of random defects independently distributed at each space-time point. To better control the perturbation theory it is useful to also modulate the number of defects in the lattice. To this aim we introduce an additional parameter: the \emph{density} $\delta$ of defects. We fixed the density because the actual arrangement of the defects does not affect appreciably the physics: one can imagine to randomly place 
$\delta x_+ x_-$ defects among the $x_+ x_-$ gates in the lattice~\eqref{eq:pictorialcorrelation}. For simplicity, however, it will be sometimes useful to imagine the defects covering a regular sublattice of~\eqref{eq:pictorialcorrelation} with lattice spacings $\nu_+$ and $\nu_-$, such that $\delta=1/\nu_+\nu_-$. For example 
\bw
\be
\braket{b_{x}|a_0(t)} =
\begin{tikzpicture}[baseline=(current  bounding  box.center), scale=0.7]
\def\eps{-0.5};
\foreach \i in {0,...,4}
\foreach \j in {3,...,-3}{
\draw[very thick] (-4.5+1.5*\j,1.5*\i) -- (-3+1.5*\j,1.5*\i);
\draw[very thick] (-3.75+1.5*\j,1.5*\i-.75) -- (-3.75+1.5*\j,1.5*\i+.75);
\draw[ thick, fill=myorange, rounded corners=2pt, rotate around ={45: (-3.75+1.5*\j,1.5*\i)}] (-4+1.5*\j,0.25+1.5*\i) rectangle (-3.5+1.5*\j,-0.25+1.5*\i);
\draw[thick, rotate around ={-45: (-3.75+1.5*\j,1.5*\i)}] (-3.75+1.5*\j,.15+1.5*\i) -- (-3.6+1.5*\j,.15+1.5*\i) -- (-3.6+1.5*\j,1.5*\i);
\draw[thick, fill=white] (1.5,1.5*\i) circle (0.1cm); 
\draw[thick, fill=white] (-4.5-3*1.5,1.5*\i) circle (0.1cm); 
\draw[thick, fill=white] (-3.75+1.5*\j,-0.75) circle (0.1cm); 
\draw[thick, fill=white] (-3.75+1.5*\j,6.75) circle (0.1cm);
}
\foreach \i in {0,...,2}
\foreach \j in {-1,...,1}{
\draw[ thick, fill=mygreen, rounded corners=2pt, rotate around ={45: (-3.75+4.5*\j,3*\i)}] (-4+4.5*\j,0.25+3*\i) rectangle (-3.5+4.5*\j,-0.25+3*\i);
\draw[thick, rotate around ={-45: (-3.75+4.5*\j,3*\i)}] (-3.75+4.5*\j,.15+3*\i) -- (-3.6+4.5*\j,.15+3*\i) -- (-3.6+4.5*\j,3*\i);
}
\draw[thick, fill=black] (1.5,6) circle (0.1cm); 
\draw[thick, fill=black] (-4.5-3*1.5,0) circle (0.1cm); 
\Text[x=2,y=6]{$b$}
\Text[x=-5-3*1.5,y=0]{$a$}
\draw [thick, decorate, decoration={brace,amplitude=4pt,mirror,raise=4pt},yshift=0pt]
(1.5,0) -- (1.5,3) node [black,midway,xshift=.55cm] {$\nu_-$};
\draw [thick, decorate, decoration={brace,amplitude=4pt,mirror,raise=4pt},yshift=0pt]
(-3.75,-0.75) -- (0.75,-0.75) node [black,midway,yshift=-.5cm] {$\nu_+$};
\Text[x=-8.25,y=3]{}
\end{tikzpicture}
\label{eq:latticewithdefects}
\,.
\ee
\ew
It follows directly from the above definitions that in both limiting cases, $\eta\to0$ or $\delta\to0$, we recover a dual-unitary circuit. The two perturbations, however, are highly inequivalent. In particular, the case of small density $\delta\ll 1$ is substantially  easier to treat than that of small strength $\eta \ll 1$ and allows for rigorous results. This difference can be appreciated through a simple combinatoric argument: While for small $\delta$ one can work with a single partition function with a small number of defects and large dual-unitary islands, separating the dual-unitary islands for non-zero $\eta$ generates a complicated sum of terms. In particular, the number of contributions at a given order $\eta^n$ corresponds to the number of ways to dispose $n$ identical objects in $x_+ x_-$ identical drawers and becomes exponentially large in the volume $x_+ x_-$ for large enough $n$. 

Remarkably, in this paper we find that --- under certain conditions on the unperturbed dual-unitary gate $W_\du$ --- the leading order contribution to the correlations \emph{can be directly computed in both cases} and, surprisingly, \emph{takes the same form}. Specifically, we observe that --- at the leading order in time --- correlations are still determined by the 1d transfer matrices $A_{\du,1}^{\mcirc\mcirc}$ and $C_{\du,1}^{\mcirc\mcirc}$ (cf. Eqs.~\eqref{eq:DUdiagram}--\eqref{eq:DUdiagram2}). The difference is that, instead of being contracted along straight lines as in \eqref{eq:DUdiagram} and \eqref{eq:DUdiagram2}, now the maps can also be contracted along zig-zag lines like  
\be
\begin{tikzpicture}[baseline=(current  bounding  box.center), scale=0.7]
\def\eps{-0.5};
\foreach \i in {1}
\foreach \j in {-3}{
\draw[very thick] (-4.5+1.5*\j,1.5*\i) -- (-3+1.5*\j,1.5*\i);
\draw[very thick] (-3.75+1.5*\j,1.5*\i-.75) -- (-3.75+1.5*\j,1.5*\i+.75);
\draw[ thick, fill=myorange, rounded corners=2pt, rotate around ={45: (-3.75+1.5*\j,1.5*\i)}] (-4+1.5*\j,0.25+1.5*\i) rectangle (-3.5+1.5*\j,-0.25+1.5*\i);
\draw[thick, rotate around ={-45: (-3.75+1.5*\j,1.5*\i)}] (-3.75+1.5*\j,.15+1.5*\i) -- (-3.6+1.5*\j,.15+1.5*\i) -- (-3.6+1.5*\j,1.5*\i);
\draw[thick, fill=white] (-3+1.5*\j,1.5*\i) circle (0.1cm); 
\draw[thick, fill=white] (-4.5+1.5*\j,1.5*\i) circle (0.1cm); 
}

\foreach \i in {3}
\foreach \j in {0}{
\draw[very thick] (-4.5+1.5*\j,1.5*\i) -- (-3+1.5*\j,1.5*\i);
\draw[very thick] (-3.75+1.5*\j,1.5*\i-.75) -- (-3.75+1.5*\j,1.5*\i+.75);
\draw[ thick, fill=myorange, rounded corners=2pt, rotate around ={45: (-3.75+1.5*\j,1.5*\i)}] (-4+1.5*\j,0.25+1.5*\i) rectangle (-3.5+1.5*\j,-0.25+1.5*\i);
\draw[thick, rotate around ={-45: (-3.75+1.5*\j,1.5*\i)}] (-3.75+1.5*\j,.15+1.5*\i) -- (-3.6+1.5*\j,.15+1.5*\i) -- (-3.6+1.5*\j,1.5*\i);
\draw[thick, fill=white] (-3+1.5*\j,1.5*\i) circle (0.1cm); 
\draw[thick, fill=white] (-4.5+1.5*\j,1.5*\i) circle (0.1cm); 
}

\foreach \i in {4}
\foreach \j in {1,...,2}{
\draw[very thick] (-4.5+1.5*\j,1.5*\i) -- (-3+1.5*\j,1.5*\i);
\draw[very thick] (-3.75+1.5*\j,1.5*\i-.75) -- (-3.75+1.5*\j,1.5*\i+.75);
\draw[ thick, fill=myorange, rounded corners=2pt, rotate around ={45: (-3.75+1.5*\j,1.5*\i)}] (-4+1.5*\j,0.25+1.5*\i) rectangle (-3.5+1.5*\j,-0.25+1.5*\i);
\draw[thick, rotate around ={-45: (-3.75+1.5*\j,1.5*\i)}] (-3.75+1.5*\j,.15+1.5*\i) -- (-3.6+1.5*\j,.15+1.5*\i) -- (-3.6+1.5*\j,1.5*\i);
\draw[thick, fill=white] (-3.75+1.5*\j,1.5*\i-0.75) circle (0.1cm); 
\draw[thick, fill=white] (-3.75+1.5*\j,1.5*\i+0.75) circle (0.1cm); 
}

\foreach \i in {2}
\foreach \j in {-2,...,-1}{
\draw[very thick] (-4.5+1.5*\j,1.5*\i) -- (-3+1.5*\j,1.5*\i);
\draw[very thick] (-3.75+1.5*\j,1.5*\i-.75) -- (-3.75+1.5*\j,1.5*\i+.75);
\draw[ thick, fill=myorange, rounded corners=2pt, rotate around ={45: (-3.75+1.5*\j,1.5*\i)}] (-4+1.5*\j,0.25+1.5*\i) rectangle (-3.5+1.5*\j,-0.25+1.5*\i);
\draw[thick, rotate around ={-45: (-3.75+1.5*\j,1.5*\i)}] (-3.75+1.5*\j,.15+1.5*\i) -- (-3.6+1.5*\j,.15+1.5*\i) -- (-3.6+1.5*\j,1.5*\i);
\draw[thick, fill=white] (-3.75+1.5*\j,1.5*\i-0.75) circle (0.1cm); 
\draw[thick, fill=white] (-3.75+1.5*\j,1.5*\i+0.75) circle (0.1cm); 
}

\foreach \j in {-1}
\foreach \i in {0,..., 1}{
\draw[very thick] (-4.5+4.5*\j,3*\i) -- (-3+4.5*\j,3*\i);
\draw[very thick] (-3.75+4.5*\j,3*\i-.75) -- (-3.75+4.5*\j,3*\i+.75);
\draw[ thick, fill=mygreen, rounded corners=2pt, rotate around ={45: (-3.75+4.5*\j,3*\i)}] (-4+4.5*\j,0.25+3*\i) rectangle (-3.5+4.5*\j,-0.25+3*\i);
\draw[thick, rotate around ={-45: (-3.75+4.5*\j,3*\i)}] (-3.75+4.5*\j,.15+3*\i) -- (-3.6+4.5*\j,.15+3*\i) -- (-3.6+4.5*\j,3*\i);
}
\foreach \j in {0}
\foreach \i in {1,..., 2}{
\draw[very thick] (-4.5+4.5*\j,3*\i) -- (-3+4.5*\j,3*\i);
\draw[very thick] (-3.75+4.5*\j,3*\i-.75) -- (-3.75+4.5*\j,3*\i+.75);
\draw[ thick, fill=mygreen, rounded corners=2pt, rotate around ={45: (-3.75+4.5*\j,3*\i)}] (-4+4.5*\j,0.25+3*\i) rectangle (-3.5+4.5*\j,-0.25+3*\i);
\draw[thick, rotate around ={-45: (-3.75+4.5*\j,3*\i)}] (-3.75+4.5*\j,.15+3*\i) -- (-3.6+4.5*\j,.15+3*\i) -- (-3.6+4.5*\j,3*\i);
}
\foreach \j in {1}
\foreach \i in {2}{
\draw[very thick] (-4.5+4.5*\j,3*\i) -- (-3+4.5*\j,3*\i);
\draw[very thick] (-3.75+4.5*\j,3*\i-.75) -- (-3.75+4.5*\j,3*\i+.75);
\draw[ thick, fill=mygreen, rounded corners=2pt, rotate around ={45: (-3.75+4.5*\j,3*\i)}] (-4+4.5*\j,0.25+3*\i) rectangle (-3.5+4.5*\j,-0.25+3*\i);
\draw[thick, rotate around ={-45: (-3.75+4.5*\j,3*\i)}] (-3.75+4.5*\j,.15+3*\i) -- (-3.6+4.5*\j,.15+3*\i) -- (-3.6+4.5*\j,3*\i);
}
 
\draw[thick, fill=white] (-3.75-4.5,3+0.75) circle (0.1cm); 
\draw[thick, fill=white] (-4.5-4.5,3) circle (0.1cm); 
\draw[thick, fill=white] (-3.75,3-0.75) circle (0.1cm); 
\draw[thick, fill=white] (-3,3) circle (0.1cm); 
\draw[thick, fill=white] (-3.75-4.5,0-0.75) circle (0.1cm); 
\draw[thick, fill=white] (-3-4.5,0) circle (0.1cm);
\draw[thick, fill=white] (-3.75,6+0.75) circle (0.1cm); 
\draw[thick, fill=white] (-4.5,6) circle (0.1cm);  
\draw[thick, fill=white] (4.5-3.75,6+0.75) circle (0.1cm); 
\draw[thick, fill=white] (4.5-3.75,6-0.75) circle (0.1cm);

\draw[thick, fill=black] (1.5,6) circle (0.1cm); 
\draw[thick, fill=black] (-4.5-3*1.5,0) circle (0.1cm); 
\Text[x=1.5,y=6.4]{$b$}
\Text[x=-5-3*1.5+0.5,y=-0.35]{$a$}
\Text[x=-8.25,y=3]{}
\end{tikzpicture}\!\!\!
\label{eq:skeletonexample}
\ee
which we dub ``skeleton diagrams''. In particular, the correlation between two arbitrary points in the causal light cone is obtained by summing the contributions of all skeleton diagrams connecting the two points. The turns in the diagrams are generated by the defects, this means that for $\delta<1$ all the possible positions of the turns are restricted to a sub-lattice, while for $\delta=1$ the turns can be anywhere in the lattice. Note that all skeleton diagrams with down- or left- turns are forbidden. Indeed, these diagrams are cut by the rules \eqref{eq:unitality} and do not contribute to the correlation. Such decomposition of the correlation function can be interpreted as a discrete path-integral on the 2d lattice \eqref{eq:latticewithdefects}. 

To explain how and under what conditions this simplification arises we proceed in two steps. First, in Sec.~\ref{sec:U(1)avg}, we present rigorous results in a toy setting where the wires in  \eqref{eq:latticewithdefects} are two dimensional and the doubled gates four dimensional --- we dub this case ``reduced gates''. Second, in Sec.~\ref{sec:genericnumerics}, we use a combination of analytical arguments and numerical observations to show that the phenomenology identified in the toy setting holds true for generic $d^4$-dimensional doubled gates. 

Let us summarise our results in the above two cases. At the end, for a comparison, we also provide a summary of the results we obtained for random, independently distributed perturbations (details are in Appendix~\ref{sec:averagegate}).



\subsection{Results on Reduced Gates}

We begin by noting that, since the doubled/folded gates have dimension $d^4$, one cannot reduce the dimension of these gates below $16$ by lowering the local Hilbert space dimension $d$. 
However, as explained in Sec.~\ref{sec:U(1)avg}, it is possible to reduce the problem by considering unperturbed dual-unitary gates with random $\rm U(1)$ noise (describing for example random magnetic fields in a fixed direction for spin $1/2$ degrees of freedom, $d=2$) and focussing on the correlations averaged over the noise. 
In this case each wire in \eqref{eq:latticewithdefects} becomes effectively a two-state system and we consider the following basis  
\be
\mathcal B = \{\ket{\mcirc},\ket{\mcircf}\},
\label{eq:reducedbasis}
\ee
where $\ket{\mcirc}$ corresponds to the identity operator and $\ket{\mcircf}$ is orthogonal to it (it corresponds to the only non-trivial one-site traceless operator preserved by the average). The most general averaged folded two-body gate can be expressed in the basis $\mathcal B\otimes \mathcal B$ as
\be
\label{eq:reducedW}
  w :=
  \begin{tikzpicture}[baseline=(current  bounding  box.center), scale=1]
\def\eps{0.5};
\draw (-4.25,0.5) -- (-3.25,-0.5);
\draw (-4.25,-0.5) -- (-3.25,0.5);
\draw[thick, fill=mygreen, rounded corners=2pt] (-4,0.25) rectangle (-3.5,-0.25);
\draw[thick] (-3.75,0.15) -- (-3.6,0.15) -- (-3.6,0);
\Text[x=-4.25,y=-0.65]{}
\end{tikzpicture}
=  
    \begin{pmatrix}
    1 & 0 & 0 & 0\\
0 & p \varepsilon & a & b\\
0 & c & q \varepsilon  & d \\
0 & e & f &  g
    \end{pmatrix},
\ee
where we used lower case $w$ and thin lines to highlight that the wires are just two-dimensional and the gate is hence four-dimensional. 

Since the gate $w$ is obtained via an average it is \emph{not} unitary. However, we see that \eqref{eq:reducedW} fulfils the conditions~\eqref{eq:unitality}. Moreover, we see that the gate also fulfils the conditions~\eqref{eq:dualunitality} if we set $\varepsilon=0$, namely
\be
\label{eq:reducedW0}
  w_\du :=
  \begin{tikzpicture}[baseline=(current  bounding  box.center), scale=1]
\def\eps{0.5};
\draw (-4.25,0.5) -- (-3.25,-0.5);
\draw (-4.25,-0.5) -- (-3.25,0.5);
\draw[thick, fill=myorange, rounded corners=2pt] (-4,0.25) rectangle (-3.5,-0.25);
\draw[thick] (-3.75,0.15) -- (-3.6,0.15) -- (-3.6,0);
\Text[x=-4.25,y=-0.65]{}
\end{tikzpicture}
=  
    \begin{pmatrix}
    1 & 0 & 0 & 0\\
0 & 0 & a & b\\
0 & c & 0  & d \\
0 & e & f &  g
    \end{pmatrix}.
\ee
The parameter $\varepsilon$ can then be interpreted as the dual-unitarity breaking parameter in this setting. Indeed, expressing the elements of $w$ in terms of those of $W_\eta$ (see Appendix~\ref{sec:parametersU1ave}) we find that $\varepsilon = {\cal O}(\eta)$ where $\eta$ is the strength of the perturbation in \eqref{eq:W}. On the other hand, all other parameters in $w$ are  ${\cal O}(\eta^0)$. In words this means that $\varepsilon$ vanishes for vanishing $\eta$, while all other parameters in $w$ are in general
non-zero.

More generally, from the explicit parametrisation of Appendix~\ref{sec:parametersU1ave} we also find that $w_{ij}\in[-1,1]$ and that     the gate becomes a \emph{bistochastic} matrix (see the definition \eqref{eq:Wbistochastic}) when conjugated with $H\otimes H$, where 
\be
H:=\frac{1}{\sqrt 2}
\begin{pmatrix}
1 & \phantom{-}1 \\
1 & {-}1
\end{pmatrix}\!,
\label{eq:Hadamard}
\ee  
is the Hadamard transformation. As shown in Appendix~\ref{sec:markovchains} this means that the evolution generated by $w$ can be mapped to that of a Markov circuit defined on a chain of two-state systems. Remarkably, we also observe the opposite, every local propagator of a Markov circuit on two-state systems can be mapped into a $\rm U(1)$-averaged unitary gate. This means that all our results on reduced gates are also applicable to Markov circuits.

For reduced gates all non-trivial correlations are proportional to $\braket{\mcircf_{x} |\mcircf_0(t)}$ and we find the following three main results:

\subsubsection{Exactly solvable cases}
Apart from the dual-unitary point, $\varepsilon=0$, the gate \eqref{eq:reducedW} has \emph{four} additional non-trivial exactly solvable points: 

\begin{align}
{\rm (i)}&\quad p=0; & {\rm (ii)}&\quad q=0;\notag\\
{\rm (iii)}& \quad  b=d=0; & {\rm (iv)}&\quad e=f=0;\notag
\end{align}
where the parameters that are not set to zero can take arbitrary values. As shown in Sec.~\ref{sec:exactlysolvablereducedgates} in all these cases the correlation functions are exactly given by the sum of skeleton diagrams. In particular, considering a regular sub-lattice of defects as in Eq.~\eqref{eq:latticewithdefects} we find

\bw
\be
\braket{\mcircf_{x}|\!\mcircf_0(t)}=
\begin{cases}
\displaystyle  a^{x_+}\delta_{x_--1}+\sum_{n=1}^{\bar n_{1}}
(p q \varepsilon^2)^n \binom{\tilde x_{+}}{n} \binom{\tilde x_{-} -2}{n-1} a^{x_{+} -n} c^{x_{-}  - n-1} & x \in \mathbb Z\\
\displaystyle q \varepsilon \sum_{n=0}^{\bar n_{2}}
(p q \varepsilon^2)^n \binom{\tilde x_{+}-1}{n} \binom{\tilde x_{-} -1}{n} a^{x_{+} -n-1} c^{x_{-}  - n-1} & x \in \mathbb Z+1/2
\end{cases}\,,
\label{eq:skeletoncontribution}
\ee
\ew
where we introduced the integers ${\bar n_{1}={\rm min}(\tilde x_{+}, \tilde x_{-}-1)}$, ${\bar n_{2}={\rm min}(\tilde x_{-}-1,\tilde x_{+}-1)}$~\cite{note} and the rescaled light cone coordinates
\be
\tilde x_\pm = \lfloor x_\pm/\nu_\pm \rfloor\,.
\label{eq:rescaledxpm}
\ee
Here the symbol $\lfloor \cdot \rfloor$ denotes the \emph{floor function}, such that $\lfloor x \rfloor \in \mathbb Z$ and $x-1< \lfloor x \rfloor \leq x$ for any $x\in \mathbb R$. 

Note that the ray along which \eqref{eq:skeletoncontribution} exhibits the slowest asymptotic decay in time is generically different from the light ray (i.e. the unit speed ray)~\cite{KBP:gSFF}, a representative example is reported in Fig.~\ref{fig:LCasy}.   

We stress that, generically, the gates fulfilling either of (i)--(iv) generate highly complex dynamics, e.g., they efficiently scramble quantum information. For example, we numerically computed a standard dynamical complexity indicator for locally interacting systems --- the so called local-operator entanglement~\cite{Zana01, BKP:OEsolitons, BKP:OEergodicandmixing, AlDM19, PL:OE, JHN:OperatorEntanglement, PP:OEXY, PP:OEIsing, PZ07} --- observing a linear growth. The key property leading to the simple form \eqref{eq:skeletoncontribution} is that the dynamics generated by the gates (i)--(iv) are not time-reversal symmetric. In particular, it is true that evolving forward (backward) in time the support of local operators can grow, forming larger and larger strings of local operator products. Yet these large strings cannot shrink back and do not contribute to the overlap with ultra-local operators. The fact that very large strings do not contribute much to the correlations of local operators is expected to hold quite generally. For example, a similar idea has recently been invoked in Ref.~\cite{NumericalOperatorEvolution} to devise a numerical method able to access the late-time regime. The key point is that this property becomes exact  in the cases (i)--(iv).\\

\begin{figure}[t!]
\includegraphics[scale=.9]{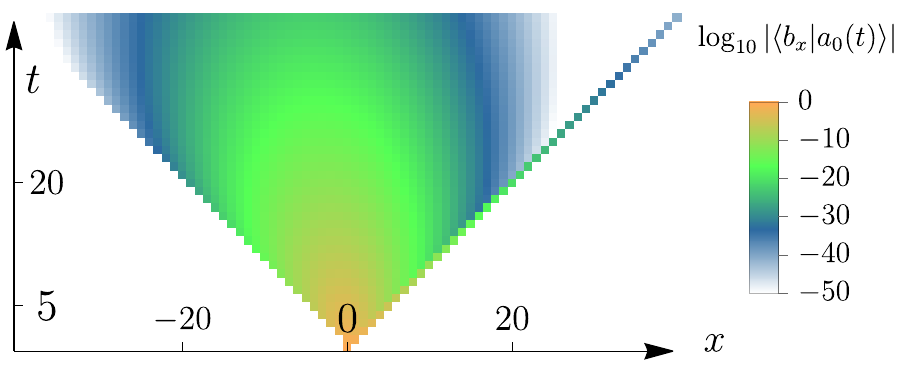}
\caption{
Correlations between integer sites given by Eq.~\eqref{eq:skeletoncontribution}. The correlations are non-zero within the whole light cone and the slowest decay is along $|\zeta^*|<1$. We used defects at all sites ($\delta=1$) and parameters $\varepsilon=0.4$, $p=q=1$, $a= 0.0945626$ and $c= 0.195892$ (as in gate 3 in Table~\ref{tab:U1gates}).
}
\label{fig:LCasy}
\end{figure}

\subsubsection{Low density limit}
\label{subsub2} 
Two simple conditions on the spectrum of horizontal and vertical transfer matrices  constructed with the reduced dual-unitary gate~\eqref{eq:reducedW0} (see Sec.~\ref{sec:PTdensity}) allow us to prove that the expression \eqref{eq:skeletoncontribution} is the dominant contribution to the correlation function at low density $\delta$ and for any $\varepsilon$. More precisely, if the horizontal (vertical) transfer matrix fulfils the aforementioned conditions,  \eqref{eq:skeletoncontribution} is dominant in the limit
\be
x_{+(-)},\bar\nu_{+(-)}\to \infty \qquad \tilde x_{+(-)}={\rm fixed},
\label{eq:lowdensitylimits}
\ee
where $\bar \nu_{+(-)}$ is the minimal separation among the defects in the horizontal (vertical) direction, and the relative error decays exponentially in $\bar \nu_{+(-)}$. In Sec.~\ref{sec:PTdensity} we prove that these conditions hold if the parameters of the gate $w_0$ (cf.~\eqref{eq:reducedW0}) fulfil  
\be
|a| > a^2  + \frac{|b f|}{1-\alpha}\,\qquad {\rm or}\qquad |c| > c^2  + \frac{|d e|}{1-\beta}\,,
\label{eq:conditions}
\ee 
where $\alpha$ and $\beta$, respectively, denote the largest singular values of the sub-matrices  
\be
\begin{pmatrix} c & e\\d& g \end{pmatrix} \qquad {\rm and}\qquad \begin{pmatrix} a & f\\b& g \end{pmatrix}\!.
\label{eq:alphabeta}
\ee
As explained in Sec.~\ref{sec:U(1)avg} (see in particular the discussion around Eq.~\ref{eq:onebodypropagator}), these conditions ensure that correlations propagate on paths of width 1.

\subsubsection{Small strength at density one}
\label{subsub3} 
Based on the rigorous results described in Sec.~\ref{subsub2} we argue that skeleton diagrams give the dominant contribution also for $\delta=1$ and $\varepsilon \ll1$ if \emph{both} conditions \eqref{eq:conditions} are satisfied. As explained in Sec.~\ref{sec:PTstrenght}, the main idea is that, even though \eqref{eq:skeletoncontribution} does not provide a complete perturbative expansion in $\varepsilon$, at each fixed order in perturbation theory the skeleton diagrams dominate for $x_+$ and $x_-$ \emph{both large}. This argument is tested in Sec.~\ref{sec:PTstrenght} by comparing \eqref{eq:skeletoncontribution} with the exact correlations evaluated numerically: the observed agreement is excellent. An example of the correlation pattern described by \eqref{eq:skeletoncontribution} in this case is depicted in Fig.~\ref{fig:LC}.\\

\begin{figure}[t!]
\includegraphics[scale=.9]{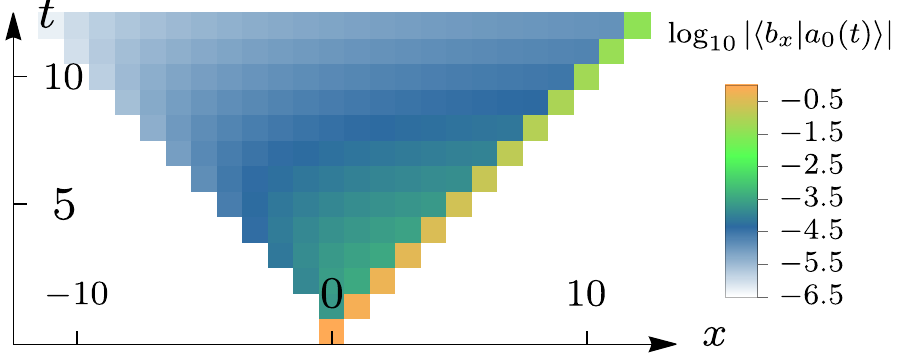}
\caption{Correlations between integer sites in a perturbed dual-unitary circuit with density of defects $\delta=1$, where the gate is defined in Table~\ref{tab:U1gates} (gate 2). We see a reminiscence of the dual-unitary behaviour with a strong peak along the light-cone edge. As opposed to the pure dual-unitary result, however, the correlations are non-zero within the whole light cone.}
\label{fig:LC}
\end{figure}

\subsection{Results on Generic Gates}

Considering quantum gates acting on qudits of generic dimension $d$ (and again focussing on a regular sub-lattice of defects) the contribution of skeleton diagrams reads as 
\bw
\be
\braket{b_{x} |a_0(t)}|_{\text{sk}} =
\begin{cases}\displaystyle
 \sum_{n=1}^{\bar n_{1}} 
 \sum_{\{l_j^+\}}'\sum_{\{l_j^-\}}'  \braket{b|(A^{\mcirc\mcirc}_{\du,1})^{l_{n+1}^+} \mathcal{E}_{1} \dots (A^{\mcirc\mcirc}_{\du,1})^{l_{2}^+}\mathcal{E}_{1} (C^{\mcirc\mcirc}_{\du,1})^{l_{1}^-}\mathcal{E}_{2} (A^{\mcirc\mcirc}_{\du,1})^{l_{1}^+}|a},  & x \in \mathbb{Z}, 
 \\
\displaystyle \sum_{n=0}^{\bar n_{2}} 
\sum_{\{l_j^+\}}' \sum_{\{l_j^-\}}'  \braket{b| (C^{\mcirc\mcirc}_{\du,1})^{l_{n}^-}\mathcal{E}_{2} \dots (A^{\mcirc\mcirc}_{\du,1})^{l_2^+} \mathcal{E}_{1} (C^{\mcirc\mcirc}_{\du,1})^{l_{1}^-} \mathcal{E}_{2} (A^{\mcirc\mcirc}_{\du,1})^{l_{1}^+}|a},  & x \in \mathbb{Z} + \frac{1}{2},
 \end{cases}
 \label{eq:generalSkeletons}
\ee
\ew
where $\bar n_{1}={\rm min}(\tilde x_{+}, \tilde x_{-}-1)$, $\bar n_{2}={\rm min}(\tilde x_{-}-1,\tilde x_{+}-1)$. We denoted by
\be
\mathcal{E}_{1} = \begin{tikzpicture}[baseline=(current  bounding  box.center), scale=1]
\def\eps{0.5};
\Wgategreen{0}{0};
\Text[x=0,y=-0.6]{}
\draw[thick, fill=white] (-.5, .50) circle (0.1cm); 
\draw[thick, fill=white] (-.5, -.50) circle (0.1cm); 
\end{tikzpicture},
\qquad\qquad 
\mathcal{E}_{2} = \begin{tikzpicture}[baseline=(current  bounding  box.center), scale=1]
\def\eps{0.5};
\Wgategreen{0}{0};
\Text[x=0,y=-0.6]{}
\draw[thick, fill=white] (.5, .50) circle (0.1cm); 
\draw[thick, fill=white] (.5, -.50) circle (0.1cm); 
\end{tikzpicture},
\ee
the ``defect-maps'' and the primed sums are subject to the constraints 
\begin{align}
&\displaystyle \sum_{j=1}^{n+1} l_j^+ = x_+ - n, \quad \sum_{j=1}^n l_j^- = x_- - n-1,\quad x \in \mathbb{Z},
\\
&\displaystyle \sum_{j=1}^{n} l_j^+ = x_+ - n, \quad \sum_{j=1}^n l_j^- = x_- - n,\quad x \in \mathbb{Z} + \tfrac{1}{2}.
\label{eq:generalConstraints}
\end{align}
The expression~\eqref{eq:generalSkeletons} is a direct generalisation of \eqref{eq:skeletoncontribution}, where one replaces numbers $a, c, p\varepsilon, q\varepsilon$ with $d^2\times d^2$ matrices $A^{\mcirc\mcirc}_{\du,1}$, $C^{\mcirc\mcirc}_{\du,1}$, ${\cal E}_1$, ${\cal E}_2$. 
 
The conditions on the spectrum of horizontal and vertical transfer matrices mentioned in Sec.~\ref{subsub2} are sufficient to rigorously prove the dominance of skeleton contributions at low densities also in the generic case (where the matrices are constructed with non-reduced dual-unitary gates). Although we cannot analytically determine the family of gates $W_\du$ for which the conditions are fulfilled, we numerically identify such a family for quantum circuits of qubits ($d=2$)  in Sec.~\ref{sec:genericnumerics}. For gates in this family the argument discussed in Sec.~\ref{subsub3} remains valid as well. Accordingly, we find that if $x_+$ and $x_-$ are both large, \eqref{eq:generalSkeletons} gives the most relevant contribution to correlations also for $\delta=1$ and $\eta\ll1$.

Let us now comment on the physical content of \eqref{eq:generalSkeletons}. The sum is composed of powers of one dimensional maps. These maps typically have non-trivial eigenvalues strictly smaller than one, resulting in exponential decay of each term in the sum. Therefore,  also the sum of all terms will generically decay exponentially in $t$. For some special choices of the parameters, however, the exponent of the decay can approach 0. For instance, if the circuit has a local conserved charge, e.g. the magnetisation, we expect diffusive correlations that decay with time as $\propto {1}/{t}$. This is in principle allowed by \eqref{eq:generalSkeletons}.

\subsection{Uncorrelated Random Perturbations}

Systematic defects (homogeneous in the space-time) have a much more drastic effect on the structure of correlations than random ones. Indeed, averages over \emph{uncorrelated random defects} \eqref{eq:defect} (with $P$ distributed according to the Gaussian Unitary Ensemble) preserve the dual-unitary form of correlations. The only effect of the defects is an additional damping factor that causes or enhances --- depending on the degree of ergodicity of the unperturbed dual-unitary circuit (see the discussion at the end of Sec.~\ref{sec:dualunitary}) --- the exponential decay of the correlations along the light cone edge (light ray). Specifically we have 
\bw
\begin{align}
\mathbb E_{\rm GUE}\!\left[\braket{b_{x+y}|a_y(t)}\right]\!\! =&  {\rm mod}(2y,2) \delta_{x+t} \! \braket{b | (C_{\du,1}^{\mcirc \mcirc})^{2t}  | a} \!\!\left[\frac{K_{\rm GUE}(\eta)-1}{d^4-1}\right]^{\tilde x_-}\!\!\!\!\!\notag\\
&+  {\rm mod}(2y+1,2) \delta_{x-t}  \braket{a |  (A_{\du,1}^{\mcirc\mcirc})^{2t} |b} \left[\frac{K_{\rm GUE}(\eta)-1}{d^4-1}\right]^{\tilde x_+}\!\!\!
\label{eq:GUEaveragedcorrelations}
\end{align}
\ew
where the transfer matrices $A_{\du,1}^{\mcirc\mcirc}$ and $C_{\du,1}^{\mcirc\mcirc}$ are defined as in \eqref{eq:transferMat} and \eqref{eq:transferMat2} but in terms of the unperturbed dual-unitary gate $W_\du$, the rescaled variables light cone coordinates are defined in \eqref{eq:rescaledxpm}, and 
\be
K_{\rm GUE}(x)\equiv \mathbb E_{\rm GUE}\left[ |{\rm tr}[e^{i x P}]|^2 \right],
\ee
is the spectral form factor of the Gaussian Unitary Ensemble ($\mathbb E_{\rm GUE}[\cdot]$ is the ensemble average).

An interesting point to note is that Gaussian defects with infinite strength allow for non-zero correlations, indeed  
\be
\lim_{\eta\to\infty}K_{\rm GUE}(\eta) = d^2\,.
\ee
This should be compared with Haar random defects, which, instead, trivialise all correlations, i.e.\ the averaged correlation function becomes Kronecker delta in space-time for any positive density $\delta > 0$ (see Appendix~\ref{sec:averagegate}). The two averages only agree in the $d\to\infty$ limit.

\section{Reduced Gates}
\label{sec:U(1)avg} 

In this section we study the problem outlined in Sec.~\ref{sec:perturbativeexpansion} considering the minimal setting where the wires in \eqref{eq:latticewithdefects} are effectively two-dimensional. 

We begin considering the quantum circuit defined in Sec.~\ref{sec:techintro} in the following special case. \\

\noindent (i) Fix $d=2$ and focus on unitary gates~\eqref{eq:Ugate} of the form
\be
\bar U = (e^{i \phi_1 \sigma^z} \otimes e^{i \phi_2 \sigma^{z}})\cdot U \cdot (e^{i \phi_3 \sigma^z} \otimes e^{i \phi_4 \sigma^z})
\label{eq:Ubar}
\ee
where $U$ is a generic $U(4)$ matrix, $\sigma^z$ is a Pauli matrix, and the phases $\phi_1, \phi_2, \phi_3, \phi_4$ are independent random variables uniformly distributed in $[0,2\pi]$. Once again, the random variables $\{\phi_j\}$ at each gate/space-time point are considered independent.\\ 

\noindent (ii) Consider two-point correlations \emph{averaged} with respect to all $\{\phi_j\}$. This corresponds to averaging over the single-site Haar measure of the $U(1)$ group. On the level of the folded gate $W$ (cf.~\eqref{eq:doublegate}) the average corresponds to a simple projection, i.e.  
\be
W \longmapsto  w =  \mathbb E_{\{\phi_j\}}[W]= (P_z\otimes P_z) W (P_z\otimes P_z)
\label{eq:averageaction}
\ee
where 
\be
P_z:= \ket{\1}\bra{\1}+\ket{\sigma^z}\bra{\sigma^z}.
\ee
In other words, the average over the Haar $U(1)$ measure projects each wire onto the subspace spanned by the diagonal matrices $\{\ket{\mcirc}\equiv \ket{\1},\ket{\mcircf}\equiv \ket{\sigma^z}\}$ (cf. \eqref{eq:operatorstate}), and, from now on, we 
conveniently use this reduced basis notation. This means that the averaged 2-qubit folded gates are effectively $4\times 4$ matrices (while they are $16\times 16$ without the average).\\

We see that, even though the $U(1)$ average simplifies the problem by reducing the size of the relevant local Hilbert space, it does not trivialise the two-folded gate and in turn the two-point correlations. This is in contrast to most averages over Haar random unitary gates that have been considered in  recent literature~\cite{VL20, RandomCircuitsEnt,Nahum:operatorspreadingRU,Keyserlingk,RPK:conservationlaws1,Kemani,Chalker,Chalker2,SPHS18, Chalker3, JHN:OperatorEntanglement, LiCF18,CBQA19,GuHu19,LiCF19,SkRN19,VPYL19,ZGWG19,BaCA20, ZN:statmech, RPK:conservationlaws, H:conservationlaws, RPK:conservationlaws2,LaAB20,Hunt19}: in the latter cases the average of the doubled gate $W$ is trivial (as in \eqref{eq:HaarUd2average}) and so are two-point functions. The minimal folded gate with a non-trivial average is the one composed of four copies of the time evolution operator (two copies of $U$ and two copies of its hermitian conjugate $U^\dag$). 

To be more specific, let us count the number of free parameters in the gate \eqref{eq:averageaction}. This can be conveniently done using the following parametrisation for the generic unitary $U$ (cf.~\eqref{eq:Ubar})~\cite{KC:para, VW:para}  
\be
U = e^{i\phi}(u_1 \otimes u_2) V[\{J_j\}]  (u_3^\dag \otimes u_4^\dag)\,.
\label{eq:gate}
\ee
Here $\phi\in \mathbb R$, 
\be
V[\{J_j\}] \!=\! \exp\Bigl[i(J_1 \,\sigma^x\! \otimes  \sigma^x + J_2 \,\sigma^y\! \otimes  \sigma^y+J_3 \,\sigma^z\! \otimes  \sigma^z)\Bigr],
\label{eq:gateV}
\ee
and $u_1,u_2,u_3,u_4$ are elements of $SU(2)$ in the fundamental representation. They can be expressed in terms of Euler angles 
\be
u_j = e^{i (\alpha_j/2) \sigma^z}  e^{i (\beta_j/2) \sigma^y} e^{i (\gamma_j/2) \sigma^z}\,.
\label{eq:parametrisationu}
\ee
From this parametrisation it is easy to see that the averaged folded gate \eqref{eq:averageaction} depends on $11$ angles: two Euler angles ($\beta_j$ and $\gamma_j$) for each single-wire unitary $u_j$, and the parameters $J_1, J_2, J_3$. The explicit form of the averaged gate in the basis \eqref{eq:reducedbasis} is given by Eq.~\eqref{eq:reducedW}, where the explicit parametrisation of the parameters $p\varepsilon,a,\dots, g$ in terms of the above angles is given in  Appendix~\ref{sec:parametersW}. 

It is important to stress that the averaging procedure described above can be carried on for any fixed value of the angles $\{\beta_j,\gamma_j,J_j\}$. In particular, setting $J_1=J_2=\pi/4$ one averages over gates in the dual-unitary sub-class~\cite{BKP:dualunitary}. In this specific case, the averaged gate, now depending on 7 parameters (see Appendix~\ref{sec:parametersW}), takes the form~\eqref{eq:reducedW0}. Comparing with \eqref{eq:reducedW} we see that the effect of requiring dual unitarity is to set $\varepsilon$ to zero. Note that after the averaging the gate continues to fulfil the dual-unitality conditions~\eqref{eq:dualunitality}.  

Before turning to the analysis of correlation functions we make a final remark. As shown in Appendix~\ref{sec:markovchains}, one can produce gates of the form \eqref{eq:reducedW} and \eqref{eq:reducedW0} considering (dual-)bistochastic gates conjugated with $H\otimes H$ (cf.\ Eq.~\eqref{eq:Hadamard}). Interestingly, there is a one-to-one correspondence between averaged (dual-)unitary gates and \mbox{(dual-)bistochastic} gates. Namely, given a legitimate averaged (dual-)unitary gate one can produce a legitimate \mbox{(dual-)bistochastic} gate by conjugating it with $H\otimes H$ and vice versa. This can be established numerically using the explicit parameterisations reported in Appendix~\ref{sec:parametersW}.

\subsection{Correlation functions as sums over paths}
\label{sec:sumofpaths}

Let us now examine Eq.~\eqref{eq:mapcorrelations} in the minimal setting under consideration in the section: two-dimensional wires and gates of the form~\eqref{eq:reducedW}. Inserting a resolution of the identity $\1 = \ket{\mcirc}\!\!\bra{\mcirc}+  \ket{\mcircf}\!\!\bra{\mcircf}$ at each wire, we can explicitly decompose~\eqref{eq:mapcorrelations} into the sum of $2^{x_+ x_-}$ terms. Each term admits a simple interpretation as a (weighted) path on the two-dimensional space, or, analogously, as the spatial configuration of a certain polymer. To see this, let us introduce a different diagrammatic representation of the weights in terms of ``tiles'' where we connect $\mcircf$s with solid lines and ignore $\mcirc$s. For example,
\be
\begin{tikzpicture}[baseline=(current  bounding  box.center), scale=0.8]
\Text[x=-5.5,y=0.0, anchor = center]{$e$}
\Text[x=-5,y=0.0, anchor = center]{$=$}
\Text[x=-2.5,y=0.0, anchor = center]{$=$}
\begin{scope}[rotate around={-45:(-3.75,0)}]
\draw (-4.25,0.5) -- (-3.25,-0.5);
\draw (-4.25,-0.5) -- (-3.25,0.5);
\draw[ thick, fill=mygreen, rounded corners=2pt] (-4,0.25) rectangle (-3.5,-0.25);
\draw[thick] (-3.75,0.15) -- (-3.6,0.15) -- (-3.6,0);
\draw[thick, fill=black] (-4.25,0.5) circle (0.1cm); 
\draw[thick, fill=black] (-3.25,0.5) circle (0.1cm); 
\draw[thick, fill=white] (-4.25,-0.5) circle (0.1cm); 
\draw[thick, fill=black] (-3.25,-0.5) circle (0.1cm); 
\end{scope}
\eGate{-1.75}{0}
\end{tikzpicture}.
\ee
The complete set of tiles (corresponding to non-zero coefficients of the gate \eqref{eq:reducedW}) is given by
\begin{align}
\begin{tikzpicture}[baseline=(current  bounding  box.center), scale=0.7]
\oGate{0}{1.5} \Text[x=1,y=1.5, anchor = center]{$=1$}
\aGate{2.5}{1.5} \Text[x={2.5+1},y=1.5, anchor = center]{$=a$}
\cGate{5}{1.5} \Text[x={5+1},y=1.5, anchor = center]{$=c$}
\gGate{0}{-1.5} \Text[x={1},y=-1.5, anchor = center]{$=g$}
\bGate{2.5}{-1.5} \Text[x={2.5+1},y=-1.5, anchor = center]{$=b\, .$}
\dGate{0}{0} \Text[x=1,y=0.0, anchor = center]{$=d$}
\fGate{2.5}{0} \Text[x={2.5+1},y=0.0, anchor = center]{$=f$}
\eGate{5}{0} \Text[x={5+1},y=0.0, anchor = center]{$=e$}
\doGate{7.5}{0} \Text[x={7.5+1},y=0.0, anchor = center]{$=p \varepsilon$}
\dtGate{7.5}{1.5} \Text[x={7.5+1},y=1.5, anchor = center]{$=q \varepsilon$}
\end{tikzpicture}
\label{eq:tiles}
\end{align}
Then, the correlation function is expressed as
\bw
\begin{align}
 \braket{\mcircf_{x} |\mcircf_0(t)} = \sum_{s_{ij} \in \text{tiles}}
\begin{tikzpicture}[baseline=(current  bounding  box.center), scale=0.65]
\foreach \i in {1,...,4}{
\foreach \j in {1,...,3}{
\sGate{\i}{4-\j}{\i}{\j}
}}
\draw[line width=\lineW] (-0.0,1) -- (0.5,1);
\draw[line width=\lineW] (4.5,3) -- (5,3);
\end{tikzpicture}
&=  
\begin{tikzpicture}[baseline=(current  bounding  box.center), scale=0.65]
\foreach \i in {1,...,4}{
\foreach \j in {1,...,3}{
\oGate{\i}{\j}
}}
\draw[line width=\lineW] (-0.0,1) -- (0.5,1);
\draw[line width=\lineW] (4.5,3) -- (5,3);
\dtGate{1}{1}
\cGate{1}{2}
\doGate{1}{3}
\aGate{2}{3}\aGate{3}{3}\aGate{4}{3}
\end{tikzpicture}
+
\begin{tikzpicture}[baseline=(current  bounding  box.center), scale=0.65]
\foreach \i in {1,...,4}{
\foreach \j in {1,...,3}{
\oGate{\i}{\j}
}}
\draw[line width=\lineW] (-0.0,1) -- (0.5,1);
\draw[line width=\lineW] (4.5,3) -- (5,3);
\aGate{1}{1}
\dtGate{2}{1}
\cGate{2}{2}
\doGate{2}{3}
\aGate{3}{3}\aGate{4}{3}
\end{tikzpicture}+\dots
\nonumber\\
&+
\begin{tikzpicture}[baseline=(current  bounding  box.center), scale=0.65]
\foreach \i in {1,...,4}{\foreach \j in {1,...,3}{\oGate{\i}{\j} }}
\draw[line width=\lineW] (-0.0,1) -- (0.5,1);
\draw[line width=\lineW] (4.5,3) -- (5,3);
\fGate{1}{1}
\cGate{1}{2}
\doGate{1}{3}
\aGate{2}{3}\bGate{3}{3}\aGate{4}{3}
\aGate{2}{1}\dtGate{3}{1}
\cGate{3}{2}
\end{tikzpicture}
+
\begin{tikzpicture}[baseline=(current  bounding  box.center), scale=0.65]
\foreach \i in {1,...,4}{\foreach \j in {1,...,3}{\oGate{\i}{\j} }}
\draw[line width=\lineW] (-0.0,1) -- (0.5,1);
\draw[line width=\lineW] (4.5,3) -- (5,3);
\fGate{1}{1}
\cGate{1}{2}
\doGate{1}{3}
\aGate{2}{3}\bGate{3}{3}\bGate{4}{3}
\aGate{2}{1}\dtGate{3}{1}
\eGate{3}{2}
\dtGate{4}{2}
\end{tikzpicture}
+\dots\, .
\label{eq:partitionF}
\end{align}
\ew
We used that all non-trivial states (i.e. those orthogonal to the ``identity'' $\ket{\mcirc}$) are proportional to $\ket{\mcircf}$. Namely, 
\be
\braket{b_{x} |a_0(t)} \propto \braket{\mcircf_{x} |\!\mcircf_0\!(t)}\!,
\ee
for all $a,b$ such that ${\rm tr }[a]={\rm tr }[b]=0$. Moreover, in writing~\eqref{eq:partitionF} we considered the case of $x,y\in \mathbb Z$. The other three possibilities correspond to different positions of the initial and final lines. Specifically, for $y\in \mathbb Z+1/2$ the initial line enters the bottom left tile from below and for $x+y\in \mathbb Z+1/2$ the final line exits the top right tile from above. 

As suggested by the diagrammatic representation in Eq.~\eqref{eq:partitionF}, the correlation function can be thought of as the sum of paths with fixed end points, where different configurations have different (possibly negative) associated weights. The paths are allowed to split and merge, with weights $e,f$ and $b,d$, respectively, but they cannot ``jump'': dangling ends of lines inside the diagram are forbidden. Finally, since  downward and leftward turns are forbidden (cf. \eqref{eq:tiles}), the paths can form loops only if they split (see, e.g., the last two diagrams on the r.h.s. of \eqref{eq:partitionF}).

\subsection{Exactly solvable cases}
\label{sec:exactlysolvablereducedgates}

By inspecting the set of allowed tiles \eqref{eq:tiles}, we identify four non-trivial solvable cases, which we will analyse in the following two subsections. 

\subsubsection{Almost dual-unitary cases}

As observed before, the case $\varepsilon=0$ corresponds to the dual-unitary point and all correlations propagate along straight paths, with no turns allowed. Depending on the initial conditions (i.e. on whether $y$ is integer or half-odd integer) the straight lines are going either upward or rightward. This is simply a path-sum reformulation of the general dual-unitary result \eqref{eq:DUcorrelations} (cf. Ref.~\cite{BKP:dualunitary}). 

Remarkably, however, the correlations remain exactly solvable also when \emph{only one} of $p \varepsilon$ or $q \varepsilon$ vanishes. Indeed, in this case we only allow for paths with a single turn and the rules~\eqref{eq:tiles} do not permit any ``dressing'', i.e. any thickening of the lines due to loops (see Sec.~\ref{sec:PTdensity} for a more precise definition). For example, choosing $p =0$ and $q \varepsilon  \neq 0$, the correlations take the form given in \eqref{eq:skeletoncontribution}, where only the first terms in each of the two lines can be non-zero (depicted in \eqref{eq:one_turn}). We see that, apart from the straight paths described above, Eq.~\eqref{eq:skeletoncontribution} establishes non-trivial correlations between integer and half-odd integer points.  

\subsubsection{Non-dual-unitary solvable cases}

Surprisingly, the tiles \eqref{eq:tiles} contain another non-trivial solvable limit that is \emph{not} dual-unitary. Indeed, there can be no loops in the paths whenever the ``split weights'' $f$ and $e$ or the ``merge weights''  $b$ and $d$ are zero. Consequently, when ${(e,f)=(0,0)}$ or ${(b,d)=(0,0)}$ the only allowed paths are the ``skeleton diagrams'' described in Sec.~\ref{sec:perturbativeexpansion} (see, e.g., the first two diagrams on the r.h.s.\ of Eq.~\eqref{eq:partitionF}). In this case we can directly evaluate the correlation function \eqref{eq:partitionF} by summing all skeleton diagrams. In particular, if the defects cover a regular sub-lattice as in \eqref{eq:latticewithdefects}, one obtains the expression~\eqref{eq:skeletoncontribution}.

The formula~\eqref{eq:skeletoncontribution} can be derived by straightforward combinatorics. Let us detail its derivation considering the case $x\in\mathbb Z$ as an example. We begin by noting that, since the path goes from the bottom left corner to the upper right one, it must involve the same number $n$ of right- and up- turns, with weights respectively given by $p  \varepsilon $ and $q  \varepsilon $. The minimum number of such turns is $0$ (it contributes only for $x_-=1$), while the maximum, ${\rm min}(\tilde x_{+},\tilde x_{-}-1)$, is set by the size of the defect's sub-lattice (the rescaled light cone coordinates $\tilde x_\pm$ are defined in Eq.~\eqref{eq:rescaledxpm}).
For each fixed $n$, one has $n$ turns, $x_+-n$ horizontal segments, contributing with a factor $a^{x_+-n}$, and $x_--1-n$ vertical ones, contributing with a factor $c^{x_--1-n}$. To count all the possible ways in which the elementary pieces can be combined, we can consider the horizontal and the vertical directions separately. In the horizontal direction we need to distribute $n$ indistinguishable pairs of turns (first up and then right) in $\tilde x_+$ positions, leading to the combinatorial factor 
\be
\binom{\tilde x_{+}}{n}.
\ee 
In the vertical direction we are instead more constrained. Indeed, the first and last turns must be in the first and last row, respectively. The other $n-1$ pairs can be distributed freely in the remaining $\tilde x_--2$ positions, leading to 
\be
\binom{\tilde x_{-}-2}{n-1}.
\ee
Putting it all together, we obtain the desired result. For defects on irregular sub-lattices the reasoning is similar but one has different combinatorial coefficients depending on the actual shape of the sub-lattice. 


\subsection{Perturbation theory around the dual-unitary point}
\label{sec:PT_DU}

Let us now consider the case of circuits that are perturbed away from the dual-unitary point. Namely, we consider the setting introduced in Sec.~\ref{sec:perturbativeexpansion}: among the $x_+ x_-$ gates in the lattice~\eqref{eq:pictorialcorrelation} there are $(1-\delta) x_+ x_-$ dual-unitary gates and $\delta x_+ x_-$ perturbations (or defects) breaking dual-unitarity (see the pictorial representation in Eq.~\eqref{eq:latticewithdefects}). As discussed in Sec~\ref{sec:perturbativeexpansion}, we use two parameters to control the perturbations: strength $\varepsilon$  (which is proportional to $\eta$ in Eq.~\eqref{eq:W} for small $\eta$) and density $\delta$. We begin by considering the case of perturbations in the density of defects, which allows for a more rigorous analysis. Later we will see that, surprisingly, most of the rigorous conclusions drawn in that case apply also for small $\varepsilon$ and arbitrary $\delta\leq1$.  

\subsubsection{Low density, unit strength}
\label{sec:PTdensity}

We begin our analysis by focusing on fixed defects with arbitrary strength $\varepsilon$ placed on a regular sublattice as in Eq.~\eqref{eq:latticewithdefects}. In this case there are regular strips (vertical and horizontal) composed only of dual-unitary gates. Whenever the widths $\nu_\pm$ (cf.~\eqref{eq:latticewithdefects}) of these strips become large enough, we can simplify the contribution by considering only the leading eigenvectors of the strips' transfer matrices $a_{\du,x}^{\mcirc\mcirc}$ and $c_{\du,x}^{\mcirc\mcirc}$. These are defined as in Eqs.~\eqref{eq:transferMat}--\eqref{eq:transferMat2} but using the $4\times4$ dual-unitary (or rather, dual bistochastic) gate $w_\du$ (cf.~\eqref{eq:reducedW0}), namely 
\begin{align}
 a_{\du,x}^{\mcirc\mcirc} = \begin{tikzpicture}[baseline={([yshift=-0.5ex] current  bounding  box.center)}, scale=0.75]
\foreach \j in {1,2}{
\draw[rotate around ={45: (-6+1.5*\j,0.5)}] (-.5-6+1.5*\j,0) -- (0.5-6+1.5*\j,1);
\draw[rotate around ={45: (-6+1.5*\j,0.5)}] (-5.4+1.5*\j,-0.1) -- (-6.6+1.5*\j,1.1);
\draw[thick, fill=myorange, rounded corners=2pt, rotate around ={45: (-6+1.5*\j,0.5)}] (-0.25-6+1.5*\j,0.25) rectangle (.25-6+1.5*\j,0.75);
\draw[thick, rotate around ={-135: (-6+1.5*\j,0.5)}] (-6+1.5*\j,0.65) -- (-5.85+1.5*\j,0.65) -- (-5.85+1.5*\j,0.5);
}
\Text[x=-1.55,y=0.5]{$\cdots$}
\foreach \j in {4,5}{
\draw[rotate around ={45: (-6+1.5*\j,0.5)}] (-.5-6+1.5*\j,0) -- (0.5-6+1.5*\j,1);
\draw[rotate around ={45: (-6+1.5*\j,0.5)}] (-5.4+1.5*\j,-0.1) -- (-6.6+1.5*\j,1.1);
\draw[thick, fill=myorange, rounded corners=2pt, rotate around ={45: (-6+1.5*\j,0.5)}] (-0.25-6+1.5*\j,0.25) rectangle (.25-6+1.5*\j,0.75);
\draw[thick, rotate around ={-135: (-6+1.5*\j,0.5)}] (-6+1.5*\j,0.65) -- (-5.85+1.5*\j,0.65) -- (-5.85+1.5*\j,0.5);
}
\draw[thick, fill=white] (-5.25,0.5) circle (0.1cm);
\draw[thick, fill=white] (2.25,0.5) circle (0.1cm);
\draw [thick, decorate, decoration={brace,amplitude=8pt, raise=4pt, mirror},yshift=0pt]
(-4.75,-0.1) -- (1.75,-0.1) node [black,midway,yshift=-0.95cm, anchor = south] {$x$};
\Text[x=2.3,y=2.2]{}
\end{tikzpicture},
\label{eq:DUtransferMatcc}
\\
c_{\du,x}^{\mcirc\mcirc} = \begin{tikzpicture}[baseline={([yshift=-0.5ex] current  bounding  box.center)}, scale=0.75]
\foreach \j in {1,2}{
\draw[rotate around ={45: (-6+1.5*\j,0.5)}] (-.5-6+1.5*\j,0) -- (0.5-6+1.5*\j,1);
\draw[rotate around ={45: (-6+1.5*\j,0.5)}] (-5.4+1.5*\j,-0.1) -- (-6.6+1.5*\j,1.1);
\draw[thick, fill=myorange, rounded corners=2pt, rotate around ={45: (-6+1.5*\j,0.5)}] (-0.25-6+1.5*\j,0.25) rectangle (.25-6+1.5*\j,0.75);
\draw[thick, rotate around ={-45: (-6+1.5*\j,0.5)}] (-6+1.5*\j,0.65) -- (-5.85+1.5*\j,0.65) -- (-5.85+1.5*\j,0.5);
}
\Text[x=-1.55,y=0.5]{$\cdots$}
\foreach \j in {4,5}{
\draw[rotate around ={45: (-6+1.5*\j,0.5)}] (-.5-6+1.5*\j,0) -- (0.5-6+1.5*\j,1);
\draw[rotate around ={45: (-6+1.5*\j,0.5)}] (-5.4+1.5*\j,-0.1) -- (-6.6+1.5*\j,1.1);
\draw[thick, fill=myorange, rounded corners=2pt, rotate around ={45: (-6+1.5*\j,0.5)}] (-0.25-6+1.5*\j,0.25) rectangle (.25-6+1.5*\j,0.75);
\draw[thick, rotate around ={-45: (-6+1.5*\j,0.5)}] (-6+1.5*\j,0.65) -- (-5.85+1.5*\j,0.65) -- (-5.85+1.5*\j,0.5);
}
\draw[thick, fill=white] (-5.25,0.5) circle (0.1cm);
\draw[thick, fill=white] (2.25,0.5) circle (0.1cm);
\draw [thick, decorate, decoration={brace,amplitude=8pt, raise=4pt, mirror},yshift=0pt]
(-4.75,-0.1) -- (1.75,-0.1) node [black,midway,yshift=-0.95cm, anchor = south] {$x$};
\Text[x=2.3,y=2.2]{}
\end{tikzpicture}.
\label{eq:DUtransferMatBcc}
\end{align}
To treat these matrices we make use of the following rigorous result, proved in Appendix~\ref{sec:spectrum}.
\begin{property}
\label{prop:P2}
The matrices $a_{\du,x}^{\mcirc \mcirc}, c_{\du,x}^{\mcirc \mcirc}$ take the following block diagonal form 
\begin{align}
& a_{\du,x}^{\mcirc\mcirc}  = p^{\mcirc}_{x,0} + a \sum_{k=1}^x  p^{\mcirc}_{x,k} +  r_{1,x}, \label{eq:specdec1}\\
& c_{\du,x}^{\mcirc\mcirc}  = p^{\mcirc}_{x,0} + c \sum_{k=1}^x  p^{\mcirc}_{x,k} +  r_{2,x}, \label{eq:specdec2}
\end{align}
where we defined 
\begin{align}
p^{\mcirc}_{x,0} &:= \ket{\mcirc}^{\otimes x} \bra{\mcirc}^{\otimes x},\nonumber\\
p^{\mcirc}_{x,k} &:= \ket{\underbrace{\mcirc\ldots\mcirc\mcircf}_{k}\mcirc\ldots\mcirc}\!\!\bra{\underbrace{\mcirc\ldots\mcirc\mcircf}_{k}\mcirc\ldots\mcirc},
\label{eq:proj}
\end{align}
and the ``reminders'' $r_{1,x}, r_{2,x}$ are non-zero only in the subspaces defined by the projectors $\1-\sum_{j=0}^x p^{\mcirc}_{x,j}$. Moreover,  
\be
\| r_{1,x}\| \leq a^2  + \frac{|b f|}{1-\alpha},\qquad \text{for}\qquad |a| > a^2  + \frac{|b f|}{1-\alpha},
\label{eq:boundR1}
\ee
and analogously  
\be
\| r_{2,x}\| \leq c^2  + \frac{|d e|}{1-\beta},\qquad \text{for}\qquad |c| > c^2  + \frac{|d e|}{1-\beta},
\ee
where $\alpha,\beta\in[0,1]$ are, respectively, the operator norms (largest singular values) of the matrices 
\be
\begin{pmatrix} c & e\\d& g \end{pmatrix}, \quad \begin{pmatrix} a & f\\b& g \end{pmatrix}.
\ee
\end{property}

\noindent Consider now a vertical strip $x_-\times \nu_+$ with horizontal transfer matrix $a_{\du,x_-}^{\mcirc\mcirc}$. Property~\ref{prop:P2} guarantees that for  
\be
|a| > a^2  + \frac{|b f|}{1-\alpha}\,,
\label{eq:condition}
\ee
we can make the replacement
\be
a_{\du,x_-}^{\mcirc \mcirc} \longmapsto \!p^{\mcirc}_{x_-,0} \!+ \!a\! \sum_{k=1}^{x_-}  p^{\mcirc}_{x_-,k} 
\label{eq:replacement}
\ee
with an error  bounded in operator norm $\| r_{1,x_{-}}\|$ by $a^2  + {|b f|}/{(1-\alpha)}$. The replacement \eqref{eq:replacement} makes the calculation of correlations extremely easy. 
Let us illustrate this considering the diagram on the r.h.s. of \eqref{eq:latticewithdefects}, which becomes 
\bw
\begin{align}
\braket{\mcircf_{x}|\!\!\mcircf_0\!\!(t)} \!=\! a^{x_+-3}\sum_{j_1,j_2}
\begin{tikzpicture}[baseline=(current  bounding  box.center), scale=0.7]
\def\eps{-0.5};
\foreach \i in {0,...,4}
\foreach \j in {-3}{
\draw (-4.5+1.5*\j,1.5*\i) -- (-3+1.5*\j,1.5*\i);
\draw (-3.75+1.5*\j,1.5*\i-.75) -- (-3.75+1.5*\j,1.5*\i+.75);
\draw[ thick, fill=myorange, rounded corners=2pt, rotate around ={45: (-3.75+1.5*\j,1.5*\i)}] (-4+1.5*\j,0.25+1.5*\i) rectangle (-3.5+1.5*\j,-0.25+1.5*\i);
\draw[thick, rotate around ={-45: (-3.75+1.5*\j,1.5*\i)}] (-3.75+1.5*\j,.15+1.5*\i) -- (-3.6+1.5*\j,.15+1.5*\i) -- (-3.6+1.5*\j,1.5*\i);
\draw[thick, fill=white] (-3-3*1.5,1.5*\i) circle (0.1cm); 
\draw[thick, fill=white] (-4.5-3*1.5,1.5*\i) circle (0.1cm); 
\draw[thick, fill=white] (-3.75+1.5*\j,-0.75) circle (0.1cm); 
\draw[thick, fill=white] (-3.75+1.5*\j,6.75) circle (0.1cm);
}
\foreach \i in {0,...,2}
\foreach \j in {-1}{
\draw[ thick, fill=mygreen, rounded corners=2pt, rotate around ={45: (-3.75+4.5*\j,3*\i)}] (-4+4.5*\j,0.25+3*\i) rectangle (-3.5+4.5*\j,-0.25+3*\i);
\draw[thick, rotate around ={-45: (-3.75+4.5*\j,3*\i)}] (-3.75+4.5*\j,.15+3*\i) -- (-3.6+4.5*\j,.15+3*\i) -- (-3.6+4.5*\j,3*\i);
}
\draw[thick, fill=black] (-3-3*1.5,3) circle (0.1cm); 
\draw[thick, fill=black] (-4.5-3*1.5,0) circle (0.1cm); 
\draw [thick, decorate, decoration={brace,amplitude=4pt,mirror,raise=4pt},yshift=0pt]
(-7.5,0) -- (-7.5,3) node [black,midway,xshift=.55cm] {$j_1$};
\Text[x=-8.25,y=-1]{}
\foreach \i in {0,...,4}
\foreach \j in {-1}{
\draw (-4.5+1.5*\j,1.5*\i) -- (-3+1.5*\j,1.5*\i);
\draw (-3.75+1.5*\j,1.5*\i-.75) -- (-3.75+1.5*\j,1.5*\i+.75);
\draw[ thick, fill=myorange, rounded corners=2pt, rotate around ={45: (-3.75+1.5*\j,1.5*\i)}] (-4+1.5*\j,0.25+1.5*\i) rectangle (-3.5+1.5*\j,-0.25+1.5*\i);
\draw[thick, rotate around ={-45: (-3.75+1.5*\j,1.5*\i)}] (-3.75+1.5*\j,.15+1.5*\i) -- (-3.6+1.5*\j,.15+1.5*\i) -- (-3.6+1.5*\j,1.5*\i);
\draw[thick, fill=white] (-3+\j*1.5,1.5*\i) circle (0.1cm); 
\draw[thick, fill=white] (-4.5+\j*1.5,1.5*\i) circle (0.1cm); 
\draw[thick, fill=white] (-3.75+1.5*\j,-.75) circle (0.1cm); 
\draw[thick, fill=white] (-3.75+1.5*\j,6.75) circle (0.1cm);
}
\foreach \i in {0,...,2}
\foreach \j in {0}{
\draw[ thick, fill=mygreen, rounded corners=2pt, rotate around ={45: (-4.5-0.75+4.5*\j,3*\i)}] (-4.75-0.75+4.5*\j,0.25+3*\i) rectangle (-4.25-0.75+4.5*\j,-0.25+3*\i);
\draw[thick, rotate around ={-45: (-4.5-0.75+4.5*\j,3*\i)}] (-4.5-0.75+4.5*\j,.15+3*\i) -- (-4.35-0.75+4.5*\j,.15+3*\i) -- (-4.35-0.75+4.5*\j,3*\i);
}
\draw[thick, fill=black] (-3-1.5,4.5) circle (0.1cm); 
\draw[thick, fill=black] (-4.5-1.5,3) circle (0.1cm); 
\draw [thick, decorate, decoration={brace,amplitude=4pt,mirror,raise=4pt},yshift=0pt]
(-4.5,0) -- (-4.5,4.5) node [black,midway,xshift=.5cm]{};
\draw [thick, decorate, decoration={brace,amplitude=4pt,raise=4pt},yshift=0pt]
(-6,0) -- (-6,3) node [black,midway,xshift=-.5cm]{};
\foreach \i in {0,...,4}
\foreach \j in {1}{
\draw (-4.5+1.5*\j,1.5*\i) -- (-3+1.5*\j,1.5*\i);
\draw (-3.75+1.5*\j,1.5*\i-.75) -- (-3.75+1.5*\j,1.5*\i+.75);
\draw[ thick, fill=myorange, rounded corners=2pt, rotate around ={45: (-3.75+1.5*\j,1.5*\i)}] (-4+1.5*\j,0.25+1.5*\i) rectangle (-3.5+1.5*\j,-0.25+1.5*\i);
\draw[thick, rotate around ={-45: (-3.75+1.5*\j,1.5*\i)}] (-3.75+1.5*\j,.15+1.5*\i) -- (-3.6+1.5*\j,.15+1.5*\i) -- (-3.6+1.5*\j,1.5*\i);
\draw[thick, fill=white] (-3+\j*1.5,1.5*\i) circle (0.1cm); 
\draw[thick, fill=white] (-4.5+\j*1.5,1.5*\i) circle (0.1cm); 
\draw[thick, fill=white] (-3.75+1.5*\j,-.75) circle (0.1cm); 
\draw[thick, fill=white] (-3.75+1.5*\j,6.75) circle (0.1cm);
}
\foreach \i in {0,...,2}
\foreach \j in {1}{
\draw[ thick, fill=mygreen, rounded corners=2pt, rotate around ={45: (-3.75-4*0.75+4.5*\j,3*\i)}] (-4-4*0.75+4.5*\j,0.25+3*\i) rectangle (-3.5-4*0.75+4.5*\j,-0.25+3*\i);
\draw[thick, rotate around ={-45: (-3.75-4*0.75+4.5*\j,3*\i)}] (-3.75-4*0.75+4.5*\j,.15+3*\i) -- (-3.6-4*0.75+4.5*\j,.15+3*\i) -- (-3.6-4*0.75+4.5*\j,3*\i);
}
\draw[thick, fill=black] (-3+1.5,6) circle (0.1cm); 
\draw[thick, fill=black] (-4.5+1.5,4.5) circle (0.1cm); 
\draw [thick, decorate, decoration={brace,amplitude=4pt,raise=4pt},yshift=0pt]
(-3,0) -- (-3,4.5) node [black,midway,xshift=-.5cm] {$j_2$};
\Text[x=-8.25,y=-1]{}
\end{tikzpicture}
= a^{x_+-3}\sum_{j_1,j_2}
\begin{tikzpicture}[baseline=(current  bounding  box.center), scale=0.7]
\def\eps{-0.5};
\foreach \i in {0,...,2}
\foreach \j in {-3}{
\draw (-4.5+1.5*\j,1.5*\i) -- (-3+1.5*\j,1.5*\i);
\draw (-3.75+1.5*\j,1.5*\i-.75) -- (-3.75+1.5*\j,1.5*\i+.75);
\draw[ thick, fill=myorange, rounded corners=2pt, rotate around ={45: (-3.75+1.5*\j,1.5*\i)}] (-4+1.5*\j,0.25+1.5*\i) rectangle (-3.5+1.5*\j,-0.25+1.5*\i);
\draw[thick, rotate around ={-45: (-3.75+1.5*\j,1.5*\i)}] (-3.75+1.5*\j,.15+1.5*\i) -- (-3.6+1.5*\j,.15+1.5*\i) -- (-3.6+1.5*\j,1.5*\i);
\draw[thick, fill=white] (-3-3*1.5,1.5*\i) circle (0.1cm); 
\draw[thick, fill=white] (-4.5-3*1.5,1.5*\i) circle (0.1cm); 
\draw[thick, fill=white] (-3.75+1.5*\j,-0.75) circle (0.1cm); 
\draw[thick, fill=white] (-3.75+1.5*\j,3.75) circle (0.1cm);
}
\foreach \i in {0,...,1}
\foreach \j in {-1}{
\draw[ thick, fill=mygreen, rounded corners=2pt, rotate around ={45: (-3.75+4.5*\j,3*\i)}] (-4+4.5*\j,0.25+3*\i) rectangle (-3.5+4.5*\j,-0.25+3*\i);
\draw[thick, rotate around ={-45: (-3.75+4.5*\j,3*\i)}] (-3.75+4.5*\j,.15+3*\i) -- (-3.6+4.5*\j,.15+3*\i) -- (-3.6+4.5*\j,3*\i);
}
\draw[thick, fill=black] (-3-3*1.5,3) circle (0.1cm); 
\draw[thick, fill=black] (-4.5-3*1.5,0) circle (0.1cm); 
\draw [thick, decorate, decoration={brace,amplitude=4pt,mirror,raise=4pt},yshift=0pt]
(-7.5,0) -- (-7.5,3) node [black,midway,xshift=.55cm] {$j_1$};
\Text[x=-8.25,y=-1]{}
\foreach \i in {2,...,3}
\foreach \j in {-1}{
\draw (-4.5+1.5*\j,1.5*\i) -- (-3+1.5*\j,1.5*\i);
\draw (-3.75+1.5*\j,1.5*\i-.75) -- (-3.75+1.5*\j,1.5*\i+.75);
\draw[ thick, fill=myorange, rounded corners=2pt, rotate around ={45: (-3.75+1.5*\j,1.5*\i)}] (-4+1.5*\j,0.25+1.5*\i) rectangle (-3.5+1.5*\j,-0.25+1.5*\i);
\draw[thick, rotate around ={-45: (-3.75+1.5*\j,1.5*\i)}] (-3.75+1.5*\j,.15+1.5*\i) -- (-3.6+1.5*\j,.15+1.5*\i) -- (-3.6+1.5*\j,1.5*\i);
\draw[thick, fill=white] (-3+\j*1.5,1.5*\i) circle (0.1cm); 
\draw[thick, fill=white] (-4.5+\j*1.5,1.5*\i) circle (0.1cm); 
\draw[thick, fill=white] (-3.75+1.5*\j,2.25) circle (0.1cm); 
\draw[thick, fill=white] (-3.75+1.5*\j,5.25) circle (0.1cm);
}
\foreach \i in {1,...,1}
\foreach \j in {0}{
\draw[ thick, fill=mygreen, rounded corners=2pt, rotate around ={45: (-4.5-0.75+4.5*\j,3*\i)}] (-4.75-0.75+4.5*\j,0.25+3*\i) rectangle (-4.25-0.75+4.5*\j,-0.25+3*\i);
\draw[thick, rotate around ={-45: (-4.5-0.75+4.5*\j,3*\i)}] (-4.5-0.75+4.5*\j,.15+3*\i) -- (-4.35-0.75+4.5*\j,.15+3*\i) -- (-4.35-0.75+4.5*\j,3*\i);
}
\draw[thick, fill=black] (-3-1.5,4.5) circle (0.1cm); 
\draw[thick, fill=black] (-4.5-1.5,3) circle (0.1cm); 
\draw [thick, decorate, decoration={brace,amplitude=4pt,mirror,raise=4pt},yshift=0pt]
(-4.5,0) -- (-4.5,4.5) node [black,midway,xshift=.5cm]{};
\draw [thick, decorate, decoration={brace,amplitude=4pt,raise=4pt},yshift=0pt]
(-6,0) -- (-6,3) node [black,midway,xshift=-.5cm]{};
\foreach \i in {3,...,4}
\foreach \j in {1}{
\draw (-4.5+1.5*\j,1.5*\i) -- (-3+1.5*\j,1.5*\i);
\draw (-3.75+1.5*\j,1.5*\i-.75) -- (-3.75+1.5*\j,1.5*\i+.75);
\draw[ thick, fill=myorange, rounded corners=2pt, rotate around ={45: (-3.75+1.5*\j,1.5*\i)}] (-4+1.5*\j,0.25+1.5*\i) rectangle (-3.5+1.5*\j,-0.25+1.5*\i);
\draw[thick, rotate around ={-45: (-3.75+1.5*\j,1.5*\i)}] (-3.75+1.5*\j,.15+1.5*\i) -- (-3.6+1.5*\j,.15+1.5*\i) -- (-3.6+1.5*\j,1.5*\i);
\draw[thick, fill=white] (-3+\j*1.5,1.5*\i) circle (0.1cm); 
\draw[thick, fill=white] (-4.5+\j*1.5,1.5*\i) circle (0.1cm); 
\draw[thick, fill=white] (-3.75+1.5*\j,3.75) circle (0.1cm); 
\draw[thick, fill=white] (-3.75+1.5*\j,6.75) circle (0.1cm);
}
\foreach \i in {2,...,2}
\foreach \j in {1}{
\draw[ thick, fill=mygreen, rounded corners=2pt, rotate around ={45: (-3.75-4*0.75+4.5*\j,3*\i)}] (-4-4*0.75+4.5*\j,0.25+3*\i) rectangle (-3.5-4*0.75+4.5*\j,-0.25+3*\i);
\draw[thick, rotate around ={-45: (-3.75-4*0.75+4.5*\j,3*\i)}] (-3.75-4*0.75+4.5*\j,.15+3*\i) -- (-3.6-4*0.75+4.5*\j,.15+3*\i) -- (-3.6-4*0.75+4.5*\j,3*\i);
}
\draw[thick, fill=black] (-3+1.5,6) circle (0.1cm); 
\draw[thick, fill=black] (-4.5+1.5,4.5) circle (0.1cm); 
\draw [thick, decorate, decoration={brace,amplitude=4pt,raise=4pt},yshift=0pt]
(-3,0) -- (-3,4.5) node [black,midway,xshift=-.5cm] {$j_2$};
\Text[x=-8.25,y=-1]{}
\end{tikzpicture},
\label{eq:correlationsreplacement}
\end{align}
\ew
where in the second step we used the unitality relations \eqref{eq:unitality} to contract the vertical lines. Using the explicit form of the one-dimensional transfer matrices 
\begin{align}
&\!\!\!\!a_{1}^{\mcirc\mcirc}\!\!=\begin{tikzpicture}[baseline=(current  bounding  box.center), scale=0.7]
\def\eps{-0.5};
\foreach \i in {0}
\foreach \j in {-3}{
\draw (-4.5+1.5*\j,1.5*\i) -- (-3+1.5*\j,1.5*\i);
\draw(-3.75+1.5*\j,1.5*\i-.75) -- (-3.75+1.5*\j,1.5*\i+.75);
\draw[ thick, fill=mygreen, rounded corners=2pt, rotate around ={45: (-3.75+1.5*\j,1.5*\i)}] (-4+1.5*\j,0.25+1.5*\i) rectangle (-3.5+1.5*\j,-0.25+1.5*\i);
\draw[thick, rotate around ={-45: (-3.75+1.5*\j,1.5*\i)}] (-3.75+1.5*\j,.15+1.5*\i) -- (-3.6+1.5*\j,.15+1.5*\i) -- (-3.6+1.5*\j,1.5*\i);
\draw[thick, fill=white] (-3-3*1.5,1.5*\i) circle (0.1cm); 
\draw[thick, fill=white] (-4.5-3*1.5,1.5*\i) circle (0.1cm); 
}
\Text[x=-8.25,y=-0.9]{}
\end{tikzpicture}= \begin{bmatrix} 1 & 0\\ 0& c \end{bmatrix}, & 
&\!\!\!a_{1}^{\mcircf\mcirc}\!\!=\begin{tikzpicture}[baseline=(current  bounding  box.center), scale=0.7]
\def\eps{-0.5};
\foreach \i in {0}
\foreach \j in {-3}{
\draw (-4.5+1.5*\j,1.5*\i) -- (-3+1.5*\j,1.5*\i);
\draw (-3.75+1.5*\j,1.5*\i-.75) -- (-3.75+1.5*\j,1.5*\i+.75);
\draw[ thick, fill=mygreen, rounded corners=2pt, rotate around ={45: (-3.75+1.5*\j,1.5*\i)}] (-4+1.5*\j,0.25+1.5*\i) rectangle (-3.5+1.5*\j,-0.25+1.5*\i);
\draw[thick, rotate around ={-45: (-3.75+1.5*\j,1.5*\i)}] (-3.75+1.5*\j,.15+1.5*\i) -- (-3.6+1.5*\j,.15+1.5*\i) -- (-3.6+1.5*\j,1.5*\i);
\draw[thick, fill=white] (-3-3*1.5,1.5*\i) circle (0.1cm); 
\draw[thick, fill=black] (-4.5-3*1.5,1.5*\i) circle (0.1cm); 
}
\Text[x=-8.25,y=-0.9]{}
\end{tikzpicture}= \begin{bmatrix} 0 & 0\\ q \varepsilon  & d \end{bmatrix}\!,\\
&\!\!\!\!a_{1}^{\mcirc\mcircf}\!\!=\begin{tikzpicture}[baseline=(current  bounding  box.center), scale=0.7]
\def\eps{-0.5};
\foreach \i in {0}
\foreach \j in {-3}{
\draw (-4.5+1.5*\j,1.5*\i) -- (-3+1.5*\j,1.5*\i);
\draw (-3.75+1.5*\j,1.5*\i-.75) -- (-3.75+1.5*\j,1.5*\i+.75);
\draw[ thick, fill=mygreen, rounded corners=2pt, rotate around ={45: (-3.75+1.5*\j,1.5*\i)}] (-4+1.5*\j,0.25+1.5*\i) rectangle (-3.5+1.5*\j,-0.25+1.5*\i);
\draw[thick, rotate around ={-45: (-3.75+1.5*\j,1.5*\i)}] (-3.75+1.5*\j,.15+1.5*\i) -- (-3.6+1.5*\j,.15+1.5*\i) -- (-3.6+1.5*\j,1.5*\i);
\draw[thick, fill=black] (-3-3*1.5,1.5*\i) circle (0.1cm); 
\draw[thick, fill=white] (-4.5-3*1.5,1.5*\i) circle (0.1cm); 
}
\Text[x=-8.25,y=-0.9]{}
\end{tikzpicture}=  \begin{bmatrix} 0 & p \varepsilon\\ 0& e \end{bmatrix}, & 
&\!\!\!a_{1}^{\mcircf\mcircf}\!\!=\begin{tikzpicture}[baseline=(current  bounding  box.center), scale=0.7]
\def\eps{-0.5};
\foreach \i in {0}
\foreach \j in {-3}{
\draw (-4.5+1.5*\j,1.5*\i) -- (-3+1.5*\j,1.5*\i);
\draw (-3.75+1.5*\j,1.5*\i-.75) -- (-3.75+1.5*\j,1.5*\i+.75);
\draw[ thick, fill=mygreen, rounded corners=2pt, rotate around ={45: (-3.75+1.5*\j,1.5*\i)}] (-4+1.5*\j,0.25+1.5*\i) rectangle (-3.5+1.5*\j,-0.25+1.5*\i);
\draw[thick, rotate around ={-45: (-3.75+1.5*\j,1.5*\i)}] (-3.75+1.5*\j,.15+1.5*\i) -- (-3.6+1.5*\j,.15+1.5*\i) -- (-3.6+1.5*\j,1.5*\i);
\draw[thick, fill=black] (-3-3*1.5,1.5*\i) circle (0.1cm); 
\draw[thick, fill=black] (-4.5-3*1.5,1.5*\i) circle (0.1cm); 
}
\Text[x=-8.25,y=-0.9]{}
\end{tikzpicture}= \begin{bmatrix} a & b \\ f & g \end{bmatrix}\!\!, 
\end{align}
and of the ones depicted with orange gates that are obtained by setting $\varepsilon=0$ in the above, we see that \eqref{eq:correlationsreplacement} is nothing but the sum \eqref{eq:skeletoncontribution} of skeleton diagrams. 

Taking into account the bound on the norm of $\| r_{1,x}\|$, we find that the error associated with the replacement \eqref{eq:replacement} is bounded by 
\be
{\cal O}\left(\left[ a^2  + \frac{|b f|}{1-\alpha}\right]^{\nu_+} a^{x_+-\nu_+}\right)
\ee
so that \eqref{eq:correlationsreplacement} becomes exact in the limit $x_+,\nu_+\to\infty$ with $\tilde x_+$ fixed. This bound is obtained by replacing  just one of the vertical dual-unitary strips with the remainder $r_{1,x}$.

Specifically, considering the relative error
\be
R(x_{+} , x_{-}) =\left|  \frac{\braket{\mcircf_{x}|\!\mcircf_0\!(t)}\!\!\big |_{\rm sk}-\braket{\mcircf_{x}|\!\mcircf_0\!(t)} }{\braket{\mcircf_{x}|\!\mcircf_0\!(t)}} \right|,
\label{eq:relErr}
\ee
where $\braket{\mcircf_{x}|\!\mcircf_0\!(t)}$ is the exact result and $\braket{\mcircf_{x}|\!\mcircf_0\!(t)}\!\!\big |_{\rm sk} $  is the skeleton expression (Eq.~\eqref{eq:skeletoncontribution}), we get 
\be
R(x_{+} , x_{-}) = {\cal O}\left(\Bigl[a^2  + \frac{|b f|}{1-\alpha}\Bigr]^{\nu_+} a^{-\nu_+}\right).
\label{eq:boundcorrections}
\ee
Thus, \eqref{eq:skeletoncontribution} gives a good approximation to the full result when $\nu_+$ is large enough. 

A completely analogous reasoning holds for horizontal strips $\nu_-\times x_+$, whenever 
\be
|c| > c^2  + \frac{|d e|}{1-\beta}\,,
\label{eq:conditionc}
\ee
and $\nu_-$ is large enough. In particular, for $x_-,\nu_-\to\infty$ with $\tilde x_-$ fixed we again find an exact statement. 

We stress that in the above argument we never used the fact that the defects are disposed along a regular sub-lattice: we just used that their minimal separation, $\bar\nu_\pm$, in one of the two directions becomes large. 
Provided that the latter condition applies, the correlations are sums of skeleton diagrams. 
In the case of irregular disposition of defects, however, one has to sum only over the skeleton diagrams connecting the (irregular) subset of lattice sites containing defects. Hence, the combinatorial factors differ from those in \eqref{eq:skeletoncontribution}.

Finally, we note that all this admits a simple interpretation in terms of the paths introduced in the previous subsection. The different eigenvectors of the transfer matrix can be seen as increasingly thicker horizontal lines of length 1. Their contributions --- the eigenvalues of $a^{\mcirc\mcirc}_{\du,x}$ --- are obtained by complicated combinations of the tiles~\eqref{eq:tiles}, which can be interpreted as a complicated ``dressing''. In this language, Property~\ref{prop:P2} implies that if \eqref{eq:condition} holds, the dominant weight is carried by the ``bare'' propagator
\be
\begin{tikzpicture}[baseline=(current  bounding  box.center), scale=0.7]
\aGate{0}{0}
\end{tikzpicture}\,.
\label{eq:onebodypropagator}
\ee

\subsubsection{Unit density, small strength}
\label{sec:PTstrenght}

Now let us consider a different perturbative limit: small strengths $\varepsilon\ll1$ and fixed density. In particular, for definiteness, we focus on the case of unit density $\delta=1$, which is the most interesting from the physical point of view. Namely, we take all the gates in the diagram \eqref{eq:latticewithdefects} to be green and set 
\be
  \begin{tikzpicture}[baseline=(current  bounding  box.center), scale=1]\def\eps{0.5};
\draw (-4.25,0.5) -- (-3.25,-0.5);
\draw (-4.25,-0.5) -- (-3.25,0.5);
\draw[thick, fill=mygreen, rounded corners=2pt] (-4,0.25) rectangle (-3.5,-0.25);
\draw[thick] (-3.75,0.15) -- (-3.6,0.15) -- (-3.6,0);
\Text[x=-4.25,y=-0.65]{}
\end{tikzpicture}= w_\du + \delta w_\varepsilon  
\label{eq:split}
\ee
with 
\be
w_\du :=  \begin{pmatrix}
    1 & 0 & 0 & 0\\
0 & 0 & a & b\\
0 & c & 0 & d\\
0 & e & f &  g
    \end{pmatrix}\!,
\quad 
\delta w_\varepsilon:= \varepsilon
\begin{pmatrix}
    0 & 0 & 0 & 0\\
0 & p & 0 & 0\\
0 & 0 & q  & 0 \\
0 & 0 & 0 &  0
    \end{pmatrix}\!,
    \label{eq:split2}  
\ee
where we make sure that $w_\du+\delta w_\varepsilon$ is a bona fide $\rm U(1)$-averaged unitary gate by imposing that it becomes bistochastic when conjugated with $H\otimes H$ (cf. the remark right before Sec.~\ref{sec:sumofpaths}).

The advantage of the parametrisation \eqref{eq:split} is that in this way --- in the language of Sec.~\ref{sec:sumofpaths} --- parameter $\varepsilon$ ``counts the turns''. Namely, each fixed order in perturbation theory is determined by the sum of all allowed paths with a fixed number of turns. Even though this is useful for classifying possible terms of different perturbative orders, the explicit evaluation of each order appears far from trivial. 

Let us first consider the simplest case:  the leading order that is proportional to $\varepsilon$, when the $x\in \mathbb Z + \frac{1}{2},y\in \mathbb Z$. It is given by a single skeleton diagram: 
\be
\begin{tikzpicture}[baseline=(current  bounding  box.center), scale=0.65]
\foreach \i in {0,...,5}{
\foreach \j in {1,...,3}{
\oGateO{\i}{\j}
}}
\draw[line width=\lineW] (-1,1) -- (-.5,1);
\draw[line width=\lineW] (5,3.5) -- (5,4);
\aGateO{0}{1}
\aGateO{1}{1}
\aGateO{2}{1}
\aGateO{3}{1}
\aGateO{4}{1}
\dtGate{5}{1}
\cGateO{5}{2}
\cGateO{5}{3}
\end{tikzpicture}\,,
\label{eq:one_turn}
\ee
and there are no other diagrams of the same order. Therefore, the relative error for $\varepsilon \to 0$ goes as ${\cal O}(\varepsilon)$, even for small $x_-$ or $x_+$. 

In contrast, the first non-trivial order in the case $x,y\in \mathbb Z$ and $x_+\gg x_->1$ is of order $\varepsilon^2$ and has much more complex structure. A simple set of contributions to this order is given by skeleton diagrams of the form 
\be
\begin{tikzpicture}[baseline=(current  bounding  box.center), scale=0.65]
\foreach \i in {0,...,5}{
\foreach \j in {1,...,3}{
\oGateO{\i}{\j}
}}
\draw[line width=\lineW] (-1,1) -- (-.5,1);
\draw[line width=\lineW] (5.5,3) -- (6,3);
\aGateO{0}{1}
\aGateO{1}{1}
\dtGate{2}{1}
\cGateO{2}{2}
\doGate{2}{3}
\aGateO{3}{3}\aGateO{4}{3}
\aGateO{5}{3}
\end{tikzpicture}\,.
\ee
Still, many more diagrams contribute at the same order in perturbation theory. For example, we have 
\begin{align}
&\begin{tikzpicture}[baseline=(current  bounding  box.center), scale=0.65]
\foreach \i in {0,...,5}{\foreach \j in {1,...,3}{\oGateO{\i}{\j} }}
\draw[line width=\lineW] (-1,1) -- (-0.5,1);
\draw[line width=\lineW] (5.5,3) -- (6,3);
\aGateO{0}{1}
\fGateO{1}{1}
\cGateO{1}{2}
\doGate{1}{3}
\aGateO{2}{3}\bGateO{3}{3}\aGateO{4}{3}
\aGateO{2}{1}\dtGate{3}{1}
\cGateO{3}{2}
\aGateO{5}{3}
\end{tikzpicture}\,, \nonumber\\
&\begin{tikzpicture}[baseline=(current  bounding  box.center), scale=0.65]
\foreach \i in {0,...,5}{\foreach \j in {1,...,3}{\oGateO{\i}{\j} }}
\draw[line width=\lineW] (-1.0,1) -- (-0.5,1);
\draw[line width=\lineW] (5.5,3) -- (6,3);
\aGateO{0}{1}
\fGateO{1}{1}
\cGateO{1}{2}
\doGate{1}{3}
\bGateO{2}{3}\bGateO{3}{3}\aGateO{4}{3}
\fGateO{2}{1}\dtGate{3}{1}
\cGateO{3}{2}
\cGateO{2}{2}
\aGateO{5}{3}
\end{tikzpicture}\,,
\label{eq:dressedskeletons}
\\
&\begin{tikzpicture}[baseline=(current  bounding  box.center), scale=0.65]
\foreach \i in {0,...,5}{\foreach \j in {1,...,3}{\oGateO{\i}{\j} }}
\draw[line width=\lineW] (-1.0,1) -- (-0.5,1);
\draw[line width=\lineW] (5.5,3) -- (6,3);
\aGateO{0}{1}
\fGateO{1}{1}
\cGateO{1}{2}
\doGate{1}{3}
\aGateO{2}{3}\bGateO{3}{3}\aGateO{4}{3}
\aGateO{2}{1}\dtGate{3}{1}
\dGateO{3}{2}
\aGateO{2}{2}
\eGateO{1}{2}
\aGateO{5}{3}
\end{tikzpicture}\,. \nonumber
\end{align}
The crucial observation at this point is that there exists a regime for the parameters of the gate $w_\du$, where all these ``complicated'' diagrams give negligible contribution when \emph{both} $x_-$ and $x_+$ become large. Specifically, this happens when \emph{both} of the conditions discussed in the previous section --- \eqref{eq:condition} and \eqref{eq:conditionc} --- hold. 
This observation can be understood as follows. First, we note that the diagrams in \eqref{eq:dressedskeletons} can be thought of as skeleton diagrams with ``dressed'' horizontal one-particle propagator. Then, we observe that, as we are working at a fixed order in perturbation theory, these dressed propagators are composed of dual-unitary tiles only. We can then make use of Property~\ref{prop:P2} to bound their contribution by 
\be
{\rm const} \left(a^2  + \frac{|f d|}{1-\alpha}\right)^x
\label{eq:bounderror}
\ee
where $x\gg1$ is the length of the corresponding segment of the dressed propagator. We see that for large enough $x$ this contribution is exponentially suppressed with respect to the ``bare'' line ($\propto a^x$). If both conditions, \eqref{eq:condition} and \eqref{eq:conditionc}, hold, this reasoning can be repeated at any fixed order in perturbation theory to show that the skeleton diagrams are leading when $x_+$ and $x_-$ are both large.  

To check the above reasoning, we performed a direct numerical evaluation of the correlation function \eqref{eq:pictorialcorrelation} and compared it with the prediction~\eqref{eq:skeletoncontribution}, obtained by summing all skeleton diagrams. The comparison is extremely encouraging: the deviations are typically undetectable on the scale of the plot, see Fig.~\ref{fig:SkeletonComparison0} for a representative example. The technical details of numerical simulations are discussed in Appendix~\ref{sec:numerics}.

Turning to a more quantitative analysis of the agreement, we considered the relative error~\eqref{eq:relErr} where $\braket{\mcircf_{x}|\!\mcircf_0\!(t)}$ is calculated numerically with no approximations while $\braket{\mcircf_{x}|\!\mcircf_0\!(t)}\!\!\big |_{\rm sk}$ is calculated using~\eqref{eq:skeletoncontribution}. The results are reported in Figs.~\ref{fig:SkeletonComparison}\,--\,\ref{fig:SkeletonComparison3}. Specifically, Fig.~\ref{fig:SkeletonComparison} concerns the case of large $x_+ = x_-$. We see that when the ``unperturbed gate'' $w_\du$ fulfils the conditions \eqref{eq:condition} and \eqref{eq:conditionc}, the relative error is always very small and appears to \emph{vanish with $\varepsilon$}. 
Interestingly, even if the predictions \eqref{eq:skeletoncontribution} for ${x\in\mathbb Z}$ and ${x\in\mathbb Z+1/2}$ are of different orders in $\varepsilon$, we observe almost the same relative errors in the two cases. 
Lastly, an important point highlighted by Fig.~\ref{fig:SkeletonComparison} is that  the relative error is of order one for any $\varepsilon$ when the conditions \eqref{eq:condition} and \eqref{eq:conditionc} are violated. 

\begin{figure}
\centering
    \includegraphics[scale=0.81]{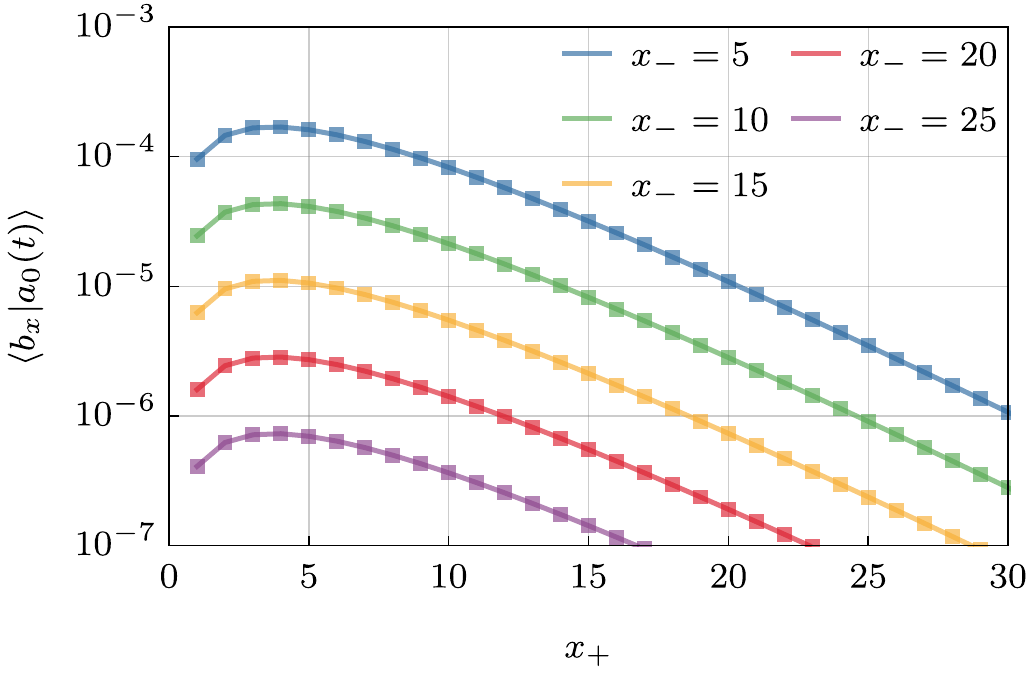}
    \caption{
Exact correlations computed numerically (solid lines) and the prediction of Eq.~\eqref{eq:skeletoncontribution} (squares) for $x\in\mathbb Z$ for a quantum circuit with elementary gate given by gate 2 of Table \ref{tab:U1gates} and maximal density of defects $\delta=1$.}
    \label{fig:SkeletonComparison0}
\end{figure}

\begin{figure}
\centering
    \includegraphics[scale=0.555]{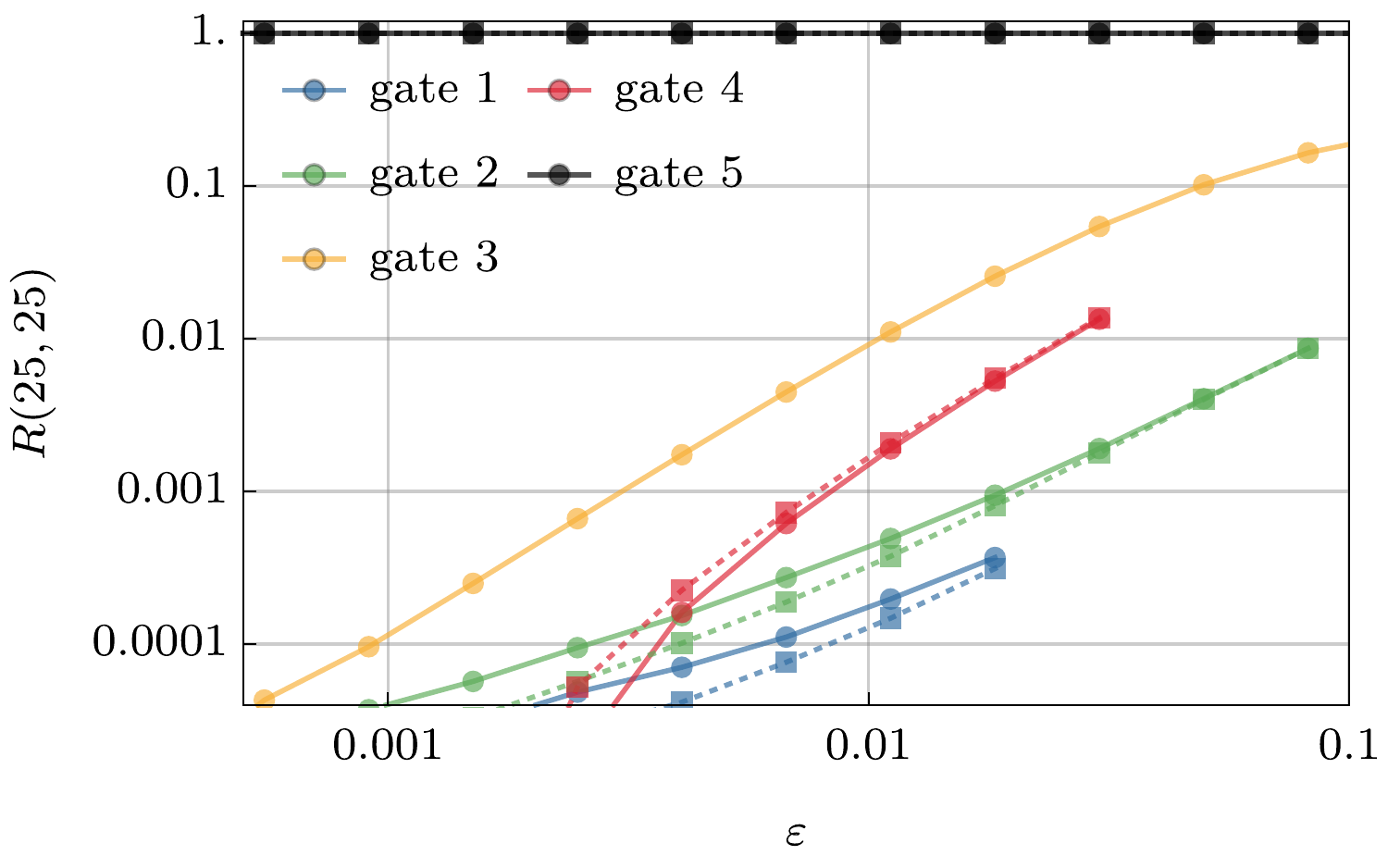}
    \caption{Relative error $R(x_{+} , x_{-})$ (cf.~\eqref{eq:relErr}) for $x_+=x_-$ as a function of $\varepsilon$, with maximal density of defects $\delta=1$.
    Full and dashed lines respectively correspond to integer and half-odd-integer endpoints ($x\in\mathbb Z,\mathbb Z+\tfrac{1}{2}$), while different colours correspond to different gates (specified in Table~\ref{tab:U1gates}).  Note that the gate corresponding to the black line does not fulfil the condition \eqref{eq:condition}. 
    For each gate, we stop at the value of $\varepsilon$, at which the gate ceases to be bistochastic.
    }
    \label{fig:SkeletonComparison}
\end{figure}

Our argument above relied on the fact that $x_-$ and $x_+$ are both large. 
When one of the two, say $x_-$, is fixed, we expect the relative error to be ${\cal O}(\varepsilon^0)$ in the case $x,y \in \mathbb{Z}$ (${\cal O}(\varepsilon)$ for the case of integer and half-integer endpoints), and bounded by \eqref{eq:bounderror} with $y=x_-$. For small $x_-$ this can be directly verified by computing the exact correlations through Eq.~\eqref{eq:TMcorrelations}. For example, in the case of $x_-=2$ and $x\in\mathbb Z$ we find 
\begin{align}
\braket{\mcircf_{x}|\!\mcircf_0\!(t)}=& p q \varepsilon^2 \binom{x_+}{1} a^{x_+ - 1} \left(1 - \frac{bf}{a^2 + bf - a}\right)\notag\\
&- \frac{p q \varepsilon^2 bf (a^{x_+} - (a^2 + bf)^{x_+})}{(a^2 + bf - a)^2}\,.
\end{align}
This result shows that the leading correction is a ``dressing'' of the skeleton diagram in the ``short direction'' and gives 
\be
R(x_{+} ,2) \approx \frac{|bf|}{(a^2 + bf - a)}, \qquad x_+ \gg 1\,.
\ee
This point is confirmed by our numerical results, as illustrated in Fig.~\ref{fig:SkeletonComparison2}. For increasing values of $x_-$ the order zero contribution becomes increasingly small and is eventually dominated by the ${\cal O}(\varepsilon)$ contributions. 

\begin{figure}
\centering
    \includegraphics[scale=0.555]{R_vsD_1000_v2.pdf}
    \caption{Relative error $R(x_{+} , x_{-})$ (cf.~\eqref{eq:relErr}) for $x_-\ll x_+$ as a function of $\varepsilon$.  
    The gate is fixed to gate 2 from Table~\ref{tab:U1gates} for $x\in\mathbb Z$ and density of defects $\delta=1$.}
    \label{fig:SkeletonComparison2}
\end{figure}

Finally, an interesting question concerns how many orders in $\varepsilon$ one has to keep in the skeleton-diagram expansion~\eqref{eq:skeletoncontribution} in order to get an accurate result. This question is considered in Fig.~\ref{fig:SkeletonComparison3}, where we compare the relative errors obtained by (i) keeping all orders in the expansion~\eqref{eq:skeletoncontribution}, (ii) keeping only the leading order in $\varepsilon$. We see that the prediction (i) is much more accurate for larger $\varepsilon$, even though the two predictions coincide for small enough $\varepsilon$.

\begin{figure}
\centering
\includegraphics[scale=0.555]{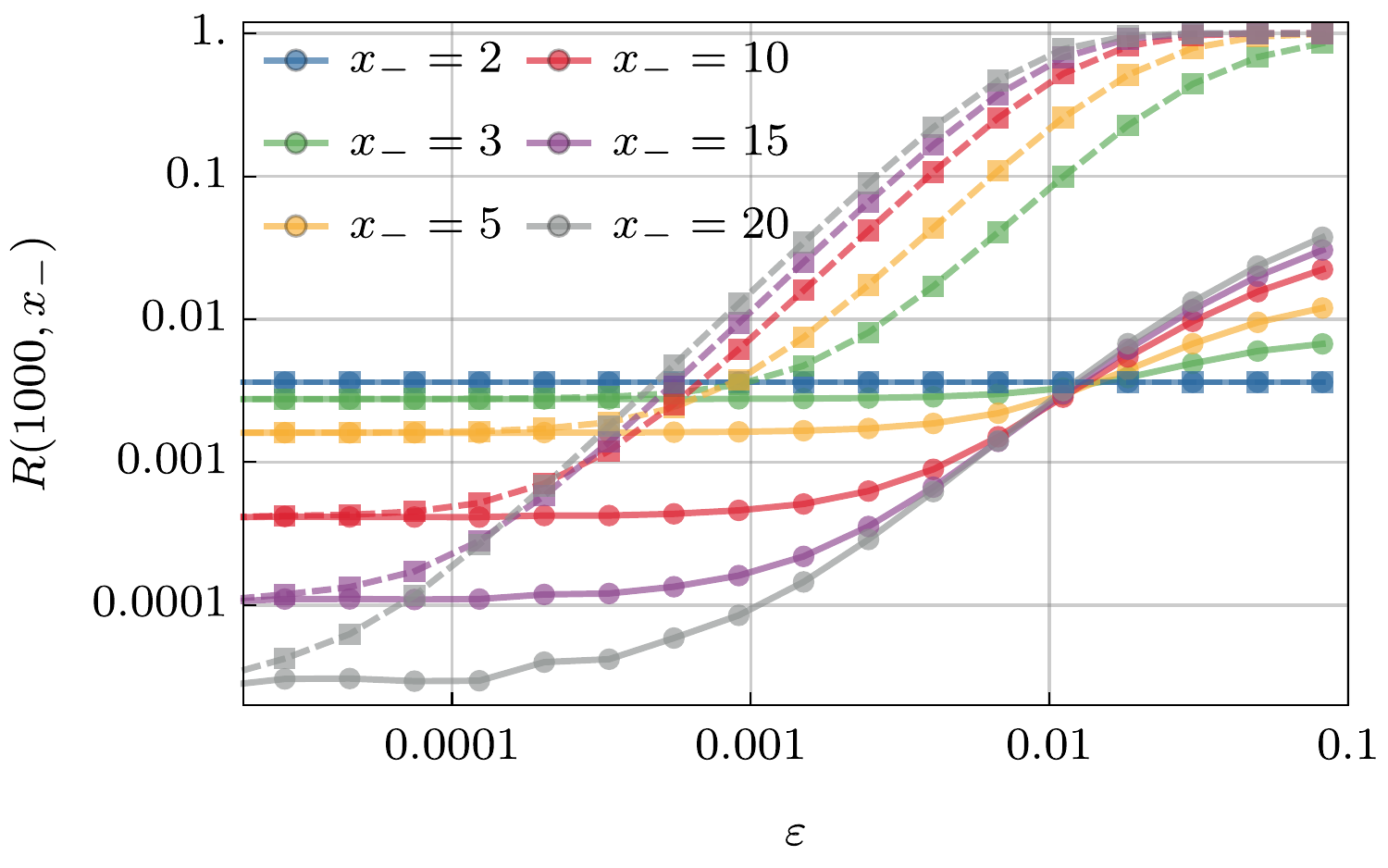}
    \caption{Relative errors $R(x_{+} , x_{-})$ (solid) and $R_1(x_{+} , x_{-})$ (dashed) vs $\varepsilon$. $R(x_{+} , x_{-})$ is computed using \eqref{eq:relErr}, while $R_1(x_{+} , x_{-})$ is computed by a modified version of \eqref{eq:relErr} where the skeleton-diagram contribution \eqref{eq:skeletoncontribution} is replaced by its first non-trivial order in $\varepsilon$. 
    The gate used for the numerical experiments is gate 2 from Table \ref{tab:U1gates} for $x\in\mathbb Z$ and density of defects $\delta=1$.}
    \label{fig:SkeletonComparison3}
\end{figure}

\section{Generic Gates}
\label{sec:genericnumerics}

\begin{figure*}[hbt!]
\centering
    \includegraphics[scale=.6]{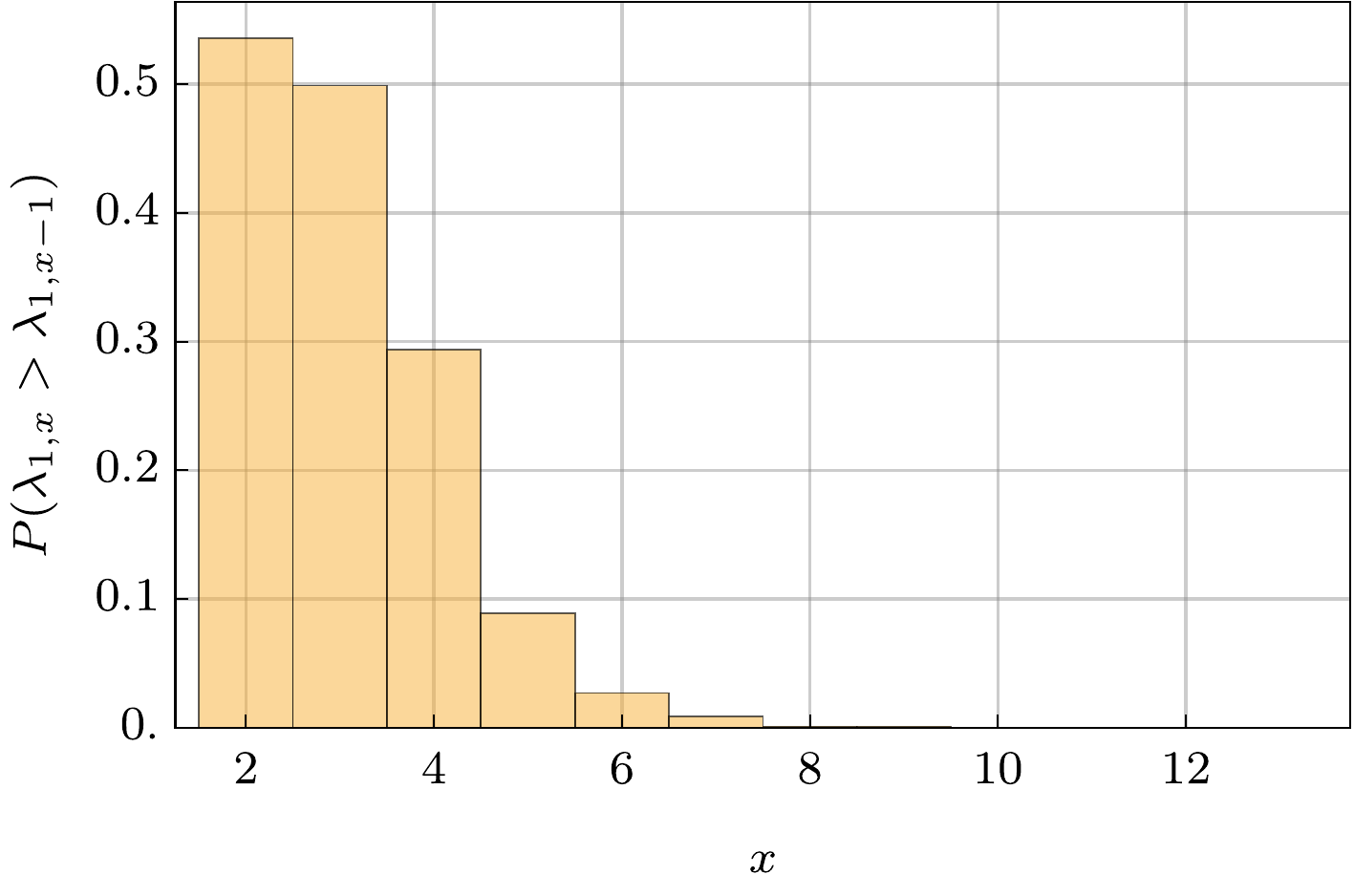}\quad\,
    \includegraphics[scale=.6]{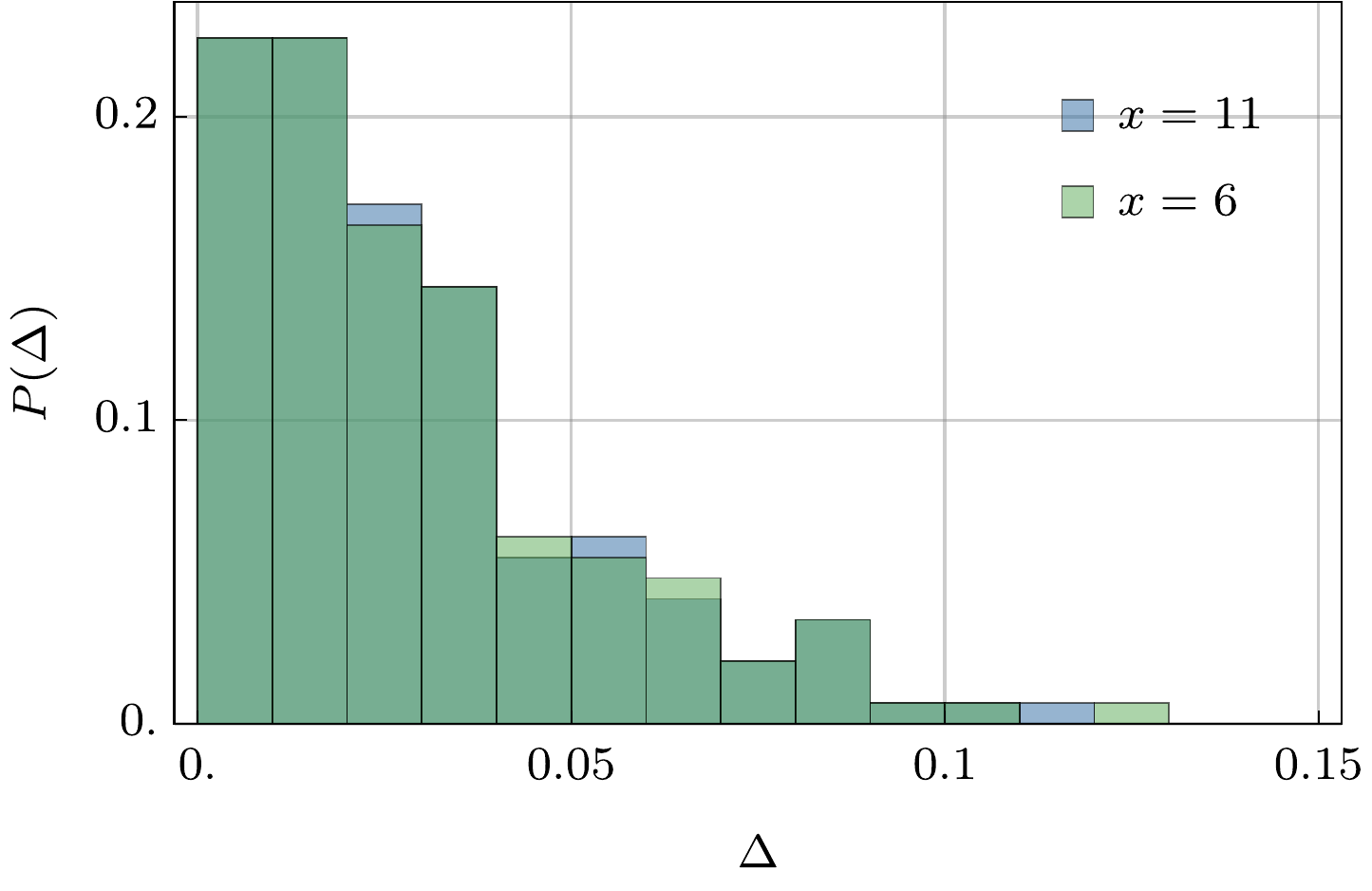}
    \caption{Left panel:  Probability of increase of the first sub-leading eigenvalue of the transfer matrix $A_{\du,x}^{\circ\circ}$ when increasing the size of $A_{\du,x}^{\circ\circ}$ from $x-1$ to $x$. The histogram was generated by considering $1000$ matrices ($100$ for $x=12,13$) at fixed $J_{1,2}=\pi/4, \ J_3=0.1$ and Haar random $U(2)$ matrices $u_i, i =\{1,2,3,4\}$  (see the parametrisation \eqref{eq:gate}), and searching for the largest eigenvalues of the transfer matrix up to $x=13$. 
Right panel: Probability distribution of the transfer matrix's gap $\Delta = \lambda_1-\lambda_2$, for the $148$ cases, which showed no increase of first sub-leading eigenvalue.        
    }
    \label{fig:Histogram}
\end{figure*}

\begin{figure*}
\centering
	\includegraphics[scale=0.81]{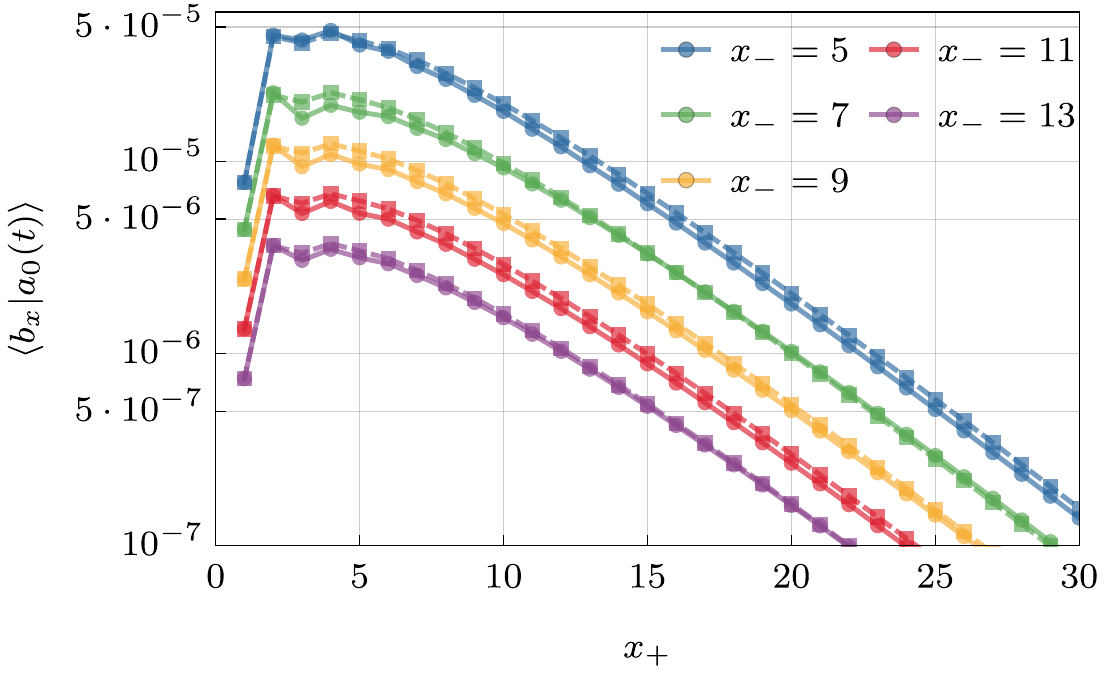}
    \includegraphics[scale=.81]{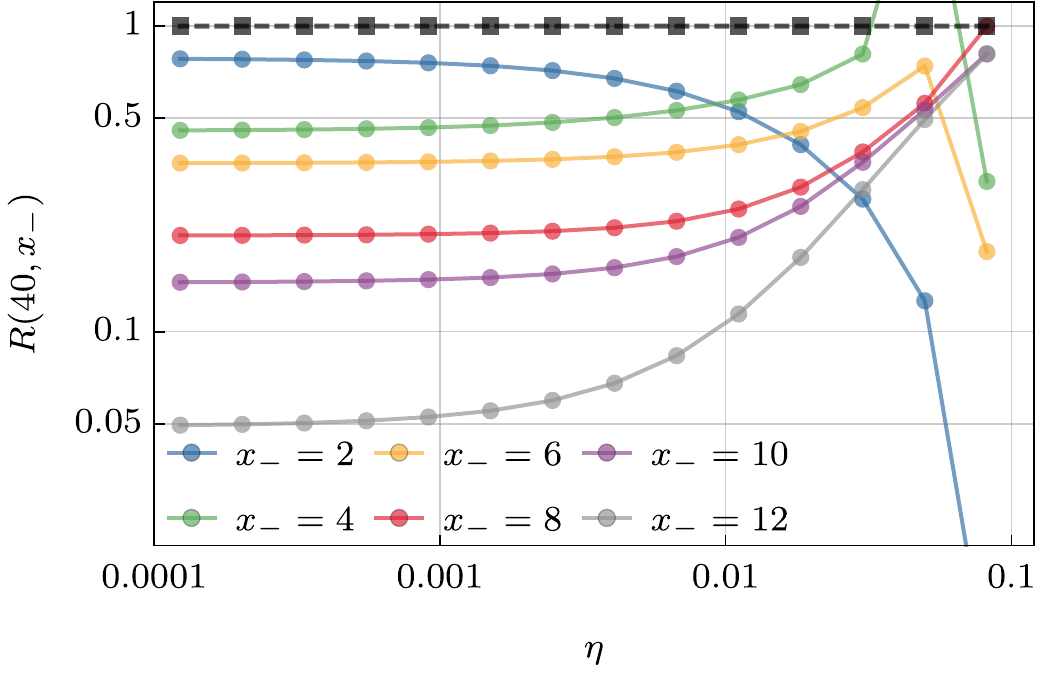}
    \caption{Left Panel: Correlation functions of operators $a=b=\sigma^x$ at integer sites versus $x_+$ at $\eta = 0.011109$. Solid lines are numerical results and dashed lines are the predictions of Eq.~\eqref{eq:generalSkeletons}. Right Panel: Relative errors versus $\eta$ at different $x_-$. A gate that does not fulfil \eqref{eq:support1ev} is drawn in black. The results are for symmetric generic gates, as specified in Tab~\ref{tab:NonAvgGate}, with maximal density of defects $\delta=1$.} 
    \label{fig:RelativeErrorNonAvg}
\end{figure*}

\begin{figure}
\centering
	\includegraphics[scale=0.56]{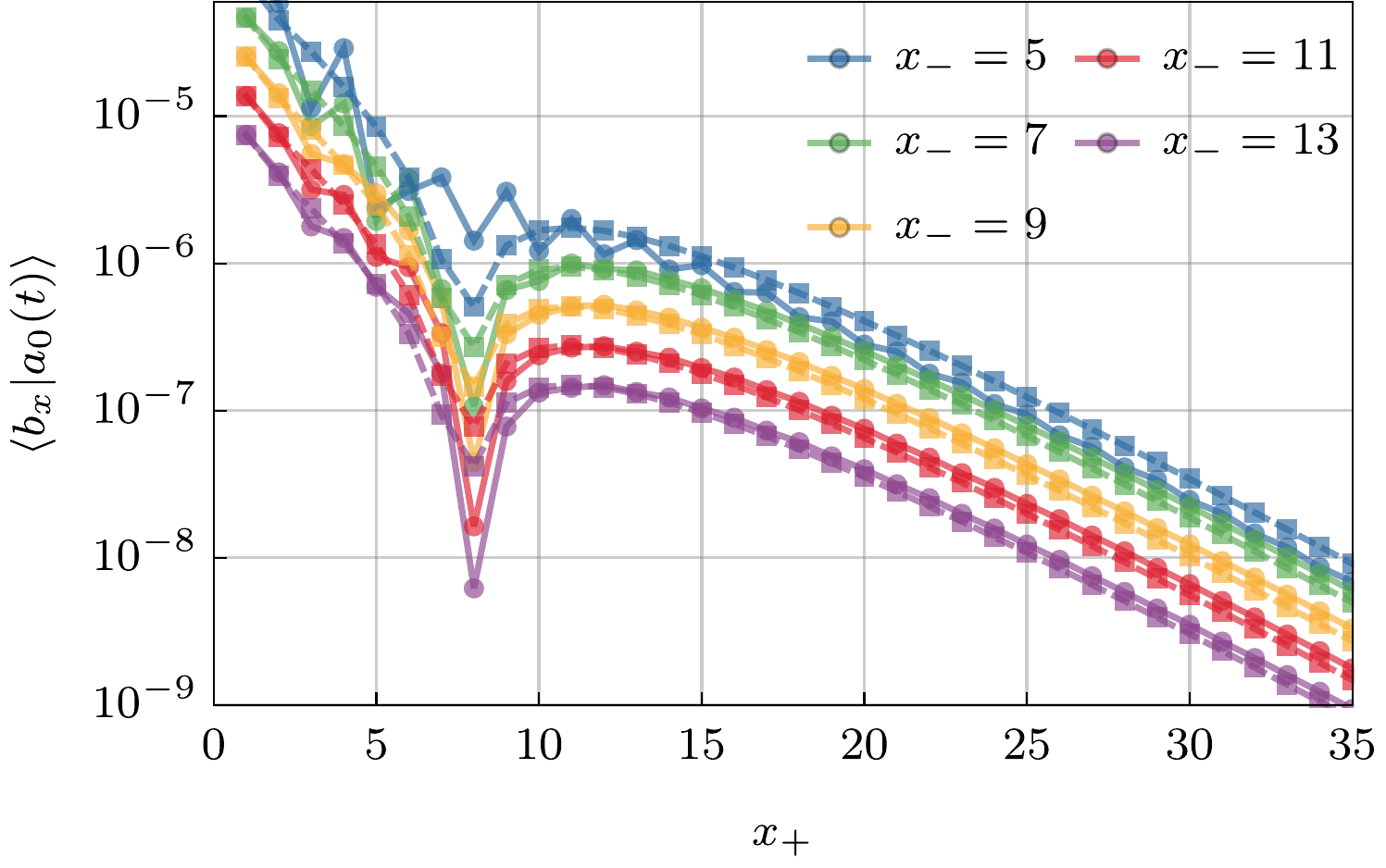}
    \caption{
    Correlation functions of operators $a=b=\sigma^x$ at integer sites versus $x_+$ for kicked Ising spin chain. Solid lines are numerical results and dashed lines are the predictions of Eq.~\eqref{eq:generalSkeletons}. The results are for kicked Ising spin chain at $J=B=\frac{\pi}{4}+0.01$, $h=1.2$, as defined in the Eq.~\eqref{eq:KI} with density of defects $\delta=1$.
    } 
    \label{fig:CorrKI}
\end{figure}

In this section we use a combination of numerical and analytical arguments to show that the main mechanisms observed in the minimal setting of the previous section carry over to the case of generic clean quantum gates. Under certain conditions on the unperturbed gate $W_\du$ the correlations computed by summing all skeleton diagrams still agree strikingly well with the exact numerical results. To simplify our numerical studies we consider the case of qubits ($d=2$), where the double gate $W$ is a $16 \times 16$ matrix.

We begin by briefly describing how to compute the contribution \eqref{eq:generalSkeletons} of all skeleton diagrams for the case of fixed defects on regular sub-lattices. The rest of the gates are dual-unitary. Firstly, we construct a single skeleton diagram by drawing a zig-zag line connecting the two operators on the lattice \eqref{eq:latticewithdefects}, and multiply the appropriate one-dimensional transfer matrices along the line (cf.~\eqref{eq:skeletonexample}). The total contribution is obtained by summing the contributions of all possible paths on the lattice connecting the two endpoints and containing no left and down turns. In particular, $l_j^+$ ($l_j^-$) in \eqref{eq:generalSkeletons} are of the lengths of each horizontal (vertical) segment and the constraints \eqref{eq:generalConstraints} come from the simple requirement that the sum of all horizontal (vertical) segments is equal to $x_+$ ($x_-$) minus the number of turns. Note that, because of the non-commutativity of the matrix product, Eq.~\eqref{eq:generalSkeletons} is considerably harder to evaluate than Eq.~\eqref{eq:skeletoncontribution}. The problem of its evaluation is addressed in Appendix~\ref{sec:evageneralSkeletons}. 

Following the analysis of the previous section we now consider small densities of defects and investigate the spectra of  horizontal and vertical transfer matrices, $A_{\du,x}^{\mcirc\mcirc}$ and $C_{\du,x}^{\mcirc\mcirc}$, composed only of dual-unitary gates $W_{\du,\eta}$. We remind the reader that, as discussed in Sec.~\ref{sec:foldedpicture}, these transfer matrices have the spectrum contained on the closed unit disk and a ``trivial'' eigenvector --- $\ket{\mcirc}^{\otimes x}$ --- corresponding to the eigenvalue $1$. 
This state, however, does not affect the correlation functions. 
One can see this by considering \eqref{eq:TMcorrelations} and observing that $\ket{\mcirc}^{\otimes x}$ has zero overlap with the states $\ket{a\mcirc\ldots\mcirc}$, $\ket{\mcirc\ldots\mcirc b}$, $A_{x}^{\mcirc b}\ket{\mcirc\ldots\mcirc}$, $C_{x}^{a\mcirc}\ket{\mcirc\ldots\mcirc}$. The relevant quantity for correlations is then the next-to-leading eigenvalue. In analogy with the discussion in Sec.~\ref{sec:PTdensity}, we conclude that if\\

\noindent (i) The next-to-leading eigenvalue $\lambda_{1}$ of $A_{\du,x}^{\mcirc\mcirc}$ \emph{has eigenvectors with support one}, i.e. it is of the form
\be
\ket{\underbrace{\mcirc\dots\mcirc}_k a \underbrace{\mcirc \dots\mcirc}_{x-k-1}},\qquad k=1,\ldots,x
\label{eq:support1ev}
\ee

\noindent (ii) There is a \emph{finite gap} (in magnitude) between $\lambda_{1}$ and the rest of the spectrum;\\

\noindent
the skeleton diagrams give a good approximation to the correlation functions for large enough $\bar \nu_+$ (minimal separation of the defects in the horizontal direction). Relative errors are bounded by 
\be
{\rm const} \left(\frac{\lambda_{2}}{\lambda_{1}}\right)^{\bar \nu_{+}}\,,
\label{eq:boundnu}
\ee
where $\lambda_{2}$ is the third leading eigenvalue of $A_{\du,x}^{\mcirc\mcirc}$. 
Note that, in the language of the previous section, this still corresponds to saying that the bare propagator carries the dominant weight. Analogous statements hold for the vertical direction, if one replaces $A_{\du,x}^{\mcirc\mcirc}$ by $C_{\du,x}^{\mcirc\mcirc}$ and $\bar \nu_{+}$ by $\bar \nu_{-}$. 

In this more complicated setting we are not able to prove a rigorous statement, such as Property~\ref{prop:P2}, to determine the set of gates for which (i) and (ii) hold. Nevertheless, investigating the spectrum numerically, we find strong evidence that a class of gates, for which the leading non-trivial eigenvectors are of the form \eqref{eq:support1ev}, exists, see the left panel of Fig.~\ref{fig:Histogram}. Moreover, focussing on this class, we isolated a subclass that shows a finite gap: an example of the probability distribution for the gap is reported in the right panel of Fig.~\ref{fig:Histogram} for three increasing values of $x$ (i.e.\ the length of the transfer matrix). Finally, we remark that the left panel of Fig.~\ref{fig:Histogram} also shows a rapid drop in the probability that a certain non-trivial eigenvector has a leading eigenvalue, with the size of its support. This suggests that if (i)
or (ii) does not hold, the dominant contribution to the correlations is still given by skeleton diagrams. In general, however, these diagrams will have ``thickened'' lines corresponding to ``dressed'' propagators: the width of the line corresponds to the support of the leading eigenvector.

Finally, we consider maximal density $\delta=1$ and small $\eta$. In complete analogy with the discussion in Sec.~\ref{sec:PTstrenght} we decompose the gate $W_\eta$ in two parts as follows 
\be
 W_\eta = \begin{tikzpicture}[baseline=(current  bounding  box.center), scale=1]
\def\eps{0.5};
\Wgategreen{-3.75}{0};
\Text[x=-3.75,y=-0.6]{}
\end{tikzpicture}= W_{\du,\eta} +\delta W_\eta  \,,
\label{eq:splitW}
\ee
where $W_{\du,\eta}$ is dual-unitary and the only non-zero elements of $\delta W_{\eta}$ (18 matrix elements in total) are the ``turns''
\be  
\begin{tikzpicture}[baseline=(current  bounding  box.center), scale=1]
\def\eps{0.5};
\Wgategreen{0}{0};
\draw[thick, fill=white] (.5, .50) circle (0.1cm); 
\draw[thick, fill=white] (.5, -.50) circle (0.1cm); 
\draw[thick, fill=black] (-.5, .50) circle (0.1cm); 
\draw[thick, fill=black] (-.5, -.50) circle (0.1cm); 
\Text[x=-.75,y=0.6]{$a$}
\Text[x=-.75,y=-0.6]{$b$}
\end{tikzpicture},
\qquad
\begin{tikzpicture}[baseline=(current  bounding  box.center), scale=1]
\def\eps{0.5};
\Wgategreen{0}{0};
\draw[thick, fill=white] (-.5, .50) circle (0.1cm); 
\draw[thick, fill=white] (-.5, -.50) circle (0.1cm); 
\draw[thick, fill=black] (.5, .50) circle (0.1cm); 
\draw[thick, fill=black] (.5, -.50) circle (0.1cm); 
\Text[x=.75,y=0.6]{$a$}
\Text[x=.75,y=-0.6]{$b$}
\end{tikzpicture},
\qquad a,b=\sigma^x,\sigma^y,\sigma^z\,.
\ee
Note that both, the elements of $W_{\du,\eta}$ and those of $\delta W_{\eta}$, will in general depend on $\eta$. The difference is that for small $\eta$ the former are ${\cal O}(\eta^0)$, while the latter are ${\cal O}(\eta)$.
 
Equipped with the definition \eqref{eq:splitW}, we are now in a position to repeat the argument of Sec.~\ref{sec:PTstrenght} by formally considering a perturbative expansion in the number of turns (which at the first order is equivalent to that in $\eta$). We then conclude that, if the points (i) and (ii) above hold for both $A_{\du,x}^{\mcirc\mcirc}$ and $C_{\du,x}^{\mcirc\mcirc}$, Eq.~\eqref{eq:generalSkeletons} gives the leading contribution at each fixed order in $\eta$ when $x_+$ and $x_-$ are both large. This is in agreement with our numerical findings: representative examples are reported in Figs.~\ref{fig:RelativeErrorNonAvg} -- \ref{fig:CorrKI}. Specifically, the left panel of Fig.~\ref{fig:RelativeErrorNonAvg} and Fig.~\ref{fig:CorrKI} show excellent agreement between the prediction of Eq.~\eqref{eq:generalSkeletons} and the exact numerical results for two different examples:\\

\noindent 1) A symmetric quantum circuit with local gate
\be
U = (u \otimes u) V\left[\frac{\pi}{4}+\eta,\frac{\pi}{4}+\eta,0.1\right] (v \otimes v),
\ee
where $V[\{J_i\}]$ is defined in Eq~\eqref{eq:gateV}, while $u$ and $v$ are reported in Table~\ref{tab:NonAvgGate}. This gate can be brought to the form \eqref{eq:defect} with dual-unitary part and perturbation explicitly given by
\begin{align}
&U_{\rm du} = (u \otimes u) V\left[\frac{\pi}{4},\frac{\pi}{4},0.1\right] (v \otimes v),\\
&P = (v^\dag \sigma^x v) \otimes (v^\dag \sigma^x v) + (v^\dag \sigma^y v) \otimes (v^\dag \sigma^y v).
\end{align}
We verified numerically that $A_{\du,x}^{\mcirc\mcirc}$ (equal to $C_{\du,x}^{\mcirc\mcirc}$ as the gate is symmetric) constructed with the dual unitary part of the gate fulfils (i) and (ii).\\ 

\noindent 2) The kicked Ising chain~\cite{kickedIsing1, kickedIsing2}. Namely we consider the local gate~\cite{BKP:dualunitary, GoLa19}
\be
U[J,B,h]=U_I[J,h] e^{-i B( \sigma^x \otimes \1 + \1 \otimes \sigma^x)}U_I[J,h],
\ee
with 
\be
U_I[J,h]=e^{-i J (\sigma^z \otimes \sigma^z) - i h/2( \sigma^z \otimes \1 + \1 \otimes \sigma^z)},
\label{eq:KI}
\ee
The model is dual-unitary for ${B=J=\pi/4}$, while here we considered ${B=J=\pi/4+\eta}$. In this case the gate can again be written in the form \eqref{eq:defect} with $U_{\rm du} = U[\pi/4,\pi/4,h]$ (we do not report the explicit form of $P$ because it is unwieldily). Interestingly we found that, depending on parameter $h$, the transfer matrix $A_{\du,x}^{\mcirc\mcirc}$ ($=C_{\du,x}^{\mcirc\mcirc}$) can either fulfil (i) or not. Moreover, the additional structure of the kicked Ising model (see Sec.~5.3 of Ref.~\cite{BKP:OEergodicandmixing}) generates additional eigenvectors associated to $\lambda_{1}$ with support strictly larger than one (although small). This means, even when (i) holds, (ii) is not strictly valid. However, it is reasonable to assume that the eigenvector with support 1 gives the leading contribution as it has the largest overlap with local operators. This is confirmed by our numerical results (see Fig.~\ref{fig:CorrKI}): considering a value of $h$ for which (i) holds we find excellent agreement between the prediction Eq.~\eqref{eq:generalSkeletons} and exact numerics.\\

More detailed information on the comparison between Eq.~\eqref{eq:generalSkeletons} and the numerical results is reported in the right panel of Fig.~\ref{fig:RelativeErrorNonAvg} 
and in Fig.~\ref{fig:RelativeErrorNonAvg2}. In particular the latter shows that for fixed $\eta$ and large $x_+$ the relative error decreases with $x_-$. This is in agreement with our expectations (cf.~\eqref{eq:bounderror} and \eqref{eq:boundnu}) of an error bounded by 
\be
{\rm const} \left(\frac{\lambda_{2}}{\lambda_{1}}\right)^{x_-}\,.
\label{eq:boundxm}
\ee
In fact Fig.~\ref{fig:RelativeErrorNonAvg2} shows an even faster decay, suggesting that the bound is not tight. Finally, from the black line in the right panel of Fig.~\ref{fig:RelativeErrorNonAvg} we see that if the conditions (i) and (ii) are violated, the agreement is immediately absent.

\begin{figure}
\centering
    \includegraphics[scale=0.91]{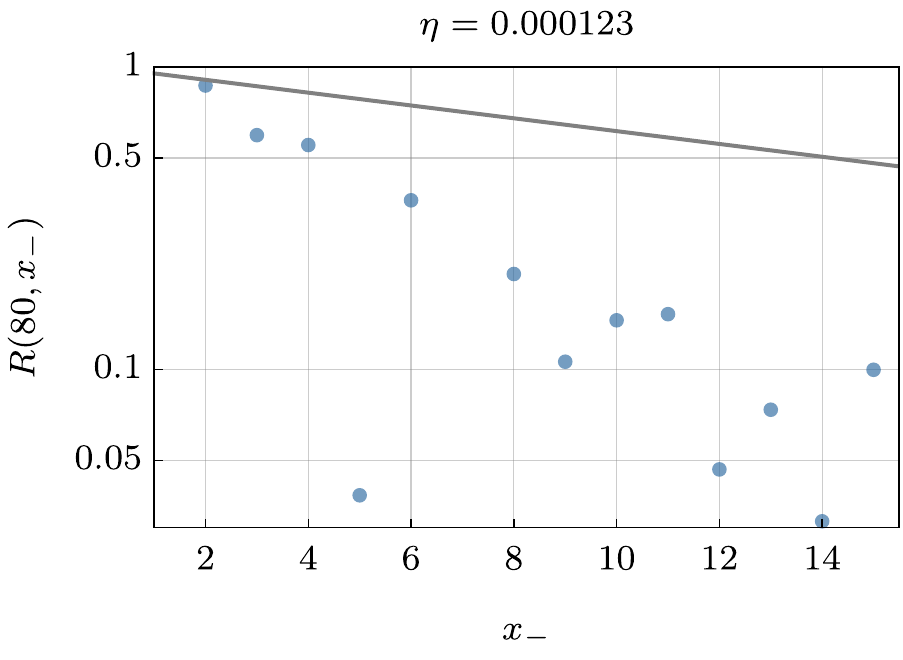}
    \caption{Relative error for the correlations among $\sigma^x$ operators at integer sites versus $x_-$ at large $x_+$ and small $\eta$. The grey line reports the upper bound \eqref{eq:boundxm}. The results are for symmetric generic gate fulfilling \eqref{eq:support1ev}, as specified in Tab.~\ref{tab:NonAvgGate}, with density of defects $\delta=1$. The point at $x_-=7$ has relative error smaller than $0.3\%$ and is not shown.}
    \label{fig:RelativeErrorNonAvg2}
\end{figure}

\section{Conclusions and perspectives}
\label{sec:conclusions}

In this paper we studied correlation functions in perturbed dual-unitary circuits, or, equivalently, in perturbed dual-bistochastic Markov chains. We considered the problem for three increasingly more realistic (increasingly complex) settings --- clean systems with random defects, noisy systems with fixed defects, and clean systems with fixed defects --- 
and identified a class of dual-unitary circuits that is stable under perturbations. More precisely, when these systems are perturbed, their correlations continue to be given in terms of one-dimensional transfer matrices (single qudit maps) and hence to (generically) decay exponentially. 
These exponentially decaying correlations are continuous with respect to dual-unitarity-breaking perturbations demonstrating the structural stability. 
The main qualitative change induced by the perturbations is that they allow the correlations to spread through the whole causal light cone and not just along its edge, as it occurs in the pure dual-unitary case. One can intuitively think of the perturbations as defects that permit the correlations ``to turn'' in spacetime. Remarkably, the quantitative features of the correlations are extremely well captured by a simple ``path integral'' formula, corresponding to the sum of correlations over all one-dimensional paths within the causal light cone that connect the endpoints.  
In conclusion, this paper presents, to our knowledge, the first set of theoretical results on the stability of the ergodic regime of quantum many-body systems under generic perturbations and opens the research arena of quantum many-body ergodic theory.

The case of noisy systems with fixed defects turned out to be an intriguing minimal model. It shows all the physical features of the generic setting but, simultaneously allows for rigorous derivations. This is not an isolated case  (see, e.g., \cite{BP:tripartite}): introducing a small degree of randomness to isolate minimal settings appears to be a fruitful route for accessing non-trivial information about interacting many-body systems. In the minimal setting we identified four additional classes of circuits --- beside the dual-unitary ones --- where correlation functions are exactly solvable --- i.e.\ the aforementioned ``path integral'' formula applies exactly. Systems in these classes are generically strongly interacting and generate highly complex dynamics. An obvious direction for future research is to study them further, understanding, for example, their spectral statistics and their non-equilibrium dynamics. Moreover, it would be very interesting to understand whether these classes can be defined in generic quantum circuits.  

Another outstanding question raised by our work concerns what happens when one perturbs dual-unitary circuits that are not in the stable class. One possibility, which seems to be hinted by our numerical results, is that the correlation functions continue to have a ``path integral'' form but the paths are ``thickened''~\cite{noteAdam}. In other words they are computed in terms of transfer matrices for a chain of width $n$ rather than of width $1$. If $n$ does not scale with time --- as we seem to observe in general --- the physical picture remains very similar to the one studied here and, in particular, all correlations continue to decay exponentially. However, it would be interesting to understand whether there exists a class of circuits for which $n$ grows with time. This could lead to a phase transition in the behaviour of correlations and therefore to richer physics. Importantly, this would allow for power-law decaying correlations signalling non-trivial conservation laws.  

We can immediately propose two generalisations of our setup. First, instead of unitary quantum circuits, one can consider circuits of completely positive maps. These include the class of circuits with projective measurements that is currently attracting substantial attention as it displays measurement driven phase transitions \cite{SkRN19, Ludwig, VPYL19, ZGWG19}. Specifically, it is straightforward to generalise the concept
of dual bistochasticity to {\em dual quantum bistochasitity}. Second, our treatment can be directly extended to perturbed dual-unitary (or dual-bistochastic) circuits in higher spatial dimensions where generic circuits display non-trivial complexity transitions \cite{Harrow}.
In this case we again expect the correlations to be written as sums over one-dimensional paths.
\medskip

\section*{Acknowledgements}
The work has been supported by ERC Advanced grant 694544 -- OMNES and the program P1-0402 of Slovenian Research Agency. 

\appendix 
\bw

\section{Dual-Bistochastic Markov Circuits}
\label{sec:markovchains}

In this appendix we show that  the object \eqref{eq:mapcorrelations} can be interpreted as the correlation function of a Markov circuit, i.e. a discrete time classical Markov chain where at each half time step the time evolution couples only nearest neighbours with bistochastic matrices. To see this let us regard  \eqref{eq:mapcorrelations} as the fundamental object, forgetting its origin in terms of the U(1) noise averaged quantum circuit and focus on the following setting:\\

\noindent (i) Each wire in the tensor network \eqref{eq:mapcorrelations} has generic dimension $N\in\mathbb N$, not restricted to squares of positive integers. In particular, we choose a certain basis 
\be
\{\ket{\alpha}:\quad \alpha=1,\ldots,N\}\,,
\label{eq:bistochasticbasis}
\ee
and interpret each state as a possible state of a classical spin (or any abstract configuration of a classical system). Thus, we now view our physical system as a chain of $2L$ classical spins. The probability distributions over configurations of such chain can be formally expanded in the product basis 
\be
\{\ket{\alpha_1,\ldots,\alpha_{2L}}:\quad \alpha_i=1,\ldots,N\}\,.
\label{eq:bistochasticbasisL}
\ee
Namely, we can write 
\be
\ket{p(t)} = \sum_{\{\alpha_j\}\in\{1\ldots N\}^{2L}} p(t; {\{\alpha_j\}}) \ket{\alpha_1,\ldots,\alpha_{2L}}\,,
\ee
where the coefficient $p(t; {\{\alpha_j\}})$ gives the probability that the system is in the configuration ${\alpha_1,\ldots,\alpha_{2L}}$ at time $t$, and hence  
\be
p(t; {\{\alpha_j\}})\in[0,1] \qquad \text{and}\quad  \sum_{\{\alpha_j\}\in\{1\ldots N\}^{2L}}  \!\!\!p(t; {\{\alpha_j\}}) =1\,.
\label{eq:normalisation}
\ee

\noindent (ii) The state $\ket{\mcirc}$ has the following expansion in the basis \eqref{eq:bistochasticbasis}
\be
\ket{\mcirc} : = \frac{1}{\sqrt{N}}  \sum_{\alpha =1}^N \ket{\alpha}\!.
\label{eq:circbistobasis}
\ee
Namely, $\ket{\mcirc}$ is the flat sum of all possible configurations of a single spin and, apart from a normalisation factor, it represents the flat probability distribution. In the context of classical stochastic processes, such state is known as the \emph{maximal entropy state} and is typically denoted by $\ket{\omega}$. Therefore, restoring the correct normalisation we have 
\be
\ket{\mcirc}= \sqrt{N} \ket{\omega}\,.
\ee 

\noindent (iii) The local two-body gate $W$ is not unitary but \emph{bistochastic} in the tensor product of two bases \eqref{eq:bistochasticbasis}. Specifically,
\be
\label{eq:Wbistochastic}
\begin{split}
&0 \leq \braket{\alpha\beta|W|\alpha'\beta'} \leq 1,\;\; \forall \alpha,\beta,\alpha',\beta'\\
&\!\!\!\sum_{\alpha,\beta=1}^N \!\!\!\braket{\alpha\beta|W|\alpha'\beta'}= 1=\!\!\!\sum_{\alpha,\beta=1}^N \!\!\!\braket{\alpha'\beta'|W|\alpha \beta}\!,\,\,\; \forall \alpha',\beta'.
\end{split}
\ee
Therefore, $\mathbb W$ in \eqref{eq:Wevolution} is now a Markov chain propagator evolving probability distributions of the configurations of a chain of $2L$ classical spins (or any other discrete variable degrees of freedom). More precisely, one can define 
\be
\ket{p(t+1)} =\mathbb W \ket{p(t)} 
\ee
as the time-evolved probability distribution. 

The bistochastic property \eqref{eq:Wbistochastic} of the elementary local gate $W$ implies that the maximal entropy state is stationary: $W\ket{\omega}\!\otimes\!\ket{\omega}=\ket{\omega}\!\otimes\!\ket{\omega}$. This means that the  unitality relations \eqref{eq:unitality} are satisfied even if the gate is not unitary and we recover the natural light-cone causal structure leading to \eqref{eq:pictorialcorrelation} in the thermodynamic limit.

In this language, \eqref{eq:mapcorrelations} is the correlation function in the maximal entropy state of two (diagonal) local observables
\be
A = \sum_{\alpha} A_\alpha \ket{\alpha}\!\!\bra{\alpha}, \qquad B = \sum_{\alpha} B_\alpha \ket{\alpha}\!\!\bra{\alpha}\,, 
\ee
such that 
\be
A\ket{\mcirc}=\ket{a},\quad B\ket{\mcirc}=\ket{b}\!,
\ee
where $\ket{a}=\sum_\alpha A_\alpha \ket{\alpha}$ and $\ket{b}=\sum_\beta B_\beta\ket{\beta}$ are the states appearing in \eqref{eq:mapcorrelations}. In other words, we can rewrite the r.h.s.\ of \eqref{eq:mapcorrelations} as  
\be
\braket{b_{x+y} |a_y(t)} = N^{2 L} \braket{\omega\ldots \omega|B_{x+y} \mathbb W^t A_{y}|\omega\ldots \omega}\,,
\ee
which is the expression for dynamical correlation functions in classical Markov chains and classical cellular automata~\cite{Bobenko2,Bobenko3,Katja,Katja2}. 

Moreover, defining the ``dual'' local Markov gate $\tilde W$ by 
\be
 \braket{\alpha\beta|\tilde W|\alpha'\beta'} = \braket{\beta' \beta|W|\alpha' \alpha}
\ee
and requiring it to be bistochastic in the basis ${\{\ket{\alpha\beta}\equiv\ket{\alpha}\otimes\ket{\beta}\}}$, the dual-unitality relations~\eqref{eq:dualunitality} are also satisfied. This immediately implies that --- if both $W$ and $\tilde W$ are bistochastic --- the correlations take the simple form \eqref{eq:DUcorrelations}. Indeed, the simplification of the correlation functions is only based on the diagrammatic relations \eqref{eq:unitality} and \eqref{eq:dualunitality} without utilising the unitarity of the gates. We will refer to bistochastic gates $W_\du$ with this special property as {\em dual bistochastic}.  

\subsection{Reduced gates as Markov circuits with $N=2$}

Hereby, we can now establish a direct connection with the U(1)-averaged unitary gates studied in Sec.~\ref{sec:U(1)avg}. Consider the minimal case where each classical spin takes only $N=2$ values and define $\ket{\mcircf}$ as the state orthogonal to $\ket{\mcirc}$, i.e. 
\be
\ket{\mcircf} := \frac{1}{\sqrt{2}}  (\ket{1}-\ket{2}).
\label{eq:circfbistobasis}
\ee
Using the second requirement in \eqref{eq:Wbistochastic} it is immediate to verify that

\begin{property}
Let the gate $w$ be bistochastic in the basis $\{\ket{\alpha_i}\otimes\ket{\alpha_j}\}_{i,j=1}^2$, then 
\begin{itemize}
\item[(a)] $w$ takes the form \eqref{eq:reducedW} when expressed in the basis $\{\ket{\mcirc\mcirc},\ket{\mcircf\mcirc},\ket{\mcirc\mcircf},\ket{\mcircf\mcircf}\}$. The explicit parametrisation of the matrix elements is reported in Appendix~\ref{sec:parametersbisto}. 
\item[(b)] If $w$ is \emph{dual bistochastic}, it takes the form \eqref{eq:reducedW0} in the basis $\{\ket{\mcirc\mcirc},\ket{\mcircf\mcirc},\ket{\mcirc\mcircf},\ket{\mcircf\mcircf}\}$. The explicit parametrisation of the matrix elements is again reported in Appendix~\ref{sec:parametersbisto}.   
\item[(c)] The matrix implementing the change of basis from $\{\ket{\mcirc\mcirc},\ket{\mcircf\mcirc},\ket{\mcirc\mcircf},\ket{\mcircf\mcircf}\}$ to $\{\ket{\alpha_i}\otimes\ket{\alpha_j}\}$ is given by $H\otimes H$, where $H$ is the Hadamard transformation (cf.~\eqref{eq:Hadamard}).
\end{itemize}
\end{property}

\section{Parametrisation of the gates \eqref{eq:reducedW} and  \eqref{eq:reducedW0}}
\label{sec:parametersW}

In this appendix we present an explicit parametrisation of the gates \eqref{eq:reducedW} and \eqref{eq:reducedW0} depending on whether they are obtained as $U(1)$-averaged (dual-) unitary double gates or as conjugated (dual-) bistochastic ones. 

\subsection{$U(1)$-averaged (dual-) unitary double gates}
\label{sec:parametersU1ave}

To find a convenient parametrisation we note that the elements of the gate \eqref{eq:reducedW} can be computed by evaluating 
\be
\braket{o_1,o_2 |w|o_3,o_4}={\rm tr}\left[(o_1\otimes o_2) {\bar U}^\dag (o_1\otimes o_2)  \bar U \right],\qquad\qquad o_1,o_2,o_3,o_4=\1,\sigma^z\,.
\label{eq:elemetstr}
\ee
Here we defined  
\be
\bar U = (u(0,\beta_1,\gamma_1) \otimes u(0,\beta_2,\gamma_2)) V[\{J_i\}]  (u(0,\beta_3,\gamma_3)^\dag \otimes u(0,\beta_4,\gamma_4)^\dag)\,,
\ee
where $u(\alpha,\beta,\gamma) $ is defined in Eq.~\eqref{eq:parametrisationu} and $V[\{J_i\}]$ in Eq.~\eqref{eq:gateV}. Eq.~\eqref{eq:elemetstr} follows directly from the parametrisation \eqref{eq:gate} and the form \eqref{eq:averageaction} of the averaged gate. 

In particular, explicit calculations yield 
\begin{align}
p \varepsilon &= \cos(2 J_1) \cos(2 J_2) C_3^{(2)} C_3^{(4)}+ \cos(2 J_1)\cos(2 J_3) C_2^{(2)} C_2^{(4)} +  \cos(2 J_2)\cos(2 J_3) C_1^{(2)} C_1^{(4)},\label{eq:d1}\\
q \varepsilon &= \cos(2 J_1) \cos(2 J_2) C_3^{(1)} C_3^{(3)}+ \cos(2 J_1)\cos(2 J_3) C_2^{(1)} C_2^{(3)} +  \cos(2 J_2)\cos(2 J_3) C_1^{(1)} C_1^{(3)},\\
a &= \sin(2 J_1) \sin(2 J_2) C_3^{(2)} C_3^{(3)}+ \sin(2 J_1)\sin(2 J_3) C_2^{(2)} C_2^{(3)} +  \sin(2 J_2)\sin(2 J_3) C_1^{(2)} C_1^{(3)},\\
c &= \sin(2 J_1) \sin(2 J_2) C_3^{(1)} C_3^{(4)}+ \sin(2 J_1)\sin(2 J_3) C_2^{(1)} C_2^{(4)} +  \sin(2 J_2)\sin(2 J_3) C_1^{(1)} C_1^{(4)}, 
\end{align}
\begin{align}
b &= \sum_{\alpha,\beta,\gamma=1}^3 \sin(2J_\beta)\cos(2J_\gamma)  {\mathcal E}^{\alpha\beta\gamma} C^{(2)}_\alpha C^{(3)}_\beta C^{(4)}_\gamma, & d &= - \sum_{\alpha,\beta,\gamma=1}^3 \cos(2J_\beta)\sin(2J_\gamma)  {\mathcal E}^{\alpha\beta\gamma} C^{(1)}_\alpha  C^{(3)}_\beta C^{(4)}_\gamma,\\
e &= - \sum_{\alpha,\beta,\gamma=1}^3 \sin(2J_\alpha)\cos(2J_\beta)   {\mathcal E}^{\alpha\beta\gamma} C^{(1)}_\alpha C^{(2)}_\beta C^{(4)}_\gamma, & f &= \sum_{\alpha,\beta,\gamma=1}^3 \cos(2J_\alpha)\sin(2J_\beta)  {\mathcal E}^{\alpha\beta\gamma} C^{(1)}_\alpha C^{(2)}_\beta C^{(3)}_\gamma,
\end{align}
\begin{align}
g =&  \sum_{\beta=1}^3 C_\beta^{(1)}C_\beta^{(2)}C_\beta^{(3)}C_\beta^{(4)}+ \cos(2 J_1)\cos(2 J_2)\!\!\sum_{\alpha\neq\beta=1}^2 \!\!\!C_\beta^{(1)}C_\alpha^{(2)}C_\beta^{(3)}C_\alpha^{(4)} + \sin(2 J_1)\sin(2 J_2)\!\!\sum_{\alpha\neq\beta=1}^2 \!\!\!C_\beta^{(1)}C_\alpha^{(2)}C_\alpha^{(3)}C_\beta^{(4)}  \notag\\
&+ 
  \cos(2 J_1)\cos(2 J_3)\!\!\sum_{\alpha\neq \beta=1,3} \!\!\!C_\beta^{(1)}C_\alpha^{(2)}C_\beta^{(3)}C_\alpha^{(4)} + \sin(2 J_1)\sin(2 J_3)\!\!\sum_{\alpha\neq \beta=1,3} \!\!\!C_\beta^{(1)}C_\alpha^{(2)}C_\alpha^{(3)}C_\beta^{(4)} \notag\\
&+ 
  \cos(2 J_2)\cos(2 J_3)\!\!\sum_{\alpha\neq\beta=2,3} \!\!\!C_\beta^{(1)}C_\alpha^{(2)}C_\beta^{(3)}C_\alpha^{(4)} + \sin(2 J_2)\sin(2 J_3)\!\!\sum_{\alpha\neq \beta=2,3} \!\!\!C_\beta^{(1)}C_\alpha^{(2)}C_\alpha^{(3)}C_\beta^{(4)}\,, \label{eq:g}
\end{align}
where ${\mathcal E}^{\alpha\beta\gamma}$ is the tree-dimensional Levi-Civita symbol and we defined 
\begin{align}
C^{(i)}_1  &= \sin(\beta_i)\cos(\gamma_i), & C^{(i)}_2&= \sin(\beta_i)\sin(\gamma_i), & C_3^{(i)}&= \cos(\beta_i)\,,& &i=1,2,\\
C^{(i)}_1  &= -\sin(\beta_i)\cos(\gamma_i), & C^{(i)}_2&= \sin(\beta_i)\sin(\gamma_i), & C_3^{(i)}&= \cos(\beta_i)\,,& &i=3,4.
\end{align}
The parametrisation of  \eqref{eq:reducedW0} follows by replacing $J_1$ and $J_2$ by ${\pi}/{4}$. Explicitly we have  
\begin{align}
p \varepsilon &=0=q \varepsilon,\label{eq:deltaanglesDUint}\\
a &=  C_3^{(2)} C_3^{(3)}+ \sin(2 J_3) \left(C_2^{(2)} C_2^{(3)} + C_1^{(2)} C_1^{(3)}\right), & c &= C_3^{(1)} C_3^{(4)}+\sin(2 J_3) \left( C_2^{(1)} C_2^{(4)} + C_1^{(1)} C_1^{(4)}\right),\\
b &= \cos(2J_3) C^{(4)}_3 (C^{(2)}_1 C^{(3)}_2 -C^{(2)}_2 C^{(3)}_1), & d &=  \cos(2J_3)C^{(3)}_3 (C^{(1)}_1   C^{(4)}_2-C^{(1)}_2   C^{(4)}_1),\\
e &= \cos(2J_3)C^{(2)}_3(C^{(1)}_1  C^{(4)}_2-C^{(1)}_2  C^{(4)}_1), & f &=\cos(2J_3)C^{(1)}_3 ( C^{(2)}_1  C^{(3)}_2-C^{(2)}_2  C^{(3)}_1),
\end{align}
\begin{align}
g =&  \sum_{\beta=1}^3 C_\beta^{(1)}C_\beta^{(2)}C_\beta^{(3)}C_\beta^{(4)}+\!\!\!\!\sum_{\alpha\neq\beta=1}^2 \!\!\!C_\beta^{(1)}C_\alpha^{(2)}C_\alpha^{(3)}C_\beta^{(4)} \!\!\!+ \sin(2J_3)\!\!\left[\sum_{\alpha\neq\beta=1,3} \!\!\!C_\beta^{(1)}C_\alpha^{(2)}C_\alpha^{(3)}C_\beta^{(4)}\!\!+\!\!\!\!\! \sum_{\alpha\neq\beta=2,3} \!\!\!\!\!C_\beta^{(1)}C_\alpha^{(2)}C_\alpha^{(3)}C_\beta^{(4)}\right]\!\!.\label{eq:ganglesDUint}
\end{align}  
Noting that 
\begin{align}
&(u(0,\beta_1,\gamma_1) \otimes u(0,\beta_2,\gamma_2)) V[\{\pi/4,\pi/4,J_3\}]  (u(0,\beta_3,\gamma_3)^\dag \otimes u(0,\beta_4,\gamma_4)^\dag =\notag\\
&= (u(0,\beta_1,0) \otimes u(0,\beta_2,0)) V[\{\pi/4,\pi/4,J_3\}]  (u(0,\beta_3,\gamma_3-\gamma_2)^\dag \otimes u(0,\beta_4,\gamma_4-\gamma_1)^\dag
\end{align}
we can set 
\be
\gamma_1=\gamma_2=0
\ee
and \eqref{eq:deltaanglesDUint}--\eqref{eq:deltaanglesDUint} further simplify to 
\begin{align}
p\varepsilon &=0=q \varepsilon,\label{eq:deltaanglesDU}\\
a &=  \cos(\beta_2)\cos(\beta_3)- \sin(2 J_3) \sin(\beta_2)\sin(\beta_3)\cos(\gamma_3), & c &= \cos(\beta_1)\cos(\beta_1)-\sin(2 J_3)\sin(\beta_1)\sin(\beta_4)\cos(\gamma_4),\\
b &= \cos(2J_3) \cos(\beta_4) \sin(\beta_2)\sin(\beta_3)\sin(\gamma_3), & d &=  \cos(2J_3) \cos(\beta_3) \sin(\beta_1)\sin(\beta_4)\sin(\gamma_4),\\
e &= \cos(2J_3)\cos(\beta_2)\sin(\beta_1)\sin(\beta_4)\sin(\gamma_4), & f &=\cos(2J_3)\cos(\beta_1)\sin(\beta_2)\sin(\beta_3)\sin(\gamma_3),
\end{align}
\begin{align}
g =&  \cos(\beta_1)\cos(\beta_2)\cos(\beta_3)\cos(\beta_4)+ \sin(\beta_1)\sin(\beta_2)\sin(\beta_3)\sin(\beta_4)\cos(\gamma_3)\cos(\gamma_4)\notag\\
&- \sin(2J_3)\!\!\left[ \sin(\beta_1)\cos(\beta_2)\cos(\beta_3)\sin(\beta_4)\cos(\gamma_4)+\cos(\beta_1)\sin(\beta_2)\sin(\beta_3)\cos(\beta_4)\cos(\gamma_3)\right]\!.\label{eq:ganglesDU}
\end{align}

\subsection{Conjugated (dual-) bistochastic gates}
\label{sec:parametersbisto}

Considering a $4\times4$ bistochastic matrix 
\be
B = \begin{pmatrix}
x_{11} & x_{12} & x_{13}& 1-\sum_{j} x_{1j}\\
x_{21} & x_{22} & x_{23} & 1-\sum_{j} x_{2j}\\
x_{31} & x_{32} & x_{33} & 1-\sum_{j} x_{3j}\\
1-\sum_{i} x_{i1} & 1-\sum_{i} x_{i2}  & 1-\sum_{i} x_{i3}  & \sum_{ij} x_{ij} -2\\
\end{pmatrix}
\label{eq:bistohastic0}
\ee
parametrized by elements of a convex set
\be
\mathcal R_{\rm b} = \left \{  (x_{11},x_{12},\ldots,x_{33});\quad x_{ij} \in [0,1],\quad \left( 0\leq \sum_{i} x_{ij} \leq1 \right) \wedge\left(0\leq \sum_{i} x_{ji} \leq 1 \right),\quad j=1,2,3\right\}\!.
\ee
Conjugating \eqref{eq:bistohastic0} with $H\otimes H$, we obtain a gate of the form \eqref{eq:reducedW} with coefficients
\begin{align}
p \varepsilon = x_{11}+x_{13}+x_{31}+x_{33}-1, & & a=x_{11}+x_{12}+x_{31}+x_{32}-1, & & b= 1-x_{12}-x_{13}-x_{32}-x_{33}, \label{eq:Pbisto1}\\
 c = x_{11}+x_{13}+x_{21}+x_{23}-1, & & q  \varepsilon  = x_{11}+x_{12}+x_{21}+x_{22}-1, & & d= 1-x_{12}-x_{13}-x_{22}-x_{23}, \label{eq:Pbisto2}\\
 e = 1-x_{21}-x_{23}-x_{31}-x_{33}, & & f = 1-x_{21}-x_{22}-x_{31}-x_{32}, & & g= x_{22}+x_{23}+x_{32}+x_{33}-1. 
\label{eq:Pbisto3}
\end{align}
Instead, a dual bistochastic matrix is written as  
\be
B_{\rm db} = \begin{pmatrix}
x_{11} & x_{12} & x_{13}& 1-\sum_{j} x_{1j}\\
x_{21} & 1-x_{21}-x_{12}-x_{11}  & x_{23} & x_{11}+x_{12}-x_{23}\\
x_{31} & x_{32} & 1-x_{31}-x_{13}-x_{11} & x_{11}+x_{13}-x_{32}\\
1-\sum_{i} x_{i1} & x_{11}+x_{21}-x_{32}  & x_{11}+x_{31}-x_{23} & x_{23}+x_{32}-x_{11}, \\
\end{pmatrix}
\ee
the set of which is is parametrized by a 7-dimensional convex set
\be
\mathcal R_{\rm db}
= \left \{
(x_{11},x_{12},x_{21},x_{13},x_{31},x_{23},x_{32});\; 
(x_{11},x_{12}, x_{13},x_{21},1\!-\!x_{12}\!-\!x_{21}\!-\!x_{11},x_{23},x_{31},x_{32},1\!-\!x_{13}\!-\!x_{32}\!-\!x_{11} \in \mathcal R_{\rm b})\right\}.
\ee
In this case, the parametrisation \eqref{eq:Pbisto1}--\eqref{eq:Pbisto3} simplifies to
\begin{align}
p \varepsilon &= 0, &  a&=x_{11}+x_{12}+x_{31}+x_{32}-1, &  b&= x_{31}+x_{11}-x_{12}-x_{32}, \label{eq:deltabistoDU}\\
 c &= x_{11}+x_{13}+x_{21}+x_{23}-1, &  q \varepsilon &= 0, & d&= x_{21}+x_{11}-x_{13}-x_{23}, \\
 e &= x_{13}+x_{11}-x_{21}-x_{23}, & f&= x_{12}+x_{11}-x_{31}-x_{32}, &  g&= 1-\sum_k (x_{k1}+x_{1k})+x_{23}+x_{32}. \label{eq:gbistoDU}
\end{align}

\section{GUE-Averaged Gate}
\label{sec:averagegate}

In this appendix we treat analytically the problem outlined in Sec.~\ref{sec:perturbativeexpansion} considering correlations averaged over \emph{random} defects. Specifically, we take $P$ in Eq.~\eqref{eq:defect} to be random matrices independently distributed at \emph{each} space-time point according to the Gaussian Unitary Ensemble, i.e. 
\be
P_{\rm GUE}(P)= \mathcal N_d \, e^{-\frac{d^2}{2}{\rm tr}[(P-\mu \1)(P-\mu \1)^\dag]}\,,
\label{eq:GUEaverage}
\ee
where $\mathcal N_d$ is a normalisation constant, $d^2$ is the dimension of $P$, $\mu$ is the disorder mean, while the variance of matrix elements $\sigma^2$ is fixed to unity. This is not a restriction as the ``strength'' of the disorder is already controlled by the parameter $\eta$ and all physical quantities depend on $\eta$ and $\sigma$ only through $\eta/\sigma$. 

We start by proving the following property. 

\begin{property}
The average of the doubled gate $W_\eta$ reads as   
\be
\!\!\!\mathbb E_{\rm GUE}\!\left[ W_\eta \right]\!= 
\!\!\!\!\begin{tikzpicture}[baseline=(current  bounding  box.center), scale=1]
\def\eps{0.5};
\draw[very thick] (-4.25,0.5) -- (-3.25,-0.5);
\draw[very thick] (-4.25,-0.5) -- (-3.25,0.5);
\draw[ thick, fill=gray, rounded corners=2pt] (-4,0.25) rectangle (-3.5,-0.25);
\draw[thick] (-3.75,0.15) -- (-3.6,0.15) -- (-3.6,0);
\Text[x=-4.25,y=-0.75]{}
\end{tikzpicture}
\!\!\!= \frac{K_{\rm GUE}(\eta)-1}{d^4-1} W_\du\! +\frac{d^4-K_{\rm GUE}(\eta)}{d^4-1} \ket{\mcirc\mcirc}\!\!\bra{\mcirc\mcirc}\!,
\label{eq:averagedgate}
\ee
where $E_{\rm GUE}[\cdot]$ is the average over \eqref{eq:GUEaverage} and $K_{\rm GUE}(x)\equiv \mathbb E_{\rm GUE}\left[ |{\rm tr}[e^{i x P}]|^2 \right]$ is the spectral form factor of the Gaussian Unitary Ensemble. 
\end{property}
\begin{proof}
Since we have 
\be
\mathbb E_{\rm GUE}\!\left[ W_\eta \right]= \mathbb E_{\rm GUE}\left[\left(e^{-i \eta P}\otimes e^{i \eta P^T}\right)\right]W_{\rm du}\,,
\label{eq:averagedgatestep1}
\ee
we can focus on the first term in the product
\be
M\equiv\mathbb E_{\rm GUE}\left[\left(e^{-i \eta P}\otimes e^{i \eta P^T}\right)\right]\,.
\ee
We now observe that $M$ is invariant under conjugation with $U\otimes U^*$, where $U$ is a $d^2\times d^2$ unitary matrix and $({\cdot})^*$ represents complex conjugation in the canonical basis~\eqref{eq:localbasis}. By Shur Lemma, this immediately implies 
\be
M = \alpha P_{\rm sym} +  \beta P_{\rm anti-sym}\,,
\ee
where $P_{\rm sym}$ and $P_{\rm anti-sym}$ are the projectors on the spaces of the irreps in which the tensor product of two conjugated fundamental representations of $U(d^2)$ decomposes. Specifically, these two representations are $d^4-1$ and $1$ dimensional, and in our notations we have 
\be
P_{\rm sym} = \1-\ket{\mcirc\mcirc}\bra{\mcirc\mcirc}, \qquad P_{\rm anti-sym} = \ket{\mcirc\mcirc}\bra{\mcirc\mcirc}.
\ee
The coefficients are readily computed by multiplying $M$ by $P_{\rm sym}$ and $P_{\rm anti-sym}$, respectively, and taking the trace. This leads to 
\be
\alpha = \frac{{\rm tr}[M]-1}{{\rm tr}[P_{\rm sym}]}=\frac{K_{\rm GUE}(\eta)-1}{d^4-1}, \qquad\qquad \beta=1\,.
\ee
Plugging back into \eqref{eq:averagedgatestep1} and using $W_{\rm du} P_{\rm anti-sym}= P_{\rm anti-sym}$ we obtain \eqref{eq:averagedgate}. 
\end{proof}

Since both terms on the r.h.s. of \eqref{eq:averagedgate} fulfil the ``dual-unitality'' conditions \eqref{eq:dualunitality} and their coefficients sum to one, the averaged gate fulfils \eqref{eq:dualunitality} as well. Then the correlations immediately reduce to the form \eqref{eq:DUcorrelations}, where $\braket{a| (A_{\du,1}^{\mcirc \mcirc})^{2t}  |b}$ and $\braket{b | (C_{\du,1}^{\mcirc \mcirc})^{2t}  | a}$ are respectively replaced by  
\be
\!\!\begin{tikzpicture}[baseline=(current  bounding  box.center), scale=0.7]
\def\eps{-0.5};
\foreach \i in {0}
\foreach \j in {3,...,-3}{
\draw[very thick] (-4.5+1.5*\j,1.5*\i) -- (-3+1.5*\j,1.5*\i);
\draw[very thick] (-3.75+1.5*\j,1.5*\i-.75) -- (-3.75+1.5*\j,1.5*\i+.75);
\draw[ thick, fill=myorange, rounded corners=2pt, rotate around ={45: (-3.75+1.5*\j,1.5*\i)}] (-4+1.5*\j,0.25+1.5*\i) rectangle (-3.5+1.5*\j,-0.25+1.5*\i);
\draw[thick, rotate around ={-45: (-3.75+1.5*\j,1.5*\i)}] (-3.75+1.5*\j,.15+1.5*\i) -- (-3.6+1.5*\j,.15+1.5*\i) -- (-3.6+1.5*\j,1.5*\i);
\draw[thick, fill=white] (1.5,1.5*\i) circle (0.1cm); 
\draw[thick, fill=white] (-4.5-3*1.5,1.5*\i) circle (0.1cm); 
\draw[thick, fill=white] (-3.75+1.5*\j,-0.75) circle (0.1cm); 
\draw[thick, fill=white] (-3.75+1.5*\j, .75) circle (0.1cm);
}
\foreach \i in {0}
\foreach \j in {-1,...,1}{
\draw[ thick, fill=gray, rounded corners=2pt, rotate around ={45: (-3.75+4.5*\j,3*\i)}] (-4+4.5*\j,0.25+3*\i) rectangle (-3.5+4.5*\j,-0.25+3*\i);
\draw[thick, rotate around ={-45: (-3.75+4.5*\j,3*\i)}] (-3.75+4.5*\j,.15+3*\i) -- (-3.6+4.5*\j,.15+3*\i) -- (-3.6+4.5*\j,3*\i);
}
\draw[thick, fill=black] (1.5,0) circle (0.1cm); 
\draw[thick, fill=black] (-4.5-3*1.5,0) circle (0.1cm); 
\Text[x=2-0.5,y=-0.35]{$b$}
\Text[x=-5-3*1.5+0.5,y=-0.35]{$a$}
\draw [thick, decorate, decoration={brace,amplitude=4pt,mirror,raise=4pt},yshift=0pt] (-3.75,-0.75) -- (0.75,-0.75) node [black,midway,yshift=-.5cm] {$\nu_+$};
\Text[x=-8.25,y=1.5]{}
\end{tikzpicture},
\ee
and 
\be
\!\!\begin{tikzpicture}[baseline=(current  bounding  box.center), scale=0.7]
\def\eps{-0.5};
\foreach \i in {0}
\foreach \j in {3,...,-3}{
\draw[very thick] (-4.5+1.5*\j,1.5*\i) -- (-3+1.5*\j,1.5*\i);
\draw[very thick] (-3.75+1.5*\j,1.5*\i-.75) -- (-3.75+1.5*\j,1.5*\i+.75);
\draw[ thick, fill=myorange, rounded corners=2pt, rotate around ={45: (-3.75+1.5*\j,1.5*\i)}] (-4+1.5*\j,0.25+1.5*\i) rectangle (-3.5+1.5*\j,-0.25+1.5*\i);
\draw[thick, rotate around ={45: (-3.75+1.5*\j,1.5*\i)}] (-3.75+1.5*\j,.15+1.5*\i) -- (-3.6+1.5*\j,.15+1.5*\i) -- (-3.6+1.5*\j,1.5*\i);
\draw[thick, fill=white] (1.5,1.5*\i) circle (0.1cm); 
\draw[thick, fill=white] (-4.5-3*1.5,1.5*\i) circle (0.1cm); 
\draw[thick, fill=white] (-3.75+1.5*\j,-0.75) circle (0.1cm); 
\draw[thick, fill=white] (-3.75+1.5*\j, .75) circle (0.1cm);
}
\foreach \i in {0}
\foreach \j in {-1,...,1}{
\draw[ thick, fill=gray, rounded corners=2pt, rotate around ={45: (-3.75+4.5*\j,3*\i)}] (-4+4.5*\j,0.25+3*\i) rectangle (-3.5+4.5*\j,-0.25+3*\i);
\draw[thick, rotate around ={45: (-3.75+4.5*\j,3*\i)}] (-3.75+4.5*\j,.15+3*\i) -- (-3.6+4.5*\j,.15+3*\i) -- (-3.6+4.5*\j,3*\i);
}
\draw[thick, fill=black] (1.5,0) circle (0.1cm); 
\draw[thick, fill=black] (-4.5-3*1.5,0) circle (0.1cm); 
\Text[x=2-0.5,y=-0.35]{$a$}
\Text[x=-5-3*1.5+0.5,y=-0.35]{$b$}
\draw [thick, decorate, decoration={brace,amplitude=4pt,mirror,raise=4pt},yshift=0pt] (-3.75,-0.75) -- (0.75,-0.75) node [black,midway,yshift=-.5cm] {$\nu_-$};
\Text[x=-8.25,y=1.5]{}
\end{tikzpicture}.
\ee
Here we distributed the defects along the regular sublattice of Eq.~\eqref{eq:latticewithdefects} for a more convenient representation. The treatment of more general (e.g. random) dispositions of defects is completely analogous.

Expanding each averaged gate as in \eqref{eq:averagedgate} we produce a sum of $2^{\tilde x_\pm}$ terms (cf. \eqref{eq:rescaledxpm}). As can be directly verified using \eqref{eq:unitality}, only the term with no $\ket{\mcirc\mcirc}\bra{\mcirc\mcirc}$ gives a non-trivial contribution. Considering that term we directly find 
\begin{align}
\mathbb E_{\rm GUE}\!\left[\braket{b_{x+y}|a_y(t)}\right]\!\! =&  {\rm mod}(2y,2) \delta_{x+t} \! \braket{b | (C_{\du,1}^{\mcirc \mcirc})^{2t}  | a} \!\!\left[\frac{K_{\rm GUE}(\eta)-1}{d^4-1}\right]^{\tilde x_-}\!\!\!\!\!\notag\\
&+  {\rm mod}(2y+1,2) \delta_{x-t}  \braket{a |  (A_{\du,1}^{\mcirc\mcirc})^{2t} |b} \left[\frac{K_{\rm GUE}(\eta)-1}{d^4-1}\right]^{\tilde x_+}\!\!\!
\label{eq:eq:GUEaveragedcorrelations2}
\end{align}
Where $K_{\rm GUE}(t)$ can explicitly be written as~\cite{Mehtabook}
\be
K_{\rm GUE}(t) = -\int {\rm d}x {\rm d} y\,\, e^{i t y} \rho^c_{2,d^2}(x+y,x) + \left|\int {\rm d}x\,\, e^{i t x} \rho^c_{1,d^2}(x)\right|^2 + d^2
\ee
where  
\be
\rho^c_{1,n}(x) = K_{n}(x,x),\quad \rho^c_{2,n}(x,y) = (K_{n}(x,y))^2,\quad K_{n}(x,y)=\frac{e^{-(x^2+y^2)/2}}{ \pi^{1/2} 2^n (n-1)!} \frac{H_n(x)H_{n-1}(y)-H_n(y)H_{n-1}(x)}{x-y},
\ee
and $H_{n}(x)$ are the Hermite polynomials. In particular for $d=2$ we find
\be
K_{\rm GUE}(t) = 4 + e^{-t^2/2} \left[12 - 24 t^2 - \frac{23}{2} t^4 + \frac{8}{3} t^6 - \frac{25}{96} t^8 + \frac{1}{96}t^{10}\right]\,.
\ee
We conclude by observing that the GUE average considered here is very different from the flat average over the Haar measure on $U(d^2)$. Taking a Haar random perturbation $R$ in \eqref{eq:defect} instead of $e^{i \eta P}$ immediately gives  
\be
\mathbb E_{U(d^2)}\left[ W_\eta \right] = \mathbb E_{U(d^2)}\left[ R^\dag \otimes R^T \right] =  \ket{\mcirc\mcirc}\bra{\mcirc\mcirc}
\label{eq:HaarUd2average}
\ee
where in the second step we used the right multiplication invariance of the Haar measure, and the last step is a trivial instance of the known result on integrals over unitary groups~\cite{Collins03} (it can also be obtained from \eqref{eq:averagedgate} by replacing $K_{\rm GUE}(\eta)$ with $K_{\rm CUE}(1)=1$). A direct consequence of this result is that all correlations are trivialised by the Haar average for any density $\delta$ of defects. Interestingly, \eqref{eq:averagedgate} does not reproduce \eqref{eq:HaarUd2average} for $\eta\to\infty$ since 
\be
\lim_{\eta\to\infty}  K_{\rm GUE}(\eta) = d^2\,.
\ee  
This can be justified by noting that, even though \eqref{eq:GUEaverage} becomes flat in this limit, it gives flat weights to different variables compared to the Haar measure (roughly speaking $P$ rather than $e^{i P}$) and a flat weight for one does not imply a flat weight for the other. As expected \eqref{eq:eq:GUEaveragedcorrelations2} and \eqref{eq:HaarUd2average} agree in the limit of infinite dimensional matrices $d\to\infty$.

\section{Proof of Property~\ref{prop:P2}}
\label{sec:spectrum}

In this appendix we prove the Property~\ref{prop:P2}. We consider the horizontal transfer matrix $a_{0,x}^{\mcirc \mcirc}$, as the proof for the vertical one, $c_{\du,x}^{\mcirc \mcirc}$, is totally analogous. Let us begin by introducing the transfer matrix 
\be
    a_{\du,x}^{\mcirc\mcircf} = \begin{tikzpicture}[baseline={([yshift=-0.5ex] current  bounding  box.center)}, scale=0.75]
\foreach \j in {1,2}{
\draw[rotate around ={45: (-6+1.5*\j,0.5)}] (-.5-6+1.5*\j,0) -- (0.5-6+1.5*\j,1);
\draw[rotate around ={45: (-6+1.5*\j,0.5)}] (-5.4+1.5*\j,-0.1) -- (-6.6+1.5*\j,1.1);
\draw[thick, fill=myorange, rounded corners=2pt, rotate around ={45: (-6+1.5*\j,0.5)}] (-0.25-6+1.5*\j,0.25) rectangle (.25-6+1.5*\j,0.75);
\draw[thick, rotate around ={-135: (-6+1.5*\j,0.5)}] (-6+1.5*\j,0.65) -- (-5.85+1.5*\j,0.65) -- (-5.85+1.5*\j,0.5);
}
\Text[x=-1.55,y=0.5]{$\cdots$}
\foreach \j in {4,5}{
\draw[rotate around ={45: (-6+1.5*\j,0.5)}] (-.5-6+1.5*\j,0) -- (0.5-6+1.5*\j,1);
\draw[rotate around ={45: (-6+1.5*\j,0.5)}] (-5.4+1.5*\j,-0.1) -- (-6.6+1.5*\j,1.1);
\draw[thick, fill=myorange, rounded corners=2pt, rotate around ={45: (-6+1.5*\j,0.5)}] (-0.25-6+1.5*\j,0.25) rectangle (.25-6+1.5*\j,0.75);
\draw[thick, rotate around ={-135: (-6+1.5*\j,0.5)}] (-6+1.5*\j,0.65) -- (-5.85+1.5*\j,0.65) -- (-5.85+1.5*\j,0.5);
}
\draw[thick, fill=white] (-5.25,0.5) circle (0.1cm);
\draw[thick, fill=black] (2.25,0.5) circle (0.1cm);
\draw [thick, decorate, decoration={brace,amplitude=8pt, raise=4pt, mirror},yshift=0pt]
(-4.75,-0.1) -- (1.75,-0.1) node [black,midway,yshift=-0.95cm, anchor = south] {$x$};
\Text[x=2.3,y=2.2]{}
\end{tikzpicture},
\label{eq:DUtransferMatccf}
\ee
and noting that $\{a_{\du,x}^{\mcirc \mcirc},a_{\du,x}^{\mcirc \mcircf}\}$ fulfil the following recursive relations 
\begin{align}
a_{\du,x}^{\mcirc\mcirc} &=  a_{\du,x-1}^{\mcirc \mcirc}  \otimes \begin{pmatrix} 1 & 0\\ 0& 0 \end{pmatrix} + 
     a_{\du, x-1}^{\mcirc \mcirc} \otimes \begin{pmatrix} 0 & 0\\0& a \end{pmatrix}+  a_{\du, x-1}^{\mcirc \mcircf} \otimes \begin{pmatrix} 0 & 0\\0& b \end{pmatrix},
    \label{eq:rec1}
    \\
    a_{\du,x}^{\mcirc \mcircf} &=a_{\du,x-1}^{\mcirc \mcircf} \otimes  \begin{pmatrix} c & e\\d& g \end{pmatrix} + 
     a_{\du, x-1}^{\mcirc \mcirc} \otimes \begin{pmatrix} 0 & 0\\0& f \end{pmatrix},
    \label{eq:rec2}
\end{align}
with 
\be
  a_{\du,1}^{\mcirc\mcirc} = \begin{pmatrix} 1 & 0\\ 0& a \end{pmatrix}\,,\qquad
  a_{\du,1}^{\mcirc \mcircf} = \begin{pmatrix} 0 & 0\\0& f \end{pmatrix}\,.
\ee
These relations are directly established by plugging a resolution of the identity $\1 = \ket{\mcirc}\!\!\bra{\mcirc}+  \ket{\mcircf}\!\!\bra{\mcircf}$ in the right-most connecting wire of \eqref{eq:DUtransferMatcc} and \eqref{eq:DUtransferMatccf}. 

Using unitality \eqref{eq:unitality} and dual unitality~\eqref{eq:dualunitality}, it is straightforward to prove that $\ket{{\mcirc\ldots\mcirc}}$ is an eigenvector of $a_{\du,x}^{\mcirc\mcirc}$ with the eigenvalue $1$ and 
\be
\bigl\{ \ket{\underbrace{\mcirc\ldots\mcirc\mcircf}_k\mcirc\ldots\mcirc}\bigr\}_{k=1}^{x}
\ee 
are $x$ eigenvectors of $a_{\du,x}^{\mcirc\mcirc}$ corresponding to the eigenvalue $a$. This immediately proves the block structure of \eqref{eq:specdec1}. Therefore, to conclude the proof we just need to bound the norm of 
\be
r_{1,x}= a_{\du,x}^{\mcirc\mcirc} - p^{\mcirc}_{x,0} - a  \sum_{k=1}^x  p^{\mcirc}_{x,k}\,, 
\ee
where the projectors $\{p^{\mcirc}_{x,k}\}_{k=0}^x$ are defined in Eq.~\eqref{eq:proj}. We begin by considering  
\be
\lambda_{1,x} := \|r_{1,x} - p^{\mcirc}_{x,0}\|,
\ee
where $\|A\|$ denotes the operator norm of matrix $A$. Using \eqref{eq:rec1} leads to 
\be
 \lambda_{1,x} = {\rm max}(\lambda_{1,x-1},\| a a_{\du,x-1}^{\mcirc\mcirc} + b  a_{\du,x-1}^{\mcirc\mcircf}\|)  = {\rm max}(\lambda_{1,x-1}, |a|, \| a a_{\du,x-1}^{\mcirc\mcirc}-a p^{\mcirc}_{x-1,0} + b  a_{\du,x-1}^{\mcirc\mcircf}\|) \,,
 \label{eq:lambdax}
\ee 
where in the second step we used that $\ket{{\mcirc\ldots\mcirc}}$ is an eigenvector of $a_{\du,x-1}^{\mcirc\mcircf}$ (with eigenvalue 0). Using the triangle inequality for the operator norm on the r.h.s., we obtain 
 \be
 \| a a_{\du,x-1}^{\mcirc\mcirc}-a p^{\mcirc}_{x-1,0} + b  a_{\du,x-1}^{\mcirc\mcircf}\| \leq |a| \lambda_{1,x-1} + |b| \lambda_{x-1}^{\mcircf}\,,
\ee
where we defined
\be
\lambda_{x-1}^{\mcircf} := \|a_{\du,x}^{\mcirc\mcircf}\|\,.
\ee
Thus, applying triangle inequality to \eqref{eq:rec2}, we find
\begin{align}
     \lambda_{x}^\bullet  \leq  \alpha \lambda_{x-1}^\bullet + |f|\,,
     \label{eq:rec_bullet}
\end{align}
where we used that $\| a_{\du,x}^{\mcirc\mcirc}\|=1$ (this can be proven by the same reasoning as in Appendix A of Ref.~\cite{BKP:OEergodicandmixing}). Furthermore, we denoted by $\alpha$ the operator norm of the 2 by 2 matrix in the first term on the r.h.s. of \eqref{eq:rec2}. Explicitly, $\alpha$ can be computed by considering the largest singular value of the matrix 
which reads
\be
\alpha = \sqrt{\frac{c^2+e^2+g^2+d^2+\sqrt{(c^2+e^2-g^2-d^2)^2+4(cd +eg)^2 }}{2}}. 
\ee
By means of the parameterisations \eqref{eq:deltaanglesDU}--\eqref{eq:ganglesDU} and \eqref{eq:deltabistoDU}--\eqref{eq:gbistoDU}, one can explicitly verify that $\alpha\in[0,1]$. 

Using \eqref{eq:rec_bullet}, we see that $\lambda_{x}^\bullet$ is bounded by $y_x$, which is defined as the solution of
\begin{align}
    y_x= \alpha y_{x-1} + |f|, \qquad\qquad y_1 = |f|,
\end{align}
and reads explicitly 
\begin{align}
    y_x = \frac{|f|}{1-\alpha} (1 -  \alpha^x).
\end{align}
Hence,
\be
\| a  a_{\du,x-1}^{\mcirc\mcirc}-a P^{\mcirc}_{x-1,0} + b   a_{\du,x-1}^{\mcirc\mcircf}\| \leq  |a| \lambda_{x-1} + \frac{|b f|}{1-\alpha} (1 -  \alpha^{x-1}) )\leq |a| \lambda_{x-1} + \frac{|b f|}{1-\alpha}\,,
\label{eq:rholincombp}
\ee
where in the last step we assumed $\alpha<1$. Combining everything we get
\begin{align}
     \lambda_{1,x} \leq {\rm max}(\lambda_{1,x-1}, |a|, |a| \lambda_{1,x-1} + \frac{|b f|}{1-\alpha}).
     \label{eq:bound}
\end{align}
Considering now the range of parameters for which   
\be
 a^2 + \frac{|b f|}{1-\alpha} < |a|\,,
\label{eq:inequalityForc}
\ee
we have that for $\lambda_{1,x-1}\leq |a|$, it follows $\lambda_{1,x}\leq |a|$. In particular, since $\lambda_{1,1}= |a|$, we conclude 
\be
\lambda_{1,x} \leq |a| \qquad\qquad\forall x\,.
\ee
Moreover, from \eqref{eq:rholincombp} we conclude 
\be
\| a a_{\du,x-1}^{\mcirc\mcirc}-a p^{\mcirc}_{x-1,0} + b a_{\du,x-1}^{\mcirc\mcircf}\| \leq a^2  + \frac{|b f|}{1-\alpha},
\label{eq:usefulbound}
\ee
which combined with \eqref{eq:lambdax} implies 
\be
\lambda_{1,x} = |a| \qquad\qquad\forall x\,.
\ee 
Now we define 
\be
\lambda_{2,x}:= \|r_{1,x}\|\,,
\ee
and using \eqref{eq:rec1} we find 
\be
 \lambda_{2,x} = {\rm max}(\lambda_{2,x-1}, \| a a_{\du,x-1}^{\mcirc\mcirc}-a p^{\mcirc}_{x-1,0} + b  a_{\du,x-1}^{\mcirc\mcircf}\|) \,.
\ee
At this point we use the bound \eqref{eq:usefulbound}, which combined with $\lambda_{2,1}=0$ gives 
\be
\lambda_{2,x} \leq a^2  + \frac{|b f|}{1-\alpha}. 
\ee
This concludes the proof of the Property~\ref{prop:P2}.

Note that the bound \eqref{eq:bound} is useful even when the condition \eqref{eq:inequalityForc} is violated. Indeed, assuming $|a|<\lambda_x<1$, we have 
\be
  \lambda_{x+1} \leq {\rm max}(\lambda_{x},  |a| \lambda_{x} + \frac{|b f|}{1-\alpha}).
\ee
Since ${|a|<1}$, this equation implies ${\lambda_{x}<m^*}$ for all $x$ and some finite $m^*$. Whenever $m^*<1$, this produces a useful bound for the first non-trivial eigenvalue of $a^{\mcirc\mcirc}_{\du,x}$. 

\section{Evaluation of \eqref{eq:generalSkeletons}}
\label{sec:evageneralSkeletons}

To evaluate the Eq.~\eqref{eq:generalSkeletons}, one first needs to construct all lists 
\be
\{ l_1^+, \dots, l_{n+1}^+ \};\quad l_j^+ \geq 0,
\ee
and  
\be
\{ l_1^-, \dots, l_{n}^- \};\quad l_j^- \geq 0,
\ee
that satisfy the constraint \eqref{eq:generalConstraints}. This is done with a recursive algorithm. At the step $k$, we generate all lists 
\be
\{ p_1, \dots p_k \};\quad  p_j \geq 0,
\ee
that satisfy 
\be
\sum_{j=1}^k p_j = n_{k}.
\ee
If we are generating a list with a single element ($k=1$), this element is set by the constraint $p_1 = n_1$. Otherwise, we generate lists by choosing all possible values for the last element in $p_k \in [ 0, \dots, n_{k}]$ and solving the problem of finding  all lists $\{ p_1, \dots, p_{k-1} \}$ with the constraint   
\be
\sum_{j=1}^{k-1} p_j = n_{k}-p_k =n_{k-1}.
\ee
After all the lists are generated, we can evaluate the Eq.~\eqref{eq:generalSkeletons}. 

For large $x_\pm$ and $n$ the above procedure becomes difficult to implement. For instance, consider the $n=n_0$ in the first sum of Eq.~\eqref{eq:generalSkeletons}: this term involves the summation of  
\be
\notag
\binom{x_{+}}{n_0} \binom{x_{-} -2}{n_0-1} \approx \left (\frac{x_+ x_-}{n_0}\right)^{n_0}
\ee
contributions, which becomes eventually impractical for $x_\pm,n_0\gg 1$. This problem can be overcome by (i) truncating the expansion \eqref{eq:generalSkeletons} (this leads to accurate results for small enough $\eta$) or (ii) by approximating the higher orders by substituting $A^{\mcirc\mcirc}_{0,1}$ and $C^{\mcirc\mcirc}_{0,1}$ with the projectors on their largest non-trivial eigenvalues, which we respectively denote by $\lambda_{1,a}$ and $\lambda_{1,c}$. Thus we can rewrite the contribution of the higher orders in \eqref{eq:generalSkeletons} exactly in the form Eq.~\eqref{eq:skeletoncontribution}, with the sums running from a certain $n=n_c$ ($n=n_c-1$) instead of 1 (0). The parameters are expressed as  
\be
p \varepsilon = \bra{\lambda_{1,c}}\mathcal{E}_{1} \ket{\lambda_{1,a}},\qquad q \varepsilon = \bra{\lambda_{1,a}}\mathcal{E}_{2} \ket{\lambda_{1,c}},\qquad a = \lambda_{1,a},\qquad c=\lambda_{1,c}, 
\ee
and there is an additional factor of $\braket{b|\lambda_{1,a}}\braket{\lambda_{1,a}|a}$ ($\braket{b|\lambda_{1,c}}\braket{\lambda_{1,a}|a}$) in front of it. 
[In the latter expression, $\ket{a}$ and $\bra{b}$ refer to initial and final operators, whereas  $\ket{\lambda_{1,a}}$  and  $\ket{\lambda_{1,c}}$ refer to the leading eigenvectors ($a$ and $c$ refer to the corresponding parameters of the reduced gate).] If all $l^\pm_j\gg 1$ this replacement would give a negligible error, however, in the sum there are also terms with $l^\pm_j\approx 0$ and using the projector is strictly speaking unjustified. Nevertheless, this gives a useful approximation, especially in the limit $x_\pm \to \infty$.

\section{Numerical Methods and Parameters of the Gates}
\label{sec:numerics}

Direct (brute force) numerical evaluations have been performed by exactly contracting the $x_+\times x_-$ tensor network \eqref{eq:pictorialcorrelation}, with the help of some basic functionalities of ITensor Library~\cite{ITensor}. We represented the green gates as ITensor objects with the correct indices.
To contract \eqref{eq:pictorialcorrelation}, we started on the right side and contract the initial state by gates in the last column one by one (first and last gates in the column are also contracted with the $\mcirc$ state). Then we continued in the same fashion with the remaining columns, and end up with contracting with the final state.

The leading part of the spectrum of the dual-unitary transfer matrices was found using the power method, i.e., by iteration of the dual transfer matrix on a random initial vector, which is chosen to be (bi)orthogonal to the previously computed leading (left and right) eigenvectors.  We circumvent the need to apply the adjoint dual-unitary transfer matrix, by noticing that for the leading eigenvectors of non-tirivial support up to one, left and right eigenvectors coincide.

The parameters of the gates used in numerical experiments are given in the Tables \ref{tab:U1gates} and \ref{tab:NonAvgGate}.

\begin{table}[h]
\centering
\begin {tabular} {c | ccccccc}
 & $a$ & $b$ & $c$ & $d$ & $e$ & $f$ & $g$ \\
\hline
 \text {Gate 1} & 0.820188 & - 0.0136728 & 0.820188 & - 0.0136728 & - 
     0.0136728 & - 0.0136728 & 0.679158 \\
 \text {Gate 2} & 0.761132 & - 0.025732 & 0.761132 & - 0.025732 & - 
     0.025732 & - 0.025732 & 0.701678 \\
\text {Gate 3} & 0.0945626 & 0.0212368 & 0.195892 & 0.00603479 & \
0.0000783504 & 0.271196 & 0.443805 \\
\text {Gate 4} & - 0.714396 & - 0.0143272 & - 0.174808 & 0.139193 & - 
     0.206479 & 0.0273134 & 0.0485487 \\
\text {Gate 5} & 0.0409352 & 0.275939 & 0.0409352 & 0.275939 & \
0.275939 & 0.275939 & 0.0373261 \\
  \end {tabular}
  \caption{Parameters of the reduced gates, which were used in the figures. In the cases where $\varepsilon$ is not stated, we used $\varepsilon = -0.002660, -0.014746, 0.001$ for gates 1,2,3 respectively. We used $p=q=1$ (cf. \eqref{eq:split2}).
  }
  \label{tab:U1gates}
  \end{table}

\begin{table}[h]
\centering
\begin {tabular} {c | cc}
\text{Eigenvectors of} \\
\text{ $\lambda_1$ fulfill (\ref{eq:support1ev})} & $u$ & $v$\\
\hline
\text {Yes} & $\begin{pmatrix} -0.229466-0.562418 i & 0.507409\, -0.611202 i \\ 0.316533\, +0.728586 i & 0.377401\, -0.475959 i \end{pmatrix}$
& $\begin{pmatrix} -0.500163-0.421942 i & -0.0342946-0.755398 i \\  0.532173\, -0.537209 i & -0.653978-0.0226033 i  \end{pmatrix}$\\
\\
\text {No} & $\begin{pmatrix} -0.484203+0.476337 i & 0.207991\, +0.70384 i \\ -0.194545+0.707674 i & -0.493189-0.467026 i \\\end{pmatrix}$
& $\begin{pmatrix} 0.488082\, +0.160083 i & 0.0340135\, +0.857317 i \\ 0.197049\, +0.835058 i & 0.427305\, -0.285063 i \\  \end{pmatrix}$
\end{tabular}\\
  \caption{Parameters of the non-averaged gates, which were used in the figures.
  The gates are parametrized as $ (u \otimes u) V[\{J_i\}] (v \otimes v)$, with $V[\{J_i\}]$ defined in Eq.~\eqref{eq:gateV}. The parameters are set to $J_1=J_2=\eta + \pi/4$ and $J_3=0.1$. 
  }
  \label{tab:NonAvgGate}
  \end{table}

\ew


\begin{thebibliography}{99}
\bibitem{note} Note that in the cases (i) and (ii) only the first term on the r.h.s. contributes in the first line. Moreover, the second line is trivial in the case (ii): to find a non trivial integer-half-odd-integer correlation one has to look at half-odd-integer starting points in that case.
\end{thebibliography}
\end{document}